\documentclass[useAMS,usenatbib]{mn2e}
\usepackage{epsf}
\usepackage{epsfig}
\newcommand{\aap}{A\&A}
\newcommand{\apj}{ApJ}

\newcommand{\apjs}{ApJS}

\newcommand{\agn}{\textsc{agn}}

\newcommand{\planck}{\textit{Planck}}
\newcommand{\wmap}{\textit{WMAP}}
\newcommand{\calsim}{\textsc{bahamas}}
\newcommand{\mac}{\textsc{MACSIS}}

\def\mean#1{\left< #1 \right>}

\title[{}]{The BAHAMAS project: Calibrated hydrodynamical simulations for large-scale structure cosmology}
\author[I.~G.~McCarthy et~al.]{Ian G.~McCarthy$^1$\thanks{E-mail:i.g.mccarthy@ljmu.ac.uk}, Joop Schaye$^2$, Simeon Bird$^3$, Amandine M.~C~Le Brun$^4$ \\
$^{1}$Astrophysics Research Institute, Liverpool John Moores University, 146 Brownlow Hill, Liverpool L3 5RF\\
$^{2}$Leiden Observatory, Leiden University, P. O. Box 9513, 2300 RA Leiden, the Netherlands\\
$^{3}$Department of Physics and Astronomy, Johns Hopkins University, 3400 N.~Charles Street, Baltimore, MD 21218, USA\\
$^{4}$Laboratoire AIM, IRFU/Service d'Astrophysique -- CEA/DRF -- CNRS -- Universit\'e Paris Diderot, B\^at. 709, CEA-Saclay,\\ 91191 Gif-sur-Yvette Cedex, France}

\begin{document}

\date{Accepted ... Received ...}

\pagerange{\pageref{firstpage}--\pageref{lastpage}} \pubyear{2016}

\maketitle

\label{firstpage}

\begin{abstract}
The evolution of the large-scale distribution of matter is sensitive to a variety of fundamental parameters that characterise the dark matter, dark energy, and other aspects of our cosmological framework.  Since the majority of the mass density is in the form of dark matter that cannot be directly observed, to do cosmology with large-scale structure one must use observable (baryonic) quantities that trace the underlying matter distribution in a (hopefully) predictable way.  However, recent numerical studies have demonstrated that the mapping between observable and total mass, as well as the total mass itself, are sensitive to unresolved feedback processes associated with galaxy formation, motivating explicit calibration of the feedback efficiencies.  Here we construct a new suite of large-volume cosmological hydrodynamical simulations (called BAHAMAS, for BAryons and HAloes of MAssive Systems) where subgrid models of stellar and Active Galactic Nucleus (AGN) feedback have been calibrated to reproduce the present-day galaxy stellar mass function and the hot gas mass fractions of groups and clusters in order to ensure the effects of feedback on the overall matter distribution are broadly correct.  We show that the calibrated simulations reproduce an unprecedentedly wide range of properties of massive systems, including the various observed mappings between galaxies, hot gas, total mass, and black holes, and represent a significant advance in our ability to mitigate the primary systematic uncertainty in most present large-scale structure tests.
\end{abstract}


\section{Introduction}

The evolution of the large-scale distribution of matter is highly sensitive to a variety of fundamental cosmological parameters that control the growth rate of structure, such as the total matter density ($\Omega_m$), the amplitude ($\sigma_8$) and spectral index ($n_s$) of density fluctuations, and the evolution of dark energy.  However, since the majority of the mass density is in the form of dark matter, it is not directly observable and one must instead use observable (baryonic) quantities that trace the underlying matter distribution in some fashion.  A wide variety of such indirect probes of the matter distribution have been proposed over the years, including measurements of the Lyman-alpha forest, galaxy cluster counts, the thermal Sunyaev-Zel'dovich (SZ) effect, weak gravitational lensing of galaxies (cosmic shear) and of the cosmic microwave background (CMB lensing), galaxy clustering, and redshift space distortions.  The past few years have seen major advances in the precision with which measurements of these large-scale structure (LSS) probes are being made.

With the quality and quantity of observations of LSS rapidly improving, some interesting (possible) tensions between the analysis of the CMB and different low-redshift LSS tests have recently arisen (e.g., \citealt{Beutler2014,Planck2015a,Planck2015b}).  While there may still be relevant observational uncertainties at play (e.g., \citealt{Addison2016}), increased focus is being placed on the degree of precision with which the various LSS signatures (e.g., the predicted galaxy cluster number counts, the cosmic shear shape correlation functions) can be theoretically predicted for a given cosmology.  

Most LSS tests probe into the non-linear regime and therefore require detailed cosmological simulations to help calibrate the theoretical modelling.  The employed cosmological simulations usually only model collisionless matter.  However, it is becoming increasingly clear that to obtain precise predictions, one must model not only the dark matter but also the baryons, since they form a non-negligible fraction of the mass density.  In particular, energetic feedback processes associated with star formation and black hole growth can heat and expel gas from collapsed structures (e.g., \citealt{McCarthy2011}) and modify the large-scale distribution of matter (e.g., \citealt{vanDaalen2011,vanDaalen2014,Velliscig2014}).  Note that the extent of the effect is not simply that some fraction of the baryons are removed; there is also a corresponding large-scale expansion (or `back reaction') of the dark matter and a slowing of the accretion rate of new matter (e.g., \citealt{vanDaalen2011,Sawala2013}).

Cosmological hydrodynamical simulations are the only method which can follow all the relevant matter components and self-consistently capture the effects of feedback.  However, such simulations have had a notoriously difficult time in reproducing key observables, such as the galaxy stellar mass function.  In the context of LSS cosmology, obtaining the correct total baryon fraction (stars+gas, noting that hot gas normally dominates the baryon budget of massive systems) is arguably even more important, since this is a `zeroth order' requirement for ensuring that the feedback effects on the matter distribution are of the correct magnitude (e.g., \citealt{Semboloni2011}).  

A number of recent simulation studies have highlighted the sensitivity of the galaxy formation efficiency to the details of the subgrid prescriptions for feedback, particularly stellar feedback (e.g., \citealt{Schaye2010,Scannapieco2012,Haas2013,Puchwein2013,Vogelsberger2013,Crain2015}), while the OverWhelmingly Large Simulations (OWLS) project of \citet{Schaye2010} and cosmo-OWLS, its extension to larger volumes \citep{LeBrun2014,McCarthy2014}, have shown that a similar predicament holds for the gas content of massive dark matter haloes (groups and clusters).  On large scales and for massive haloes, the sensitivity is tied more closely to the details of the AGN feedback as opposed to that of stellar feedback (see \citealt{McCarthy2011,LeBrun2014}).  

This lack of ab initio predictive power for the stellar and gaseous fractions of collapsed systems means that in general the subgrid models for feedback must be calibrated in order to reproduce these observations\footnote{A more long-term and ultimately more desirable path is to simulate the feedback physics directly, and thus rid ourselves of the degrees of freedom in current subgrid models.  However, the dynamic range required to do this is still far too demanding at present, particularly in the context of cosmological simulations needed for the study of large-scale structure.} \citep{Schaye2015}.   Assuming this can be achieved, the realism of the model may then be tested by looking at other independent observables and at trends with redshift, environment, etc.  At present, however, we are unaware of any self-consistent cosmological hydrodynamical simulations that {\it simultaneously} reproduce the stellar and hot gas content for a representative population that spans the full range of massive haloes ($M_{\rm halo} \sim 10^{12-15}$ M$_\odot$).  [Note that some of the (cosmo)OWLS simulations approximately reproduced the observed gas fractions and stellar masses but for a smaller dynamic range.]  Constructing a simple model which can achieve this, while passing a variety of other important independent tests, is the primary aim of the present study.

In particular, we construct a new set of large-volume (400 Mpc/$h$ on a side) cosmological hydrodynamical simulations that may be used to aid the cosmological interpretation of LSS tests, building on previous work from the OWLS/cosmo-OWLS projects.  Specifically, those studies explored the effects of systematically varying the important parameters of the subgrid feedback models on the stellar and hot gas properties of haloes.  As already discussed, one arrives at the inevitable conclusion that the feedback efficiency(ies) in current simulations must be {\it calibrated} to reproduce the observed stellar and hot gas content.  Our new simulation program, dubbed \calsim~(for BAryons and HAloes of MAssive Systems), takes exactly this route.  Specifically, we calibrate a simple subgrid model of feedback to reproduce the hot gas mass fractions of local massive haloes (over the range $\log_{10}[M_{500}/{\rm M}_\odot] = 13.0-15.0$), the present-day galaxy stellar mass function (over the range $\log_{10}[M_*/{\rm M}_\odot] = 10.0-12.0$), and the amplitude of the $z=0$ stellar mass$-$black hole mass relation.  We then explore the realism of the model by comparing the predictions of the simulations to a wide range of observables over a large range of mass, spatial, and time scales.  We show that the simulations do a remarkable job at capturing the properties of massive systems.

The paper is organized as follows.  In Section 2 we describe the simulations and our calibration strategy.  In Section 3 we examine the predictions of the calibrated model for the relation between present-day galaxies and their host dark matter haloes, including the separate contributions of centrals and satellites, the spatial distributions of stellar mass, and the dynamics of satellite population in massive haloes.  Section 4 examines the evolution of basic galaxy properties, including the galaxy stellar mass function and the star formation rates.  Section 5 explores the hot gas properties of groups and clusters, including the integrated and radial X-ray and SZ effect properties.  Section 6 explores the galaxy$-$hot gas connection, comparing the simulations to recent X-ray, SZ, and weak lensing stacking analyses of massive galaxies.  Section 7 examines black hole and quasar properties.  Finally, in Section 8 we summarize our results.

For consistency with the simulations, we adopt a flat $\Lambda$CDM cosmology with WMAP 9-year based cosmological parameters throughout and halo masses are specified in M$_\odot$ (not $h^{-1} \ {\rm M}_\odot$).

\section{Simulations}

\subsection{Simulation characteristics}

As we are interested in the properties of massive dark matter haloes (corresponding to massive galaxies and groups and clusters of galaxies), large volumes are required in order to simulate representative populations.  Following cosmo-OWLS, our production runs consist of 400 Mpc/$h$ on a side periodic boxes.  We use updated initial conditions based on the maximum-likelihood cosmological parameters derived from the \wmap~9-year data \citep{Hinshaw2013} \{$\Omega_{m}$, $\Omega_{b}$, $\Omega_{\Lambda}$, $\sigma_{8}$, $n_{s}$, $h$\} = \{0.2793, 0.0463, 0.7207, 0.821, 0.972, 0.700\} and the \planck~2013 data \citep{Planck2014} = \{0.3175, 0.0490, 0.6825, 0.834, 0.9624, 0.6711\}.  We use the Boltzmann code {\small CAMB}\footnote{http://camb.info/} (\citealt{Lewis2000}; April 2014 version) to compute the transfer function and a modified version of V. Springel's software package {\small N-GenIC}\footnote{http://www.mpa-garching.mpg.de/gadget/} to make the initial conditions, at a starting redshift of $z=127$. {\small N-GenIC} has been modified by S.\ Bird to include second-order Lagrangian Perturbation Theory (2LPT) corrections and support for massive neutrinos\footnote{https://github.com/sbird/S-GenIC} (which we will consider in future studies).  We will only present the results of the \wmap~runs here, but we will comment on any significant differences in the corresponding \planck~runs.

The full production runs presented here have $2\times1024^{3}$ particles, yielding dark matter and (initial) baryon particle masses of $\approx3.85\times10^{9}~h^{-1}~\textrm{M}_{\odot}$ ($4.45\times10^{9}~h^{-1}~\textrm{M}_{\odot}$) and $\approx7.66\times10^{8}~h^{-1}~\textrm{M}_{\odot}$ ($8.12\times10^{8}~h^{-1}~\textrm{M}_{\odot}$), respectively for a \wmap-9~(\planck) cosmology.   The gravitational softening length is fixed to 4 kpc/$h$ in physical coordinates below $z=3$ and fixed in comoving coordinates at higher redshifts.

As the hydrodynamic code and the subgrid physics prescriptions used here have not been modified from those used previously for the OWLS and cosmo-OWLS projects, and are described in detail in previous papers, we present only a brief description below.  Note that while the basic subgrid modules have not been modified, we adopt different feedback parameter values here in order to calibrate the simulations to reproduce the stellar and hot gas content of dark matter haloes.  Our calibration strategy is described in Section 2.2.

The simulations were carried out with a version of the Lagrangian TreePM-SPH code \textsc{gadget-3} \citep[last described in][]{Springel2005b}, which has been extended to include new `subgrid' physics.  Radiative cooling/heating rates are computed element by element following \citet{Wiersma2009a}, interpolating the rates as a function of density, temperature and redshift from pre-computed tables generated with \textsc{cloudy} \citep[last described in][]{Ferland1998}.  The rates account for heating/cooling due to the primary CMB and a \cite{Haardt2001} ultra-violet/X-ray photoionizing background.  `Reionization' is modelled by simply switching on the background at $z=9$. Star formation (SF) is implemented following the prescription of \cite{Schaye2008}. The simulations lack the resolution and detailed physics to directly follow the cold interstellar medium (ISM), so an effective equation of state is imposed with $P\propto \rho^{4/3}$ for gas with $n_{H}>n_{H}^{*}$ where $n_{H}^{*}=0.1~\textrm{cm}^{-3}$.  Gas exceeding this density threshold is available for star formation (implemented stochastically), at a pressure-dependent rate that reproduces the observed Kennicutt-Schmidt law by construction \citep[see][]{Schaye2008}.  Stellar evolution and chemical enrichment are implemented using the model of \cite{Wiersma2009b}, which computes the timed-release of 11 elements (H, He, C, N, O, Ne, Mg, Si, S, Ca and Fe; i.e., all of the important ones for radiative cooling losses) due to Type Ia and Type II supernovae (SNe) and winds from massive stars and Asymptotic Giant Branch (AGB) stars.

Feedback from star formation is implemented using the kinetic wind model of \citet{DallaVecchia2008}.  In this model, neighbouring gas particles are given a velocity kick.  Note that kicked particles are never hydrodynamically `decoupled' the from surrounding gas.  Hence, there is the potential for the entrainment
of a large fraction of the gas surrounding the wind directly kicked particles.  Previous OWLS/cosmo-OWLS runs adopted a mass-loading factor $\eta_{w}=2$ and a wind velocity ~$v_{w}=600~\textrm{km/s}$ by default, corresponding to using approximately 40\% of the energy available from Type II supernovae assuming a \citet{Chabrier2003} IMF (which the simulations adopt).  This results in simulations {\it that neglected AGN feedback} approximately reproducing the peak of the observed cosmic star formation history at $z\sim1-3$ \citep{Schaye2010}.  As we will show below, leaving these parameter values fixed while including the effects of AGN feedback, which appears necessary to reproduce the hot gas properties of groups/clusters \citep{McCarthy2010}, results in lower-than-observed galaxy formation efficiencies for haloes with masses similar to the Milky Way's ($M_{200}\sim10^{12}$ M$_\odot$).  This is not completely unexpected, as Schaye et al.\ already showed that the inclusion of AGN feedback, while leaving the star formation feedback parameter values fixed, results in a cosmic star formation history that lies below the observed trend (see their Fig.\ 18).   Note, however, that the galaxy formation efficiency is quite sensitive to the adopted parameter values for stellar feedback; even variations in the wind mass-loading and wind velocity at fixed energy (i.e., $\eta_{w} v_{w}^2 =$ constant) can have a significant effect on the star formation histories of galaxies (see Fig.\ 14 of \citealt{Schaye2010} and Fig.\ 4 of \citealt{Haas2013}).  (This sensitivity is not so surprising, because changing the wind velocity changes how much material can escape the halo, and thus the amount of re-accretion.)  Therefore, it should be possible, at least in principle, to reproduce the galaxy formation efficiency in the presence of AGN feedback, by lowering the efficiency of stellar feedback.  We discuss this possibility further below.

Accretion onto and mergers of supermassive black holes (SMBHs) and the resulting AGN feedback is implemented using the subgrid prescription of \citet{Booth2009}, which is a modified version of the model of \citet{Springel2005a}.  Here we describe the main parameters of the model and guidance we have taken in setting their values.

At present we do not have a detailed theoretical picture of how the first massive BHs formed.  Most current models of SMBHs implemented in cosmological simulations therefore bypass this issue and simply inject BH `seed mass' particles into dark matter haloes (identified using a standard Friends-of-Friends, or FoF, algorithm) on-the-fly during the simulation, as originally employed by \citet{Springel2005a}.  Naively, specifying exactly how this is done may not seem particularly relevant as the BHs generally have negligible dynamical impact on galaxies and dark matter haloes.  However, the feedback they induce can (and generally will) have profound effects, particularly on the observable baryonic component.  Therefore, rough guidance from observations (while also taking numerical limitations into consideration) is sometimes taken to specify a minimum halo mass, or alternatively a minimum galaxy stellar mass, into which BH seeds are placed.  For example, it may be desirable to have the simulations roughly reproduce the break in the galaxy stellar mass function, in which case the BHs could be injected somewhat below the mass scale\footnote{The BH seeds can in fact be placed in much lower mass haloes, so long as the accretion model ensures inefficient growth at low halo masses (e.g., \citealt{Schaye2015}).}  corresponding to $M^*$ (i.e., $M_{\rm halo} \sim 10^{12}$ M$_\odot$).  Given that our simulations are of relatively low resolution, we cannot `resolve' haloes much lower than this mass in any case.   Here we follow \citet{Booth2009} and inject BH seed particles in FoF groups with at least 100 DM particles (corresponding to a FoF halo mass of $\sim5\times10^{11}$ M$_\odot$) but show in Appendix A the effects of increasing the minimum halo mass for BH seed injection.  

Once seeded\footnote{Note that early on in their evolution, when the BH particle mass is similar to (or smaller than) the simulation mass resolution, the BH will not dominate the local dynamics and could potentially wander from the centre of its parent halo.  In order to avoid this, \citet{Booth2009} follow the prescription of \citet{Springel2005a}, which calculates the potential energies of all of the gas particles within the vicinity of the BH and repositions the BH on top of the gas particle with the minimum potential energy (see \citealt{Booth2009} for details).  This repositioning process is halted after the mass of the BH particle exceeds 10 times the initial gas particle mass in the simulation, as the BH dominates the local potential after this point.}, BHs grow via Eddington-limited gas accretion, at a rate which is proportional to the Bondi-Hoyle-Lyttleton rate, as well as through mergers with other BHs. As the simulations do not directly model the cold ISM, they will generally underestimate the accretion rate onto the BH by a large factor.  Springel et al. and many subsequent studies that have adopted this model scaled the Bondi rate up by a constant factor $\alpha \approx 100$.  In the \citet{Booth2009} model, however,  $\alpha$ is a power-law function of the local density for gas above the SF threshold, $n_{H}^{*}$. The power-law exponent $\beta$ is set to 2 and the power-law is normalised so that $\alpha=1$ for densities equal to the SF threshold, so that at low densities, which can be resolved and for which no cold phase is expected, the accretion rate is the unmodified Bondi rate.  We use the \citet{Booth2009} model by default, but explore the effects of varying the accretion rate boost factor later.

Following \citet{Booth2009}, a fraction, $\epsilon$, of the rest mass energy of accreted gas is used to heat a number ($n_{\rm heat}$) of neighbouring gas particles, by increasing their temperature by a pre-specified level, $\Delta T_{\rm heat}$.  BHs store `accretion energy' in a reservoir until it is sufficiently large to heat the $n_{\rm heat}$ particles by $\Delta T_{\rm heat}$.  These two parameters are chosen to broadly ensure that the heated gas has a sufficiently long cooling time (and therefore does not strongly suffer from artificial numerical radiative cooling losses due to poor mass resolution; see, e.g., \citealt{DallaVecchia2012}) and that the time needed to have a feedback event is shorter than the Salpeter time for Eddington-limited accretion.  However, these criteria alone do not precisely pin down the heating temperature or heated gas mass and, as shown by \citet{LeBrun2014} (hereafter L14; see also \citealt{Schaye2015}), even relatively minor changes in $\Delta T_{\rm heat}$ can have a significant impact on the hot gas properties (particularly the gas mass fraction and quantities that depend on it, such as the X-ray luminosity and thermal Sunyaev-Zel'dovich flux) of groups and clusters.  Therefore, calibration of the heating temperature (and to a far lesser extent of the heated gas mass) is required to reproduce the observed hot gas content.  For reference, \cite{Booth2009} adopted $\Delta T_{\rm heat}=10^{8}$ K and $n_{\rm heat}=1$ which \citet{McCarthy2010} and L14 later showed does a reasonable job of reproducing the ICM gas mass fraction of groups and clusters {\it when using the fiducial stellar feedback parameter values discussed above}.

Note that for a fixed value of $n_{\rm heat}$ (i.e., fixed heated gas mass), increasing (decreasing) the heating temperature leads to more (less) bursty and energetic AGN feedback, as more (less) time is required for the BHs to accrete enough mass to heat neighbouring gas by a larger (smaller) amount.

Finally, the feedback efficiency is $\epsilon \equiv \epsilon_r \epsilon_f$, where $\epsilon_r=0.1$ is the radiative efficiency and $\epsilon_f$ is the fraction of $\epsilon_r$ that couples to neighbouring gas.  \cite{Booth2009} (see also \citealt{Booth2010}, L14, \citealt{Schaye2015}) have shown that adopting $\epsilon_f=0.15$ results in a good match between the OWLS (cosmo-OWLS) simulations and the normalisations of the $z=0$ relations between BH mass and halo mass and velocity dispersion (the slopes are naturally reproduced), as well as to the observed cosmic BH density.  In general, the choice of the efficiency is inconsequential for galaxy properties other than the black hole mass, so long as it is non-zero (see \citealt{Booth2009,Booth2010} and Appendix A of the present study).  This owes to the fact that AGN feedback quickly establishes a self-regulating scenario.

In Section 2.2 and Appendix A we show the effects of systematically varying the main parameters of the AGN accretion/feedback models (i.e., the minimum halo mass for BH seed injection, $\Delta T_{\rm heat}$, $n_{\rm heat}$, $\epsilon_f$, and the boost factor $\alpha$ applied to the Bondi accretion rate) on the galaxy stellar mass function.

Note that haloes are identified by using a standard friends-of-friends (FoF) percolation algorithm on the dark matter particles with a typical value of the linking length in units of the mean interparticle separation (b=0.2). The baryonic content of the haloes is identified by locating the nearest DM particle to each baryonic (i.e.\  gas or star) particle and associating it with the FoF group of the DM particle. Artificial haloes are removed by performing an unbinding calculation with the \textsc{subfind} algorithm \citep{Springel2001,Dolag2009}: any FoF halo that does not have at least one self-bound substructure (a `subhalo') is removed from the FoF groups list.  We define the central galaxy as the baryonic component belonging to the most massive subhalo in a FoF group, whereas satellite galaxies belong to the other (less massive) subhaloes in a FoF group.

\begin{figure}
\includegraphics[width=0.995\columnwidth]{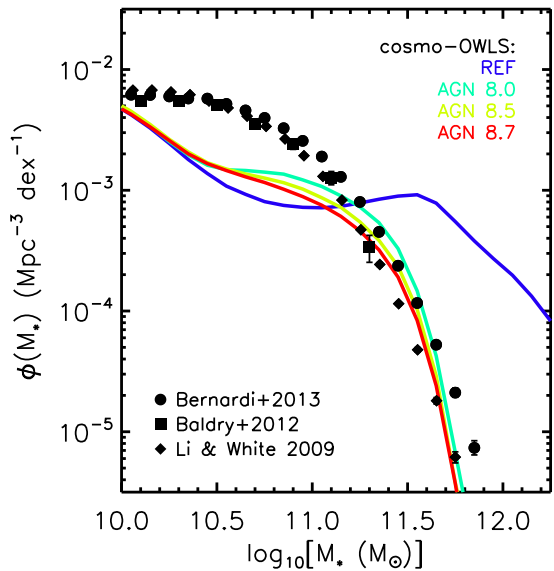}
\caption{\label{fig:gsmf_cosmoowls}
The $z=0.1$ GSMF for the cosmo-OWLS runs presented in L14 in a \planck~2013 cosmology.  A 3-D aperture of 30 kpc (physical) is adopted when calculating the stellar masses of the simulated galaxies.  All of the models have too few galaxies with $10 \la \log_{10}[M_*/{\rm M}_\odot] \la 11$, compared to recent SDSS and GAMA observations.  In addition, neglect of AGN feedback (the `REF' model) results in far too many massive galaxies.  Inclusion of AGN feedback resolves this problem, a result which is independent of the choice of AGN heating temperature (i.e., how violent/bursty the heating events are).
}
\end{figure}

\subsection{Calibration Strategy}

Recent numerical work has demonstrated the sensitivity of the predicted baryonic properties of the haloes to the implementation of subgrid prescriptions for feedback processes.  This has motivated some recent works to explicitly calibrate the subgrid feedback to reproduce key observables, while using other independent observables as tests of the realism of the model.  Two of the more notable examples of this strategy are the Illustris \citep{Vogelsberger2013} and EAGLE \citep{Schaye2015} projects.  These studies were focused on simulating the main galaxy population at high resolution and both suitably calibrated their feedback on important aspects of the galaxy population.  In the case of EAGLE, the feedback was calibrated on the local galaxy stellar mass function and the size$-$stellar mass relation of galaxies (see \citealt{Crain2015}).

As we are interested in tests of cosmology using large-scale structure, rather than simulating the galaxy population in fine detail, our calibration strategy will differ from that of EAGLE and Illustris.  In particular, the crucial property that dictates how much the total matter power spectrum (which is what large-scale structure tests probe) has been modified by baryon physics is the total baryon fraction of haloes, which is dominated by stellar mass and especially {\it hot gas}.  Our calibration strategy will therefore be aimed at reproducing the observed stellar and hot gas masses of haloes.  Note that the stellar and hot gas masses are also key for setting the magnitude of many of the other observable properties (e.g., luminosities and metallicities of galaxies, metal content and overall thermodynamic state of the ICM in groups and clusters).  To our knowledge, \calsim~is a first attempt to calibrate the feedback on the total baryon content of haloes and is the first study to explicitly calibrate the feedback on the observed properties of massive haloes.

Our approach is as follows.  We previously demonstrated that a subset of the models with AGN feedback in the OWLS/cosmo-OWLS projects reproduces a wide variety of properties of the hot gas in groups and clusters (see \citealt{McCarthy2010} and L14), as well as of the `optical' properties of the BCG.  Given this success and the fact that we now wish to carry out simulations of similar resolution, we will use these simulations as our starting point.  We will first examine the overall stellar mass distribution (the galaxy stellar mass function, GSMF) of the various cosmo-OWLS models.  As discussed above (Section 2.1), we anticipate an over-suppression of star formation in haloes with $M_{\rm halo} \sim 10^{12} $M$_\odot$ for these models.  We will examine to what extent a simple adjustment of the efficiency of stellar feedback can rectify this issue, or whether more complicated expressions for the efficiency are required.  Having calibrated the stellar feedback to reproduce the GSMF, we will investigate how the ICM properties are affected (if at all).  If there is significant ``crosstalk'' between the hot gas and stellar properties, such that adjusting the feedback parameter values to affect one has a similarly large effect on the other, then this could make simultaneous calibration an involved and expensive task.  If, on the other hand, the coupling is relatively weak, a simple re-calibration of the AGN model to fit the group/cluster gas fractions may be all that is required.

\begin{figure*}
\includegraphics[width=0.995\columnwidth]{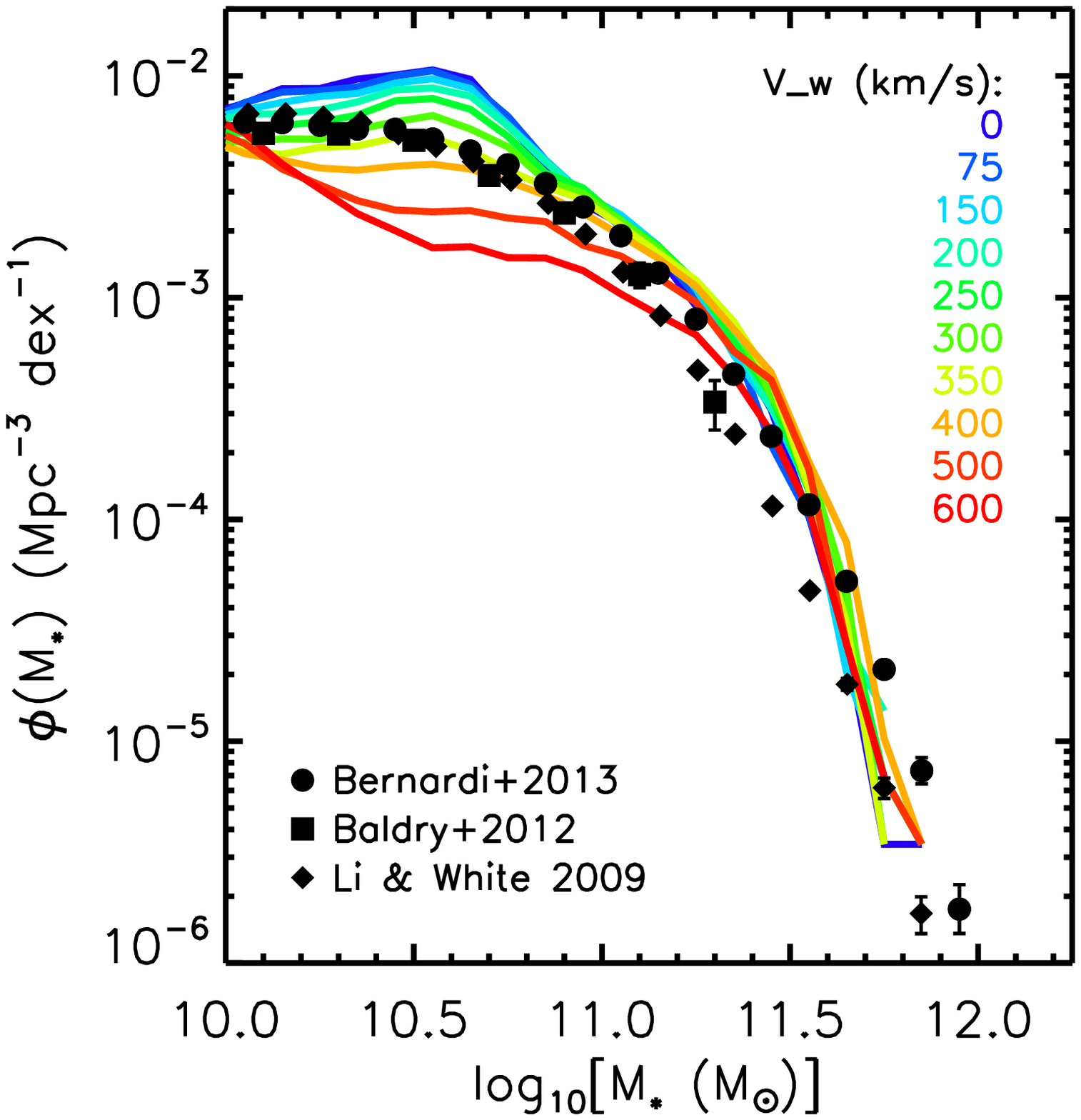}
\includegraphics[width=0.995\columnwidth]{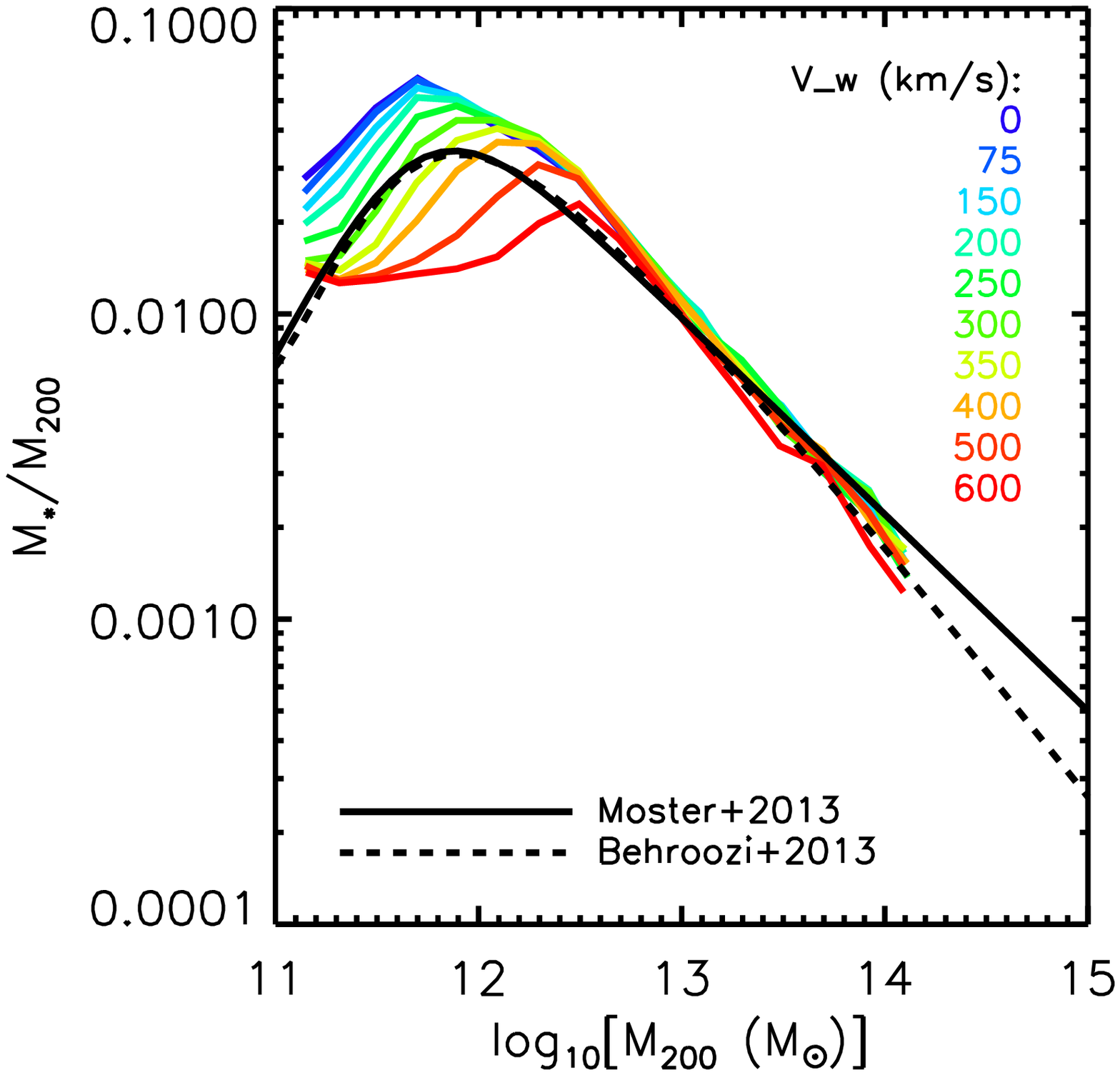}
\caption{\label{fig:gsmf_vary_vwind}
The effects of varying the wind velocity of stellar feedback on the local GSMF (left panel) and the stellar mass fraction$-$halo mass relation (right panel; for central galaxies only).  The stellar mass fraction of observed central galaxies in the right panel has been determined using abundance matching.  Dropping the wind velocity from the cosmo-OWLS value of $600$ km/s to $\approx300$ km/s resolves most of the problem with the underabundance of $\sim M^*$ galaxies.  However, it is not possible to perfectly match the data (particularly the knee of the GSMF) using a fixed velocity while leaving the parameters of the AGN feedback model at their cosmo-OWLS values. 
}
\end{figure*}

\subsection{The galaxy stellar mass function}

We begin in Fig.~\ref{fig:gsmf_cosmoowls} by examining the GSMF of the cosmo-OWLS runs presented in L14.  The GSMF is defined as the number of galaxies (including both centrals and satellites) per unit comoving Megaparsec per decade in stellar mass; i.e., $\phi(M_*) \equiv dn/d{\rm log}_{10}M_*$.  Following \citet{Schaye2015}, an aperture of 30 kpc (physical) is adopted when calculating the stellar masses of the simulated galaxies here, but we explore in Appendix B the effects of varying the aperture size and compare with recent observations that do likewise.  In short, the stellar masses of the most massive galaxies {\it are} sensitive to the choice of aperture (due to the presence of intracluster light), both in the simulations and observations, and a 30 kpc aperture is reasonable for standard `pipeline analysis'.

Fig.~\ref{fig:gsmf_cosmoowls} shows that the cosmo-OWLS models consistently have too few galaxies with $\log_{10}[M_*/{\rm M}_\odot] < 11$ compared to recent SDSS and GAMA observations.  In addition, neglect of AGN feedback (the `REF' model) results in far too many massive galaxies ($\log_{10}[M_*/{\rm M}_\odot] \ga 11.5$).  Interestingly, AGN feedback resolves this overcooling problem and the resulting GSMF matches the observations at the high-mass end very well, a result which is nearly independent of the choice of AGN heating temperature (i.e., how violent/bursty the heating events are).  This latter result contrasts with the very strong dependence of the hot gas mass fractions on the heating temperature found in L14 (see their Fig.~3); a result that we exploit later on when calibrating the AGN feedback.

The fact that all of the cosmo-OWLS models underpredict the abundance of galaxies with $\log_{10}[M_*/{\rm M}_\odot] < 11$ suggests that the stellar feedback is overly efficient, since this is the only aspect of the feedback in common between the different models.  We now seek to alter the feedback parameter values to produce a better match to the GSMF at these masses while still retaining a similarly good match to the hot gas properties of groups and clusters found by L14 for their `AGN 8.0' model.  We therefore start from this model and experiment with systematically lowering the stellar feedback wind velocity while keeping all other aspects of the model fixed, including the stellar feedback mass-loading.  Therefore, by lowering the velocity we are also lowering the fraction of the available stellar feedback energy which couples to the gas.

In the left panel of Fig.~\ref{fig:gsmf_vary_vwind} we show the results of lowering the stellar feedback wind velocity on the GSMF.  These test runs, which adopt a \wmap~9-year cosmology, were performed in a smaller 100 Mpc/$h$ on a side box, adopting the same resolution as for the full 400 Mpc/$h$ production runs presented later.  Lowering the wind velocity indeed has the desired effect of increasing the abundance of galaxies at low to intermediate stellar masses.  A wind velocity of $v_{w}\approx300~\textrm{km/s}$ does a reasonable job of reproducing the abundance of the lowest-mass galaxies under consideration ($\log_{10}[M_*/{\rm M}_\odot] \sim 10$).  However, no single value of the wind velocity results in a perfect match to the observed GSMF.  In particular, tuning to match the lowest mass galaxies results in a slight overabundance of galaxies at $\log_{10}[M_*/{\rm M}_\odot] \sim 10.5-11.5$.  This issue is more clearly visible in the right panel of Fig.~\ref{fig:gsmf_vary_vwind}, which shows that haloes with masses of $M_{200}\sim10^{12} \ {\rm M}_\odot$ have somewhat higher stellar mass fractions than inferred from abundance matching results.  We point out that we did not examine the stellar mass fractions while attempting to calibrate the feedback (we examined {\it only} the local GSMF and the hot gas fractions of groups and clusters), but found in retrospect that it more clearly demonstrates this particular issue.

To rectify the overabundance of galaxies at intermediate stellar masses we could adopt a more complicated dependence of the stellar feedback efficiency on either global or local properties, which might be appealed to on either numerical or physical grounds (see discussion in \citealt{Schaye2015}).  Alternatively, we can fix the stellar feedback wind velocity to reproduce the abundance of the lowest mass galaxies and use the freedom in the AGN feedback model to address the issue.  We opt for the latter (simpler) approach here in the first instance\footnote{Detailed comparisons to the demographics of the observed AGN population (e.g., the evolution of luminosity functions, quasar clustering, etc.) may offer an interesting set of orthogonal constraints on the AGN feedback model that may help to determine the effective halo mass where AGN feedback starts to dominate over that of stellar feedback.  We plan to examine this possibility in future work.}.

As discussed in Section 2.1, there are five main parameters in the AGN model that can be varied:  the minimum halo mass for BH seed injection, $\Delta T_{\rm heat}$, $n_{\rm heat}$, $\epsilon_f$, and the boost factor applied to the Bondi accretion rate ($\alpha$).  In Appendix A, we show that the GSMF is only very weakly dependent on the feedback efficiency  $\epsilon_f$ (see also \citealt{Booth2009}) and the heating temperature $\Delta T_{\rm heat}$ (see also Fig.~\ref{fig:gsmf_cosmoowls}), so adjustment of these parameters cannot resolve the overabundance issue at intermediate masses.  We therefore leave $\epsilon_f=0.15$, which was shown previously to result in a good match to the normalisation of various local BH scaling relations.  We also leave the heating temperature fixed at $\Delta T_{\rm heat}=10^8$ K for the moment but return to this parameter later.

\begin{figure}
\includegraphics[width=0.995\columnwidth]{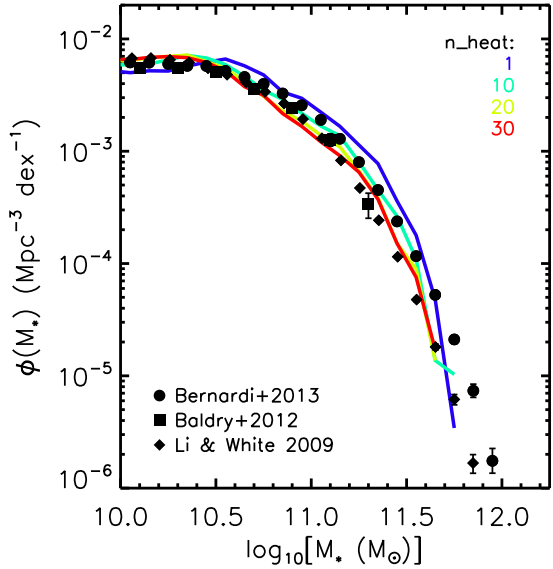}
\caption{\label{fig:gsmf_nheat}
The effect of varying the mass of gas heated (characterised by $n_{\rm heat}$, the number of gas particles heated per feedback event) by AGN feedback on the local GSMF.  Here we adopt a wind velocity of 300 km/s for the feedback from star formation.  Heating $\approx10-30$ particles, corresponding to a gas mass of $\approx1-3\times10^{10} \ {\rm M}_\odot$, yields an excellent match to the GSMF over the full range of masses considered here.  Henceforth we adopt $n_{\rm heat}=20$.
}
\end{figure}

We also show in Appendix A that the GSMF {\it is} sensitive to both the minimum halo mass for BH seeding and the scaling factor, $\alpha$, applied to the Bondi accretion rate.  Resolution considerations prevent us from exploiting the former to provide a solution to the overabundance problem, as we can only reliably {\it increase} the minimum mass scale for BH seeding, which makes the problem significantly worse.  Adopting a somewhat different density dependence to the Bondi boost factor from that used by default by \citet{Booth2009} (they adopted $\alpha \propto \rho^{\beta}$ where $\beta=2$) is a more promising possibility.  However, by changing the boost factor significantly there is a possibility that the previously obtained agreement with the observed BH scaling relations would be negatively affected.  While reproducing the amplitude of these scaling relations is not strictly necessary (it is the feedback that counts not the BH masses), it is nevertheless desirable.  We therefore retain the accretion scaling factor dependence adopted by \citet{Booth2009}.

The last AGN feedback parameter is the mass of gas heated by the AGN, expressed here in terms of the number of gas particles heated, $n_{\rm heat}$.  In Fig.~\ref{fig:gsmf_nheat} we show the effect of increasing the heated gas mass (note that $n_{\rm heat}=1$ corresponds to a heated gas mass of $\approx 1.1\times10^{9} \ {\rm M}_\odot$).  Note that by increasing the number of gas particles that are heated while keeping the heating temperature $\Delta T_{\rm heat}$ fixed, implies that the AGN heating events are more energetic as we increase $n_{\rm heat}$ (the same energy is injected into each particle but more particles are heated).   Increasing the heated gas mass to a value of $\approx1-3\times10^{10} \ {\rm M}_\odot$ ($n_{\rm heat}\approx10-30$) has the desired effect of reducing the abundance of intermediate stellar mass galaxies while only have a relatively small effect at much higher or lower masses.  Varying $n_{\rm heat}$ also has a relatively small effect on the gas mass fractions (see Appendix A).  We adopt $n_{\rm heat}=20$ henceforth.

\begin{figure}
\includegraphics[width=0.995\columnwidth]{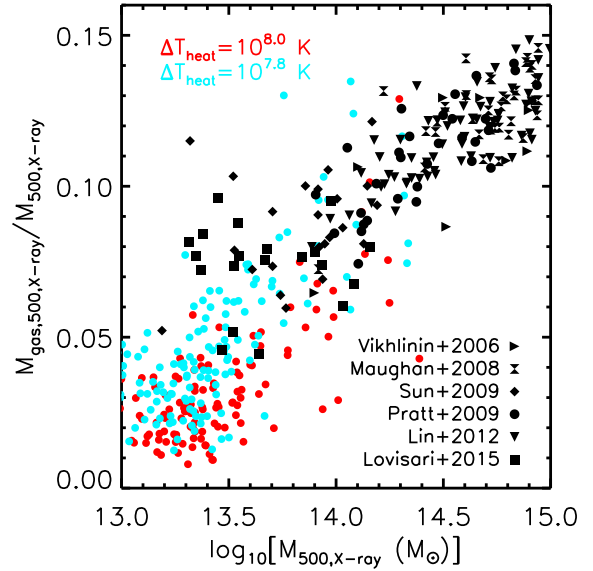}
\caption{\label{fig:fgas_vary_theat}
The effect of lowering the wind velocity from stellar feedback and increasing the mass of gas heated by AGN on the hot gas mass fractions of groups and clusters.  The gas mass fractions and halo masses for the simulations have been estimated in an observational manner meant to mimic standard X-ray analyses (see L14 for details), in order to make a like-with-like comparison to the observational data.  The gas fractions are slightly lower than observed, which is due to the fact that more of the gas has ended up in stars compared to the AGN models explored in cosmo-OWLS.  A slight reduction in the AGN heating temperature (from $10^8$ K to $10^{7.8}$ K) yields a better match to the gas fractions while leaving the quality of the match to the GSMF unchanged. 
}
\end{figure}

\subsection{Group and cluster gas fractions}

Having adjusted the stellar and AGN feedback parameter values to better reproduce the local GSMF, we ask what effect this has on the hot gas content of massive groups and clusters.  In Fig.~\ref{fig:fgas_vary_theat} we show the hot gas mass fraction as a function of halo mass ($M_{\rm 500,X-ray}$) and compare to recent X-ray measurements.  We use the synthetic X-ray pipeline described in L14 (see also Section 5.1) to process the simulations in order to make a like-with-like comparison to the X-ray data.  (Note, however, that we have not attempted to {\it select} the simulated clusters in an observational way, but instead analyse a purely mass-selected sample.)  The sparseness of the simulated sample is due to the relatively small box size of 100 Mpc/$h$ that we are using for calibration purposes.  However, it is sufficiently large to get a sense of the level of overall agreement between the new model and the observations.

\begin{figure*}
\includegraphics[width=0.995\columnwidth]{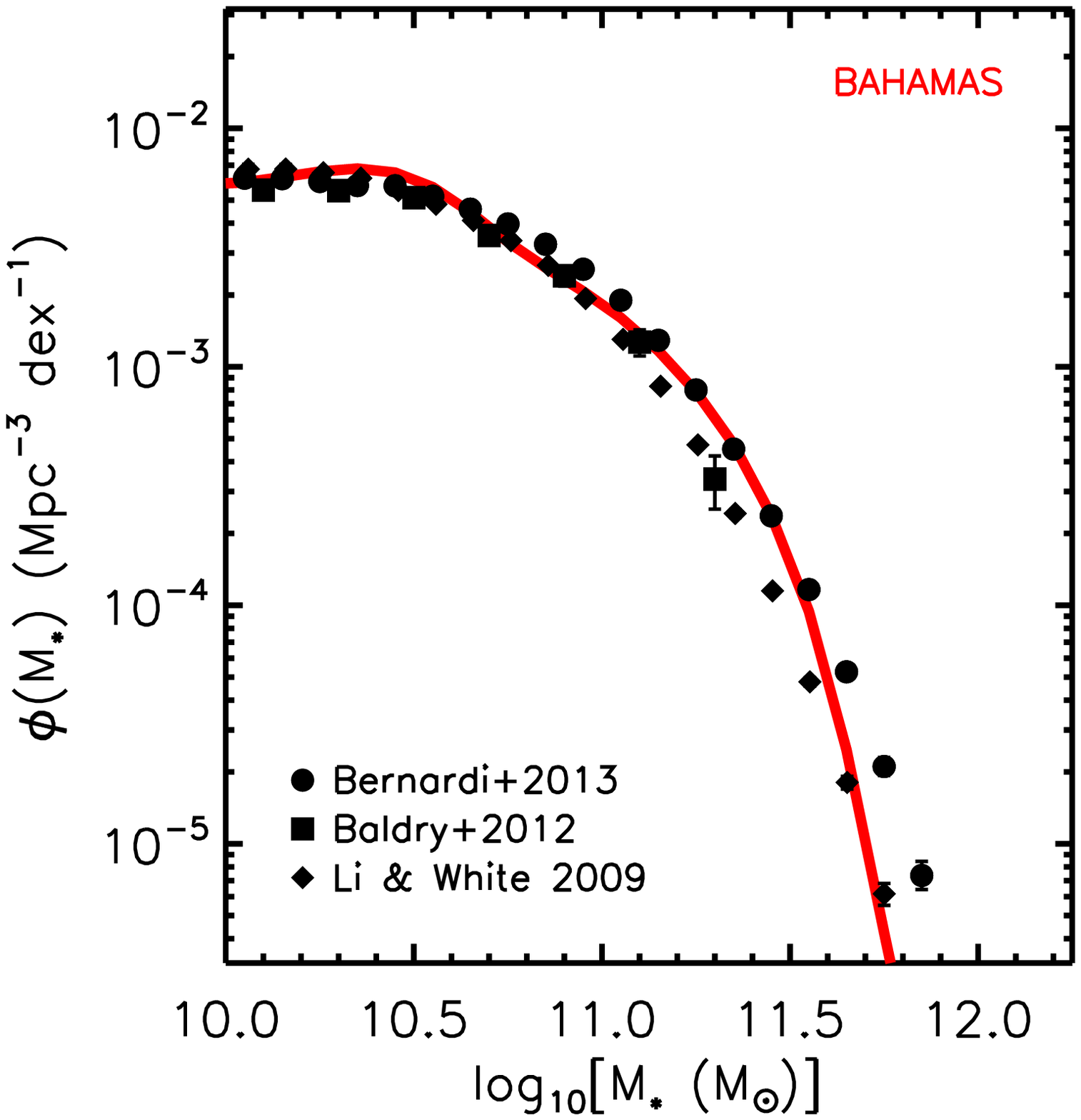}
\includegraphics[width=0.995\columnwidth]{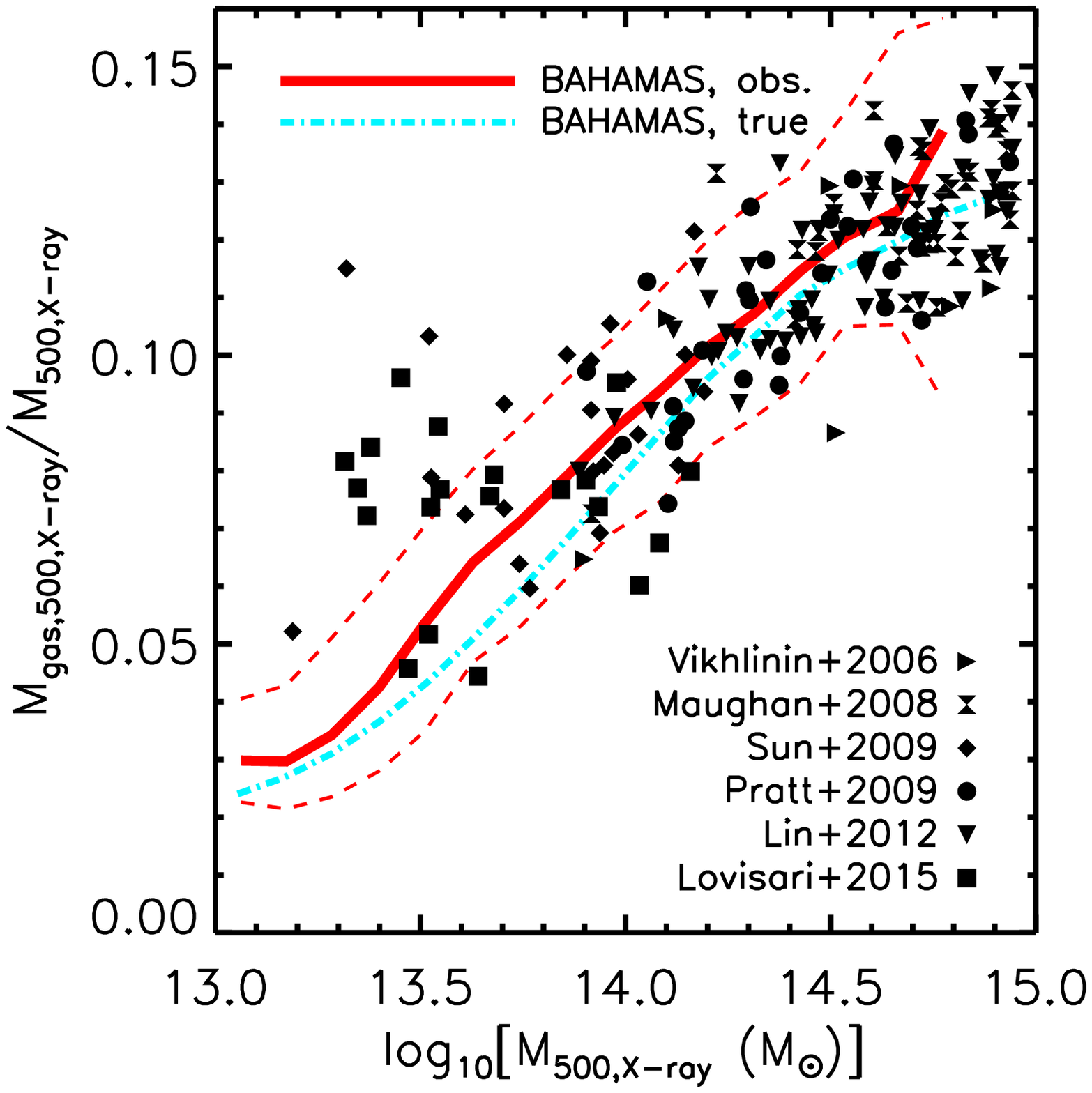}
\caption{\label{fig:calibrated}
The final calibrated local GSMF and hot gas mass fraction$-$halo mass trend, extracted from four independent 400 Mpc/$h$ box realizations.  In the right panel the red curves (solid curve represents the median, dashed curves enclose 68\% of the population) represent the relation derived from a synthetic X-ray analysis of a mass-limited sample (all haloes with $M_{\rm 500,true} > 10^{13} \ {\rm M}_\odot$).  The dot-dashed cyan curve represents the true relation (i.e., not processed through synthetic X-ray observations).  The new model reproduces both key observational diagnostics remarkably well.
}
\end{figure*}

\begin{table*}
\centering
\begin{tabular}{|l|l|l|l|l|l|r|l|l|}
         \hline
	Simulation & $v_{w}$ & $\eta_{w}$ & $\epsilon_r$ & $\epsilon_f$ & $\Delta T_{\rm heat}$ & $n_{\rm heat}$ & accretion model & min FoF mass for BH seeding\\

	\hline
        cosmo-OWLS (\agn~8.0) & 600 km/s & 2 & 0.1 & 0.15 & $10^{8.0}$ K & 1 & \citet{Booth2009} & 100 DM particles\\
        \calsim & 300 km/s & 2 & 0.1 & 0.15 & $10^{7.8}$ K & 20 & \citet{Booth2009} & 100 DM particles\\

        \hline
\end{tabular}
\caption{Comparison of the cosmo-OWLS `AGN 8.0' model parameter values and the new calibrated model.  $v_{w}$ and $\eta_{w}$ are the stellar feedback wind velocity and mass-loading, respectively.  $\epsilon_r$ is the BH radiative efficiency and $\epsilon_f$ is the fraction thereof which couples to neighbouring gas.  $\Delta T_{\rm heat}$ is the temperature jump applied to $n_{\rm heat}$ neighbouring gas particles during AGN feedback. }
\label{table:models}
\end{table*}

Using the default heating temperature of $\Delta T_{\rm heat}=10^8$ K, which worked well in \citet{McCarthy2010} and L14 in terms of comparisons to a large variety of hot gas diagnostics, we see a small over-suppression of the gas fraction compared to observations.  This is easy to understand: the reduction of the efficiency of stellar feedback (to better reproduce the low-mass end of the GSMF) has resulted in a higher fraction of baryons being locked up in stars in the progenitors of groups and clusters.  Consequently, the mass of baryons remaining in the form of hot gas has been reduced.  However, a small reduction in the AGN heating temperature to $\Delta T_{\rm heat}=10^{7.8}$ K re-establishes the good agreement with the observed gas fractions and while having essentially no effect on the GSMF (see Appendix A).

\subsection{BAHAMAS}

In Table \ref{table:models} we summarize the adjustments made to the fiducial `AGN 8.0' cosmo-OWLS model (which is identical to the OWLS model `AGN') to reproduce the local GSMF while retaining a good match to the hot gas fractions of groups and clusters.  We henceforth refer to the calibrated model as \calsim~(for BAryons and HAloes of MAssive Systems).

With a viable model in hand, we have run much larger volumes (L400N1024) in both the \wmap~9-year and \planck~2013 cosmologies.  For the \wmap~cosmology we have run four independent realizations (i.e., using different random phases when generating the initial conditions) and when presenting results for that cosmology we combine the results from the four volumes.

\begin{figure*}
\includegraphics[width=0.995\columnwidth]{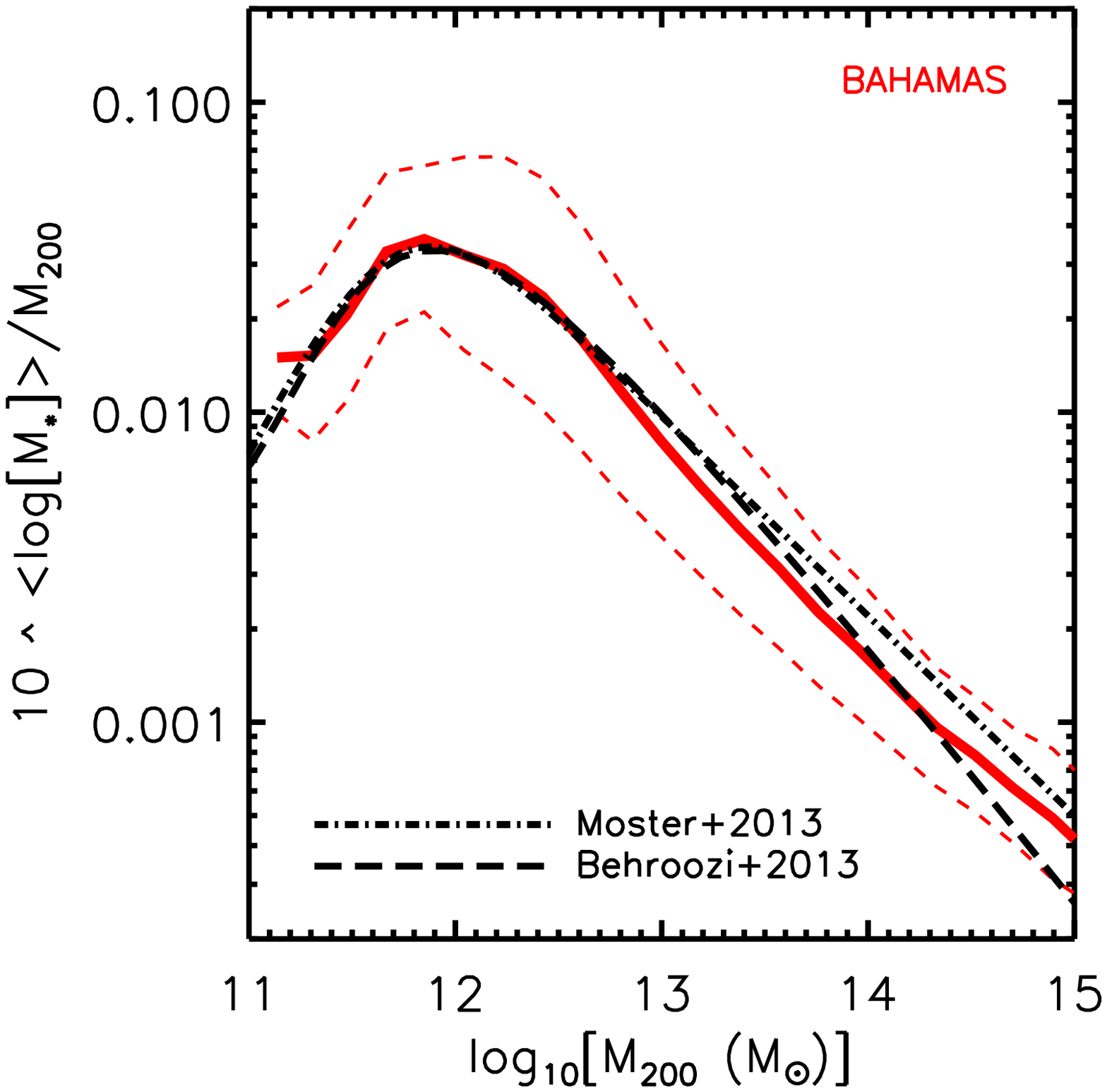}
\includegraphics[width=0.995\columnwidth]{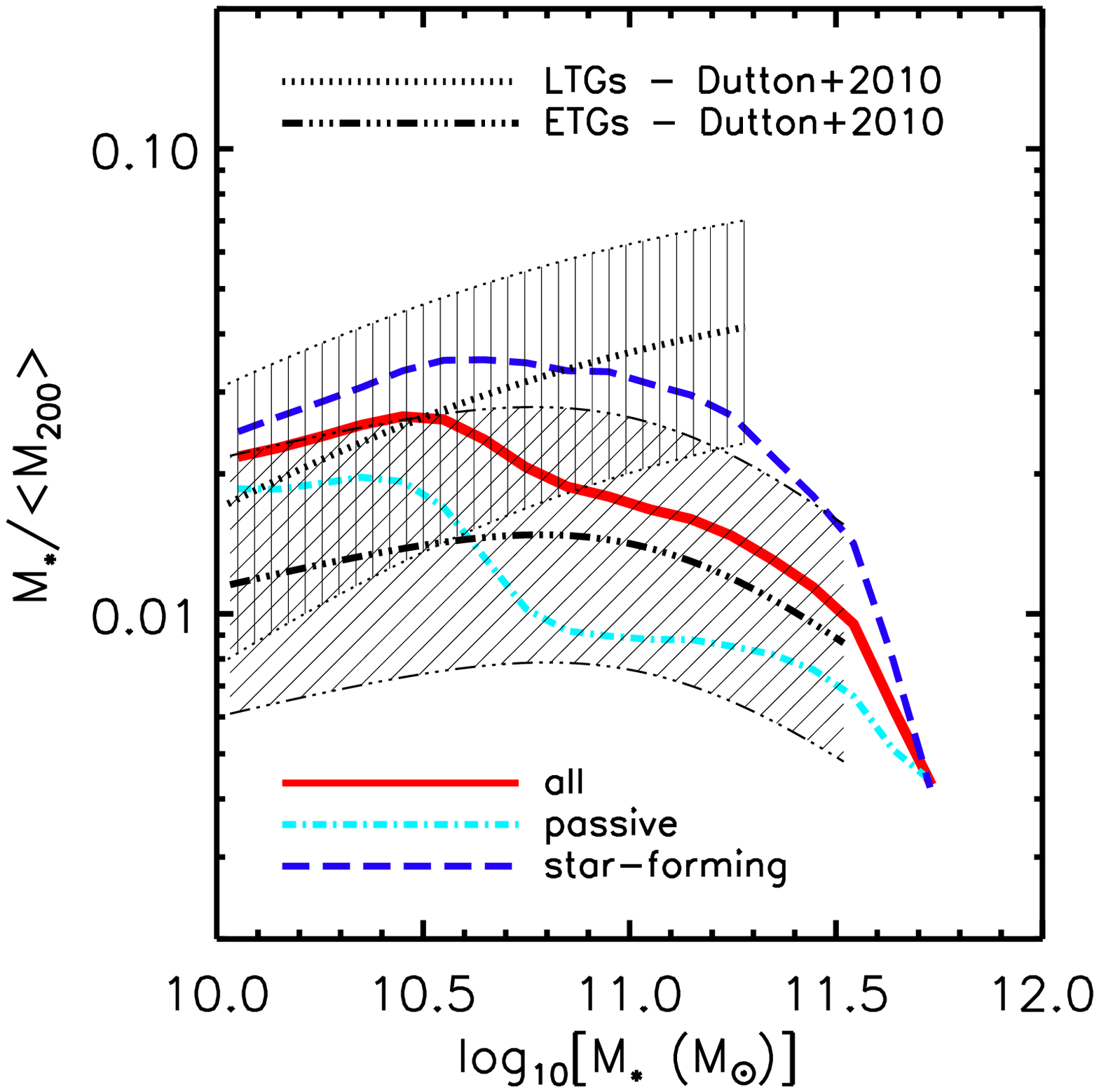}
\caption{\label{fig:fstar_centrals}
The $z=0.1$ $f_*$$-$$M_{200}$ and $f_*$$-$$M_*$ relations for central galaxies compared with abundance matching and stacked weak lensing/satellite kinematics, respectively.  Abundance matching measures $\mean{\log_{10}M_*}(M_{200})$, while stacked weak lensing/satellite kinematics measures $\mean{M_{200}}(M_*)$.  Analysed in the same way, the calibrated model reproduces the two relations well, implying the underlying $M_*$$-$$M_{200}$ (including its intrinsic scatter) is also recovered reasonably well.  The simulations also qualitatively reproduce the observed difference in halo mass at fixed stellar mass for observed early-type and late-type galaxies (i.e., ETGs have a larger mean halo mass at fixed stellar mass compared to LTGs for stellar masses of $\sim 10^{11} \ {\rm M}_\odot$).
}
\end{figure*}

In Fig.~\ref{fig:calibrated} we compare the final calibrated local GSMF and group/clusters hot gas mass fractions with the data derived from the large volume (i.e., production) runs.  For the hot gas mass fraction comparison, in addition to the results for the synthetic X-ray analysis applied to a mass-limited sample (all haloes with true $M_{\rm 500,c} > 10^{13} \ {\rm M}_\odot$; red curves), we also show the true relation (i.e., not processed through an X-ray pipeline and applied to the full mass-limited sample; dot-dashed cyan curve).   The comparison to the true relation is useful because it indicates the degree to which X-ray-inferred quantities are biased (e.g., due to the common assumption of hydrostatic equilibrium, gas clumping, spectroscopic temperature). 

The agreement with both the GSMF and the hot gas mass fractions is remarkably good\footnote{The scatter at $\sim10^{13.5} \ {\rm M}_\odot$ appears to be somewhat underestimated by the simulations, but we again note that we have not attempted to select the simulated clusters in an observational way.  Flux-limited X-ray surveys, such as those of \citet{Lovisari2015}, may preferentially select systems with higher than average gas fractions near the flux limit (i.e., in the group regime), for example.}  We stress here the simplicity of our calibrated model: the parameters governing the efficiencies\footnote{Here we use the term efficiency to refer to the overall effectiveness of the feedback.} of stellar and AGN feedback are single, fixed values.  The fact that the model closely reproduces the observed baryon content of collapsed systems over a couple of orders of magnitude in halo mass is therefore non-trivial and was certainly not guaranteed.  For example, we did not have to invoke complicated functions for the stellar or AGN feedback efficiencies to reproduce the {\it shapes} of the GSMF or the gas fraction$-$halo mass trends.  In fact, the latter appears to come out naturally from our models that include AGN feedback (i.e., it is difficult to avoid).  

We do not claim to have identified a unique solution.  Furthermore, we note that the results above regarding the parameter dependence of the stellar and hot gas fractions may not hold at much higher resolution.  However, we have achieved our main requirements at the present resolution (the baryon content of massive systems) and we can test the realism of the model against independent observations, as we do immediately below.  In Appendix C we present a numerical convergence study, showing that the simulations are not strongly affected by resolution for the massive haloes we are generally focused on here.

\section{The galaxy$-$halo connection}

In this section we examine the partitioning of stellar mass as a function of halo mass, the contribution of centrals and satellites, and the large-scale spatial and kinematic distributions of galaxies.  We compare with recent observations of the local Universe.

\subsection{Stellar mass fractions of central galaxies}

In Fig.~\ref{fig:fstar_centrals} we examine the stellar mass fractions of central galaxies as a function of halo mass (left panel) and stellar mass (right panel).  In the left panel we compare to the recent abundance matching (AM) results of \citet{Behroozi2013} and \citet{Moster2013}.  AM usually constrains the mean of the log of the stellar mass in bins of halo mass [i.e., $\mean{\log_{10}{M_*}}(M_{\rm halo})$] so this is the quantity we compute from the simulations.   Where necessary we have converted the AM halo masses to a common halo mass definition, $M_{\rm 200,c}$, by adopting the mass$-$concentration relation from the simulations\footnote{We have not fit a parametric model to the mass$-$concentration relation, but have instead computed the median concentration in bins of halo mass.  We then interpolate the concentration from this relation given a halo mass.  For a power law fit to the high-mass end of the mass$-$concentration relation from \calsim, see \citet{Henson2016}.} and assuming an NFW profile.

The agreement between the mean relation from \calsim~and those derived from the AM measurements is excellent.  Small differences are present at the high-mass end which could be due to a variety of effects, including differences in the effective apertures used to derive the stellar masses and differences in the underlying halo mass functions (AM techniques adopt mass functions from dark matter only simulations, which in general will differ from those derived from hydrodynamical simulations at the tens of percent level due to gas expulsion by stellar and AGN feedback; e.g., \citealt{Sawala2013,Cui2014,Velliscig2014,Cusworth2014}).  Note also that Fig.~\ref{fig:fstar_centrals} examines the trend for {\it central} galaxies only, while the GSMF includes both centrals and satellites.  Therefore it is possible in principle to reproduce the GSMF without reproducing the stellar mass fraction$-$halo mass trend if the satellite population differs significantly between the simulations and the observations.

Another relevant issue is that AM techniques must assume something about the intrinsic scatter in the stellar mass at fixed halo mass, which is something we have no direct control over in the simulations.  This can be particualrly important at high masses, due to the steepness of the mass function.  On this point, it is interesting to note that the scatter in the simulations (dashed red curves) appears to be significantly larger than adopted in many previous AM studies.  For example, \citet{Moster2013} adopt a fixed scatter of 0.15 dex in stellar mass, whereas the median scatter in \calsim~is 0.24 dex and declines with increasing halo mass (e.g., the scatter is $0.30$, $0.22$, and $0.20$ dex at $M_{200}/$M$_\odot=10^{13}$,$10^{14}$, and $10^{15}$).  Calibrating the models to match the GSMF therefore does not uniquely determine the scatter.  An independent constraint on the scatter can be made by comparing the predictions to measurements of galaxy clustering (which we do below) and to measurements of galaxy-galaxy lensing (which we intend to do in a future study).  Interestingly, the recent Halo Occupation Distribution (HOD) models of \citet{Leauthaud2012} and \citet{Zu2015}, which have been calibrated to reproduce these three independent and complementary observables, derive scatters of $0.23$ and $0.22$ dex, respectively, consistent with \calsim.

In the right panel of Fig.~\ref{fig:fstar_centrals} we examine the stellar mass fractions in bins of stellar mass for central galaxies and compare to recent stacked weak (galaxy-galaxy) lensing and stacked satellite kinematics, hereafter WLSK.  In contrast to AM, stacked WLSK derives the mean halo mass (or mean of the log of halo mass) in bins of stellar mass; i.e.,  $\mean{M_{\rm halo}}(M_{*})$.  This is an alternative way to compare the stellar mass fractions and one which is sensitive to the level of intrinsic scatter in the stellar mass$-$halo mass relation.

The two shaded regions in the right panel of Fig.~\ref{fig:fstar_centrals} correspond to the stellar mass fraction trends for late-type (LTGs) and early-type (ETGs) galaxies in \citet{Dutton2010}.  Dutton et al.\ compiled WLSK measurements from a variety of previous studies (WL: \citealt{Mandelbaum2006,Mandelbaum2008,Schulz2010}; SK: \citealt{Conroy2007,Klypin2009,More2011}) and took care to scale the stellar masses from these studies to a common Chabrier IMF and to adopt a common halo mass definition ($M_{200,c}$; see \citealt{Dutton2010} for more details).  The shaded regions roughly encapsulate the differences in the relations derived from the different WLSK studies and therefore gives some handle on the systematic error involved (which generally exceeds the statistical error from any individual study).  

The solid red curve represents the mean trend predicted by the simulation for all central galaxies.  However, observations show that the galaxy formation efficiency at fixed stellar mass depends on the type of galaxy being considered (i.e., disc or elliptical; see \citealt{Mandelbaum2015} for a recent comparison of different observational results).  To test whether simulated galaxies display such a trend, we split them into either `star-forming' or `passive' categories using a threshold in the specific star formation rate\footnote{The relatively low resolution of our simulations prevents us from being able to reliably classify the simulated galaxies as ETGs or LTGs on the basis of stellar morphology.  Note that in any case some of the observational WLSK studies used colour (which should closely track sSFR) rather than morphology to divide their samples.} (sSFR $\equiv$ SFR/$M_*$) of $10^{-11} {\rm yr}^{-1}$, which corresponds roughly to the dip in the observed bimodal sSFR distribution (e.g., \citealt{Wetzel2012}).  We compute the SFR within a 30 kpc aperture for each simulated galaxy.  

The dashed blue and dot-dashed cyan curves in the right panel of Fig.~\ref{fig:fstar_centrals} show the mean relations for the simulated star-forming and passive galaxies, respectively.  Here we see that, indeed, passive galaxies have larger mean halo masses (and thus lower stellar mass fractions) compared to star-forming galaxies of the same stellar mass ($M_*\sim10^{11} \ {\rm M}_\odot$).  There is also reasonably good qualitative agreement with the observed trends for ETGs and LTGs (i.e., within the systematic errors).

\subsection{Stellar mass content of groups and clusters}

\subsubsection{Integrated stellar mass fractions}

\begin{figure}
\includegraphics[width=0.995\columnwidth]{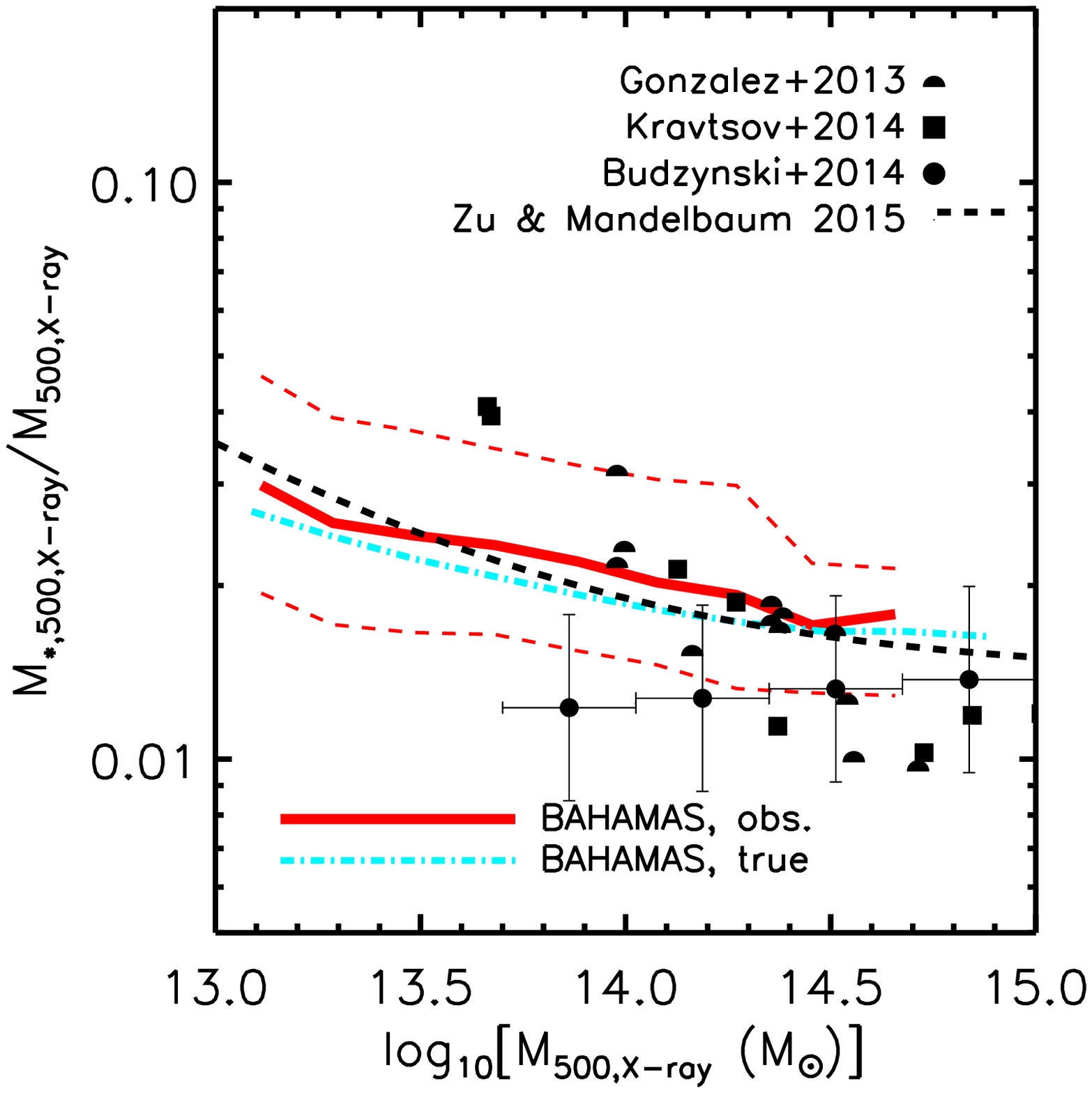}
\caption{\label{fig:fstar_integrated}
The integrated stellar mass fraction of local groups and clusters within $r_{500}$ compared with SDSS observations of nearby individual clusters with X-ray halo mass estimates \citep{Gonzalez2013,Kravtsov2014}, stacked SDSS imaging of a large sample of optically-selected groups with X-ray halo mass estimates \citep{Budzynski2014}, and HOD modelling of SDSS data \citep{Zu2015}.  The simulated groups have been processed with synthetic X-ray observations to measure a halo mass and radius in an observational manner (solid red curve represents the median and dashed red curves enclose 68\% of the population), but we also show the median relatios for the true relation (dot-dashed cyan, not processed through synthetic X-ray observations).  Note that the simulations and observations of \citet{Gonzalez2013}, \citet{Kravtsov2014}, and \citet{Budzynski2014} include intracluster light (ICL), whereas the HOD modelling results do not.  Overall the integrated stellar mass fractions are reproduced very well in terms of the normalisation.  The observational studies disagree with one another over the {\it shape} of the trend (see \citealt{Budzynski2014} for further discussion), with the simulation predictions most closely resembling the statisical HOD-derived measurements.}
\end{figure}

In Fig.~\ref{fig:fstar_integrated} we plot the integrated stellar mass fractions of local ($z\approx0.1$) galaxy groups and clusters and compare with a variety of observational measurements, assuming a Chabrier IMF throughout.  The black semi-circles and squares represent the results of \citet{Gonzalez2013} and \citet{Kravtsov2014}, respectively, who have made integrated stellar mass measurements of individual, nearby clusters with hydrostatic modelling of high-quality X-ray observations being used to estimate the halo mass.  The filled black circles correspond to the best-fit power-law relation derived by \citet{Budzynski2014} from an image stacking analysis of a large sample of optically-selected SDSS clusters.  They derived the stellar mass fractions in four halo mass bins, using an empirically-calibrated richness$-$X-ray temperature$-$hydrostatic $M_{500}$ relation to estimate halo mass.  The vertical error bars on the Budzynski et al.\ measurements correspond to the estimate of \citet{Leauthaud2012} of the (non-IMF) systematic uncertainty in the stellar mass estimates due to, e.g., differences in stellar population modelling.  Although we plot these error bars on the Budzynski data points only, they apply equally well to all other data points shown in Fig.~\ref{fig:fstar_integrated}.

Common to the \citet{Gonzalez2013}, \citet{Kravtsov2014}, and \citet{Budzynski2014} studies is the inclusion of intracluster light (ICL) and the use of X-ray observations (assuming hydrostatic equilibrium) to derive the halo mass.  The black dashed curve shows the HOD modelling results of \citet{Zu2015} for SDSS data (see also \citealt{Leauthaud2012} for HOD modelling of COSMOS data, which yields very similar results), where their HOD models have been constrained to reproduce the observed galaxy-galaxy lensing signal and galaxy clustering in bins of stellar mass, as well as the shape of the galaxy stellar mass function.  Unlike the previously mentioned studies, the halo masses here are not measured (and do not assume hydrostatic equilibrium) but are inferred from the model.  The inconsistency in the way the halo masses are derived between the studies might be a cause for concern for this comparison, but a comparison of the solid red and dashed blue curves in Fig.~\ref{fig:fstar_integrated} shows that only a small difference exists for the simulation predictions when we use true masses as opposed to hydrostatic ones.  Note that for consistency we have scaled all the stellar masses to a Chabrier IMF and have converted the halo masses of \citet{Zu2015} from $M_{200,m}$ to $M_{500,c}$ assuming an NFW profile and the mass-concentration relation from the simulations.

There is good agreement between the predictions of the simulations and the observational measurements in terms of amplitude: $f_{*,500} \approx 0.01-0.03$ for groups and clusters.  When compared with the observed hot gas mass fraction (see right panel of Fig.~\ref{fig:calibrated}), one immediately concludes that the hot gas dominates over the stellar mass in groups and clusters (see also Appendix C).  It is only when one approaches halo masses of $\sim10^{13} \ {\rm M}_\odot$ and lower that the stellar mass becomes a sigififcant fraction of the total baryon budget.

While the normalization of the stellar mass fraction$-$halo mass relation is reproduced by the simulations, the picture regarding the {\it shape} of the relation is less clear.  This is because the observational studies do not agree with one another, with the results from statistical analyses of large samples suggesting flat or mildly varying stellar mass fractions, while the studies based on individual clusters suggest a much steeper trend.  Interestingly, the simulations predict a reasonably large spread in the stellar mass fraction at fixed halo mass (thin dashed lines), with a median scatter of 0.16 dex.  Note that much of this scatter is due to the scatter in the relation between X-ray hydrostatic mass and true halo mass, as we find that the scatter in the true stellar mass fraction (within the true $r_{500}$) is only $0.07$ dex on its own.  Given that the scatter is reasonably large, this could mean that selection effects can potentially play a role for the studies based on small numbers of individual clusters and could potentially reconcile the different observational findings.  We suggest that the use of realistic mock galaxy catalogs and folding in of the precise selection functions of the different studies is a promising way to test this hypothesis.

\subsubsection{Relative contributions of centrals and satellites}

\begin{figure}
\includegraphics[width=0.995\columnwidth]{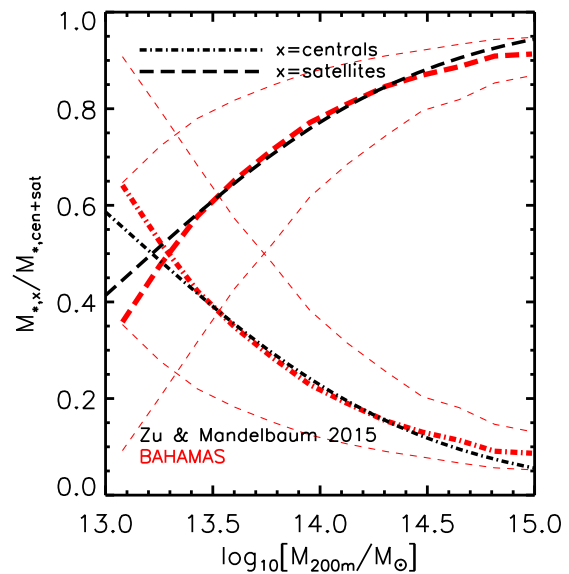}
\caption{\label{fig:fstar_sat_cen}
The fractional contribution to the total stellar mass from central and satellite galaxies as a function of halo mass (here defined as $M_{200,m}$) at $z=0.1$ compared to the HOD model results for SDSS data of \citet{Zu2015}.  The agreement between the simulations and the HOD constraints is remarkably good.}
\end{figure}

We have just shown that the predicted overall stellar content of massive dark matter haloes agrees well with observations.  Here we examine whether the simulations reproduce the relative contributions of centrals and satellites to the total stellar content.  We restrict our analysis to systems with masses exceeding $M_{200,m} > 10^{13} \ {\rm M}_\odot$, since our simulations do not have sufficient mass resolution to resolve the typical satellites (in a mass-weighted sense) of lower-mass haloes.

In Fig.~\ref{fig:fstar_sat_cen} we show the fractional contributions of central and satellite galaxies to the total stellar mass in galaxies (within $r_{200,m}$) as a function of halo mass, defined here as $M_{200,m}$, at $z=0.1$ and compare the predictions of the simulations with the HOD modelling results of \citet{Zu2015}.  For consistency, we exclude the ICL from this comparison since the observational data used to constrain the HOD model does not include this component.  Stellar masses are computed within a 30 kpc aperture for galaxies within $r_{200,m}$ of the host halo. 

The simulations predict a rapidly rising increase in the fractional contribution to the total stellar mass from satellites with increasing halo mass.  Satellites begin to dominate over centrals at a halo mass of $\log_{10}[M_{200,m}/{\rm M}_\odot] \approx 13.2-13.3$ (corresponding to $\log_{10}[M_{500,c}/{\rm M}_\odot] \approx 13.0$).  The agreement between the median relation from the simulations and the HOD modelling results is remarkably good, particularly given the fact that nothing other than the local GSMF was used to calibrate the stellar content of systems in the simulations.  The simulations also predict a large degree of system-to-system scatter in the relative contributions of centrals and satellites in the group regime which can hopefully be tested with future observations.

\subsubsection{Spatial distribution of satellites in clusters}

\begin{figure}
\includegraphics[width=0.995\columnwidth]{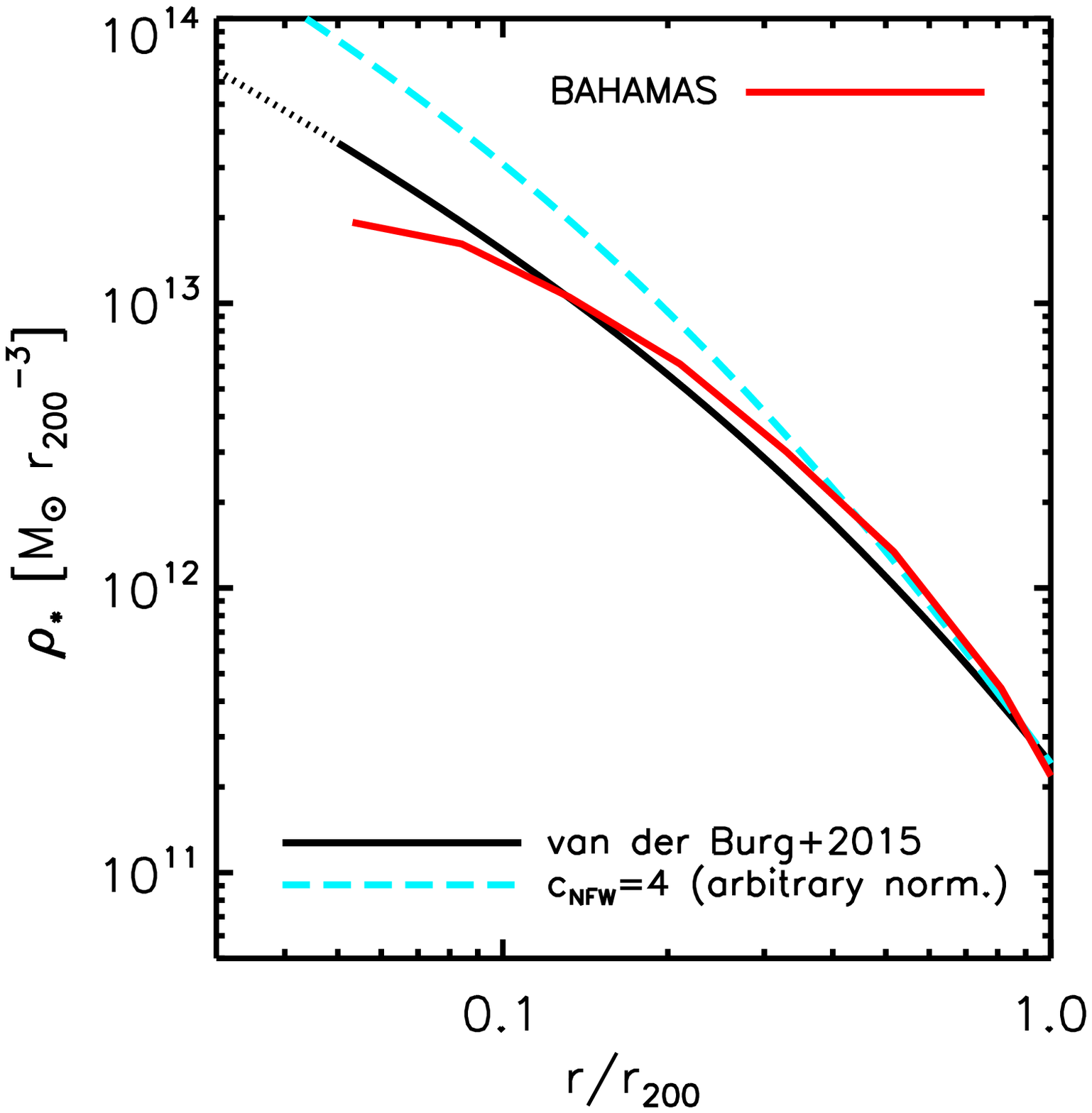}
\caption{\label{fig:rho_star_prof}
The $z=0.1$ stacked stellar mass density profiles of satellite galaxies in massive clusters, compared with the best-fitting NFW profile to the CCCP/MENeaCS sample \citep{vanderBurg2015}.  We select a subset of high-mass simulated clusters with the same mean $M_{500}$ as the observational sample.  The black curve represents the best-fit NFW profile of \citet{vanderBurg2015} ($c_{200}=2.03$), with the dotted portion of the curve indicating the region where the fit ceases to be a good description of the data.  For comparison, the long-dashed blue curve represents an NFW distribution with a concentration $c_{200}=4$ (which is typical of the underlying dark matter distribution for systems of this mass; see \citealt{Henson2016}), normalised to match the best-fit to the data at $r_{200}$.  The red curve represents the prediction for a sample simulated clusters with the same mean halo mass as the observed sample.  Similar to the case of observed clusters, the satellite distribution of the simulated clusters is more extended ($c_{200}\approx2$) than that of the underlying dark matter (typically $c_{200}\approx4-5$ at these masses).
}
\end{figure}

In the previous subsections we showed that the \calsim~simulations reproduce the observed overall stellar mass content of massive dark matter haloes reasonably well, including the breakdown by centrals vs.\ satellites.  How does the predicted {\it spatial} distribution of stellar mass in massive haloes compare with observations?

Previous observational studies have found that both the number density (typically above some luminosity threshold) and the total stellar mass density of satellite galaxies in local massive clusters can be relatively well described with an NFW distribution, but with a concentration parameter ($c_{200} \equiv r_{200}/r_s$, where $r_s$ is the scale radius) that is typically a factor of $\sim2$-$3$ lower than that predicted (and observed) for the underlying dark matter mass density profile (e.g., \citealt{Carlberg1997,Lin2004,Budzynski2012,vanderBurg2015}).  Here we compare with the recent low-redshift observational measurements of the radial distribution of the stellar mass density in satellites of massive clusters of van der Burg et al. (2015, hereafter V15) from the Multi-Epoch Nearby Cluster Survey (MENeaCS) and the Canadian Cluster Comparison Project (CCCP) cluster samples.

To make a consistent comparison to the measurements of V15 we must first select a suitable sample of simulated massive clusters, noting that the sample of V15 includes only very massive clusters.  Specifically, we use the estimated velocity dispersions of the observed clusters (see Table 1 of V15) together with stacked maxBCG velocity dispersion-weak lensing calibration (see Section 3.3 below) to estimate the mean $M_{500,c}$ for the observed sample, finding $\mean{M_{500,c}}\approx6.2\times10^{14}$ M$_\odot$.  We then simply impose a minimum halo mass cut for the simulated clusters of $M_{500,c} \approx 3.4\times10^{14}$ M$_\odot$ such that the mean value for the selected simulation population matches that of the observed sample.  This selection criterion yields 148 clusters from the four independent simulation volumes, with a maximum mass of $M_{500,c} \approx 2.4\times10^{15}$ M$_\odot$.  Following V15, we derive the mean stellar mass density by stacking the satellite catalogs of the cluster sample, normalising the satellite cluster-centric distances by $r_{200}$ prior to stacking.

In Fig.~\ref{fig:rho_star_prof} we compare the observed and predicted stacked stellar mass density profiles.  Note that we have used the best-fit NFW parameters quoted by V15 to deproject their 2D surface mass density profile into a 3-D mass density profile, for comparison with the simulations.  We adopt a mimimum satellite stellar mass of $\log_{10}{M_*/{\rm M}_\odot} > 9.5$, which is similar to the observational sample has a completeness limit (note that the result is not sensitive to this choice, so long as the mimimum mass is below the break in the galaxy stellar mass function).  Overall, the simulations reproduce the shape and normalisation of the observed stellar mass density profile reasonably well.  There are hints of a discrepancy within $\approx0.1 r_{200}$, but it is unclear if this is a real effect (e.g., due to enhanced stripping of satellites in the simulations compared to real clusters) or issues with robustly identifying substructures at such high background densities (see, e.g., \citealt{Muldrew2011}).  In any case, over the vast majority of the cluster volume the simulated satellites have a similar spatial distribution to the observed satellite population without having performed any calibration (it is not clear how you could easily calibrate this in any case).

We fit the simulated stellar mass density profile over the radial range 0.$1 \le r/r_{200} \le 1$ with an NFW distribution and, similar to what is found from observations of local clusters, infer a concentration $c_{200} \approx 1.8$.  For reference, V15 find a best-fit concentration of $c_{200} = 2.03 \pm 0.2$.

\subsection{Dynamics of cluster satellite galaxies}

\begin{figure}
\includegraphics[width=0.995\columnwidth]{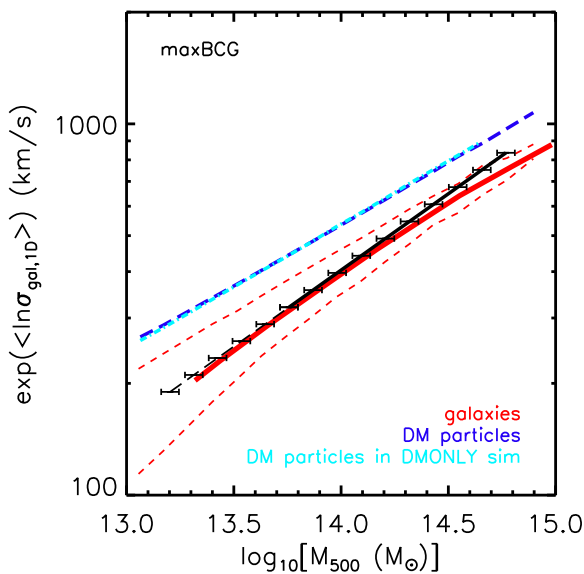}
\caption{\label{fig:sigma_mass}
The $z=0.1$ stacked velocity dispersion$-$$M_{500}$ relation of galaxy groups and clusters, compared with the best-fit power-law to SDSS observations (maxBCG).  The observed relation combines the best-fit power-law to the stacked weak lensing halo mass$-$richness relation \citep{Rozo2009} with the best-fit power-law to the stacked velocity dispersion$-$richness relation \citep{Becker2007}.  We compute and combine these relations in the same way using the simulations.  Overall, the simulated relation agrees remarkably well with the observed relation, with both showing clear evidence of a negative velocity bias with respect to the underlying dark matter distribution.
}
\end{figure}

We have so far considered the stellar content of massive systems, including the breakdown into contributions from centrals and satellites and how the satellites are distributed spatially in massive systems.  An interesting complementary test of the realism of the simulated massive systems is dynamics of the orbiting satellite population, which we now examine.

One of the largest and most well characterised group and cluster samples presently available is the optically-selected maxBCG sample \citep{Koester2007}.  We combine the best-fit power-law to the stacked velocity dispersion$-$richness relation from \citet{Becker2007} with the best-fit power-law to the stacked weak lensing$-$richness relation of \citet{Rozo2009} to derive an observed velocity dispersion$-$halo mass relation.  Note that because there is intrinsic scatter in both the mass$-$richness and velocity dispersion$-$richness relations, one must be careful to compare the same quantities for the simulations and observations.  Specifically, \citet{Becker2007} derive the mean of the log of the velocity dispersion in richness bins; i.e., $\mean{\log{\sigma_{\rm gal,1D}}}(N)$, while \citet{Rozo2009} derive the mean halo mass in richness bins; i.e., $\mean{M_{500,c}}(N)$ (see Appendix A of Rozo et al.).  To make a like-with-like comparison, we compute the stacked velocity dispersion$-$richness relation and $M_{500,c}$$-$richness relations from the simulations in the same way.  We use a simple richness estimate for the simulated clusters, which is the number of satellites with $M_* > 5\times10^{9} \ {\rm M}_\odot$ within $r_{500,c}$.  The velocity dispersion is calculated simply as the RMS of the 1D peculiar velocity distribution of these satellites.  The results are insensitive to other reasonable choices for the stellar mass threshold or host aperture (e.g., $M_* > 10^{10} \ {\rm M}_\odot$ and/or $r < r_{200,c}$).

In Fig.~\ref{fig:sigma_mass} we compare the predicted and observed $\sigma_{\rm gal,1D}$$-$$M_{500,c}$ relations.  The black solid transitioning to dashed line represents the combined stacked relations from the maxBCG studies.  The dashed portion of the curve represents an extrapolation of the stacked weak lensing mass$-$richness relation from a richness of 10 down to a richness of 3 (i.e., the stacked velocity dispersions were measured down to a richness of 3, but the weak lensing analysis was limited to richnesses $\ge 10$).  The horizontal error bars represent the 0.1 dex systematic error estimate of Rozo et al.\ on the stacked weak lensing masses.   The solid red curve represents the combined stacked relation from the simulations.  Note that for the simulations we also derived $\mean{\log{\sigma_{\rm gal,1D}}}(M_{500,c})$ (not shown), as opposed to combining $\mean{\log{\sigma_{\rm gal,1D}}}(N)$ and $\mean{M_{500,c}}(N)$, and find a virtually identical relation, implying that the precise richness definition is unimportant.  The dashed red curves enclose the central 68\% of the $\sigma_{\rm gal,1D}$ distribution in halo mass bins.  The thick dashed blue and dot-dashed cyan curves (which are nearly on top of each other) correspond to the mean velocity dispersion$-$halo mass relations using the dark matter {\it particles} within $r_{500}$ for our hydrodynamical simulations and for the corresponding dark matter-only simulation (respectively).  These relations are virtually identical to that previously derived by \citet{Evrard2008} based on a suite of dark matter-only simulations (after converting their masses from $M_{200,c}$ into $M_{500,c}$).  

Overall, the simulated relation agrees well with the observed relation, reproducing the observed trend over an order of magnitude in halo mass.  There are indications of a slight discrepancy for the highest-mass clusters, whose origin is likely tied to the adoption of pure power laws to describe the relations richness, velocity dispersion, and stacked lensing mass in the observations.  Interestingly, the simulated and observed satellite galaxy populations show clear evidence of a negative velocity bias with respect to the underlying dark matter distribution.

\subsection{Stellar mass autocorrelation function}

\begin{figure}
\includegraphics[width=0.995\columnwidth]{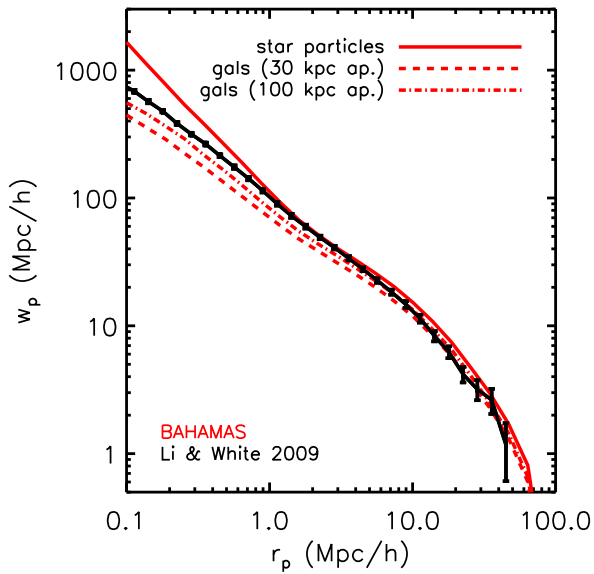}
\caption{\label{fig:stellar_acf}
The $z=0.1$ projected stellar mass autocorrelation function compared to SDSS measurements by \citet{Li2009}.  The dashed and dot-dashed red curves show the simulation predictions for different choices of aperture when computing the stellar masses of the simulated galaxies (30 and 100 kpc, respectively), while the solid red curve represents the autocorrelation derived from randomly-selected star particles.  The observed function is reproduced very well on large scales, while on small scales the level of agreement depends on the choice of aperture and tracer.}
\end{figure}

A final test we carry out on the distribution of stellar mass at low redshift is that of the projected stellar mass autocorrelation function.  This is similar to the 2-point correlation function of galaxies (`galaxy clustering'), but with a stellar mass weighting applied to each galaxy when counting galaxy pairs.  The clustering and autocorrelations serve as important independent checks on the models for a number of reasons.  First, since the clustering signal depends strongly on halo mass (with high-mass haloes being much more strongly clustered than low-mass haloes), the stellar mass autocorrelation, or galaxy clustering in bins of stellar mass, is sensitive to the stellar mass$-$halo mass relation, including its scatter and the relative contribution of centrals and satellites.  These correlation functions are also sensitive to the spatial distribution of satellites around centrals (probed by the `1-halo' term of the correlation function which dominates small projected separations), as well as to the underlying cosmology (probed by the `2-halo' term which dominates large separations).

Here we compare to the $z\approx0.1$ stellar mass autocorrelation derived from the SDSS by \citet{Li2009}.  We reproduce their methods (described in their Section 4) as closely as possible, using the same autocorrelation function estimator and method for generating the random galaxy catalog, and by adopting the same line of sight and projected distance binning strategies.

In Fig.~\ref{fig:stellar_acf} we compare the predicted and observed autocorrelations.  The red dashed and dot-dashed curves represent the autocorrelations derived from the simulated galaxy catalogs for 30 and 100 kpc apertures, respectively.  For comparison, the solid red line shows the autocorrelation of star {\it particles} in the simulation, derived by randomly selecting 5\% of the star particles in the simulation and applying the same methods used for the galaxy catalog (using the star particle masses as weights).  The black points with error bars connected by a solid black curve represent the measurements of \citet{Li2009}.

The predictions agree very well with the data at large projected separations ($r_P > 1$ Mpc/$h$) and are fairly insensitive to the choice of aperture or whether one uses the distribution of stars instead of the distribution of galaxies.  This is an important consistency check of the previous results.  \calsim~performs at least as well as previous studies based on semi-analytic models (e.g., \citealt{Campbell2015}) or subhalo abundance matching (e.g., \citealt{Conroy2006}) but without having been calibrated to do so. 

At small radii, the choice of aperture becomes important.  For our standard aperture choice of 30 kpc, for example, the predicted autocorrelation undershoots the observations somewhat (by $\approx50\%$ at 0.1 Mpc/$h$).  As this part of the function is dominated by satellite galaxies, this may signal a deficit of satellites at small projected separations, similar to that suggested by Fig.~\ref{fig:rho_star_prof}.  Whether this is a real effect (due to overly-efficient tidal stripping) or is due to deficiencies in the identification of substructures at close separations is not easy to tell.  Going to higher resolution simulations should address both of these issues at the same time.  

Interestingly, the autocorrelation of star particles (solid red) exceeds the observed autocorrelation of galaxies.  We should expect the star particle autocorrelation to provide an upper bound, since it samples all of the stellar mass in the simulation, including unidentified substructures and stellar mass erroneous assigned to the central galaxy, but also genuinely unbound (and typically not observed) stellar mass such as that is responsible for the ICL.  The fact that the star particle correlation function lies in excess of the observed autocorrelation therefore suggests that the discrepancy between the predicted and observed galaxy autocorrelations is not a fundamental one; i.e., we should expect the dashed and solid red curves to bracket the data (which they do) if the simulations have approximately the correct underlying stellar mass distribution.

\begin{figure}
\includegraphics[width=0.995\columnwidth]{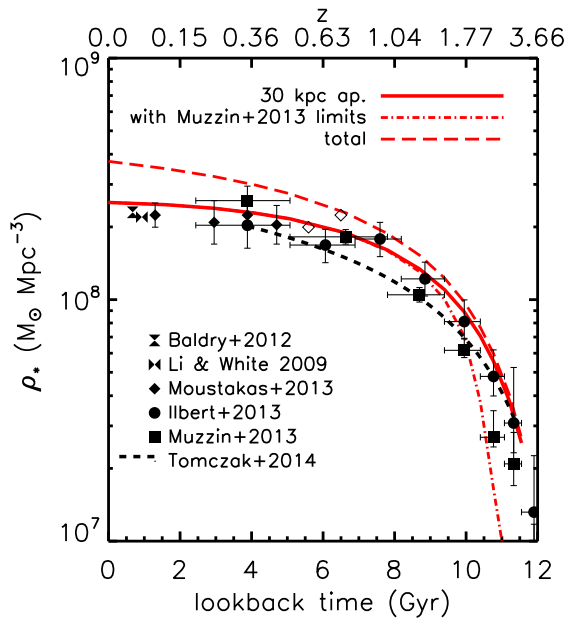}
\caption{\label{fig:rhostar_evol}
Evolution of the cosmic stellar mass density.  The solid red curve represents the predictions of the simulations using the fiducial 30 kpc aperture, while the long-dashed red curve corresponds to the total stellar mass in subhaloes.  The dot-dashed red curve represents the case of a 30 kpc aperture but with the observational mass limits of \citet{Muzzin2013} imposed on the simulations.  For comparison, we show recent GAMA \citep{Baldry2012} and SDSS \citep{Li2009,Moustakas2013} local measurements along with higher redshift data from the ZFOURGE/CANDELS \citep{Tomczak2014} and UltraVISTA/zCOSMOS \citep{Muzzin2013,Ilbert2013} surveys.  Note that for the observational data beyond $z\approx2$ (lookback time of $\sim10$ Gyr) the estimate of $\rho_*$ is really a lower limit, since the surveys are not sensitive to galaxies with $M_* \la 10^{10}$ M$_\odot$ (with the precise limit varying with redshift, whether the galaxy is star forming, and the survey details) and do not attempt to account for their contribution to the total stellar mass density.  The agreement with the observed evolution of the cosmic stellar mass density is good.
}
\end{figure}

\section{Evolution of the stellar universe}

We have constructed a simple model that reproduces many of the key diagnostics of the distribution of stellar mass in the local Universe.  While some of the diagnostics we examined were not independent of the local GSMF on which the feedback model was calibrated (such as the stellar mass fractions of central galaxies), other tests were (such as the satellite spatial distribution and kinematics, the stellar mass autocorrelation, and the contribution of centrals and satellites to the total stellar mass content).  Further independent tests of the model can be made by looking at the evolution of galaxies.  Here we focus on just a few basic tests, leaving a more detailed comparison with high-redshift measurements for future work.  In particular, we examine here the evolution of the GSMF and the overall cosmic stellar mass density (Section 4.1), as well as the evolution of the star formation rates of galaxies and the cosmic star formation rate density (Section 4.2).

\subsection{Evolution of stellar mass}

In Fig.~\ref{fig:rhostar_evol} we show the evolution of the cosmic stellar mass density, which is defined as the sum of the stellar mass of all galaxies per unit comoving volume.  We show the results for the fiducial 30 kpc aperture (solid red curve) as well as for the total (long-dashed red; i.e., all stellar mass bound to subhaloes in the simulations), integrating the simulations down to a stellar mass of $5\times10^9 \ {\rm M}_\odot$.  For comparison, we show recent local measurements from GAMA \citep{Baldry2012} and SDSS \citep{Li2009,Moustakas2013} along with high-z data from the ZFOURGE/CANDELS \citep{Tomczak2014} and UltraVISTA/zCOSMOS \citep{Muzzin2013,Ilbert2013} surveys.  Note that at $z \ga 2$ the surveys will generally miss a non-negligible fraction of the total stellar mass density due to the increasing stellar mass completeness limits with increasing redshift.  To get a rough idea of how this impacts the results, we have imposed the quoted stellar mass limits of \citet{Muzzin2013} as a function of redshift on the simulated population\footnote{Specifically, for a snapshot at a given redshift we integrate the stellar masses of all simulated galaxies above the observational stellar mass limit, where the latter is derived by interpolating the stellar mass limit vs.~redshift data of \citet{Muzzin2013}.} (dot-dashed red curves).

\begin{figure*}
\includegraphics[width=1.99\columnwidth]{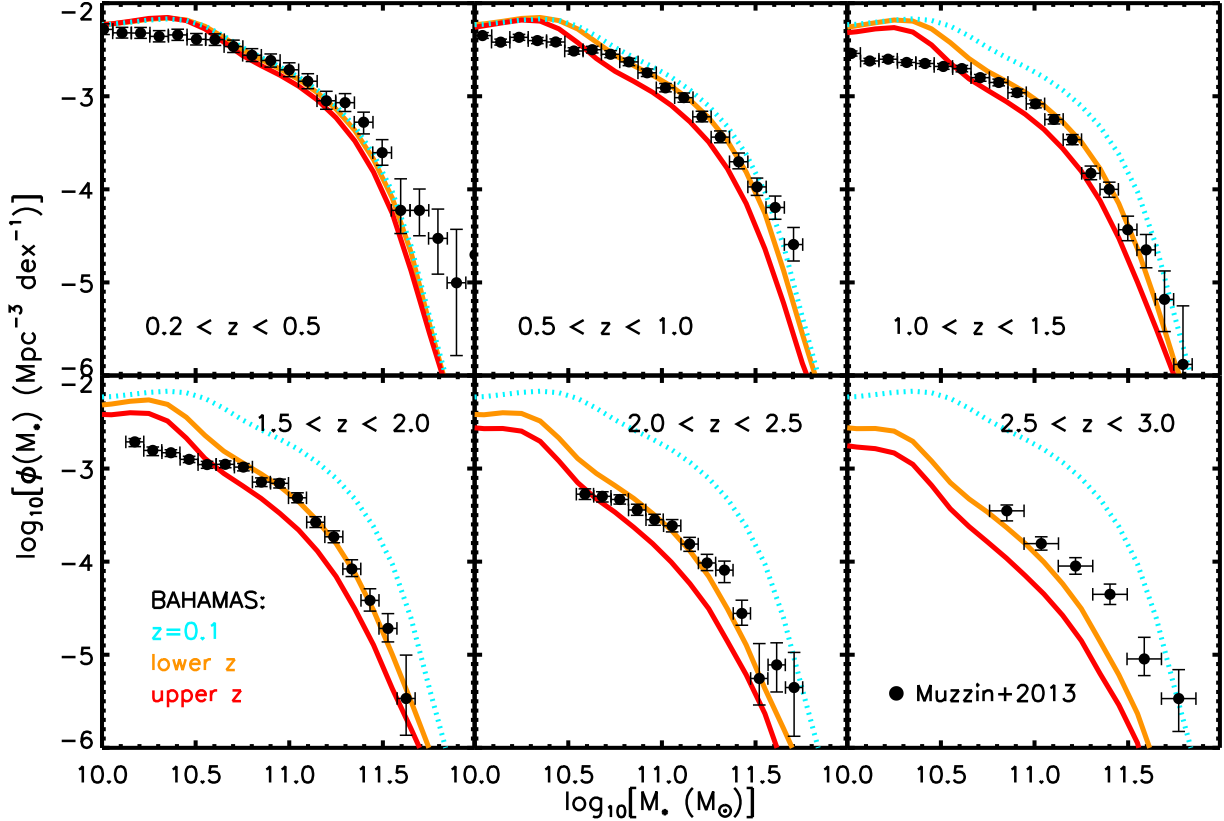}
\caption{\label{fig:GSMF_evol}
Evolution of the GSMF.  The black data points represent the UltraVISTA/zCOSMOS measurements of \citet{Muzzin2013}.  Note that observational measurements of the GSMF span redshift ranges whereas the simulation data is output at discrete redshifts (snapshots).  The solid curves represent the predictions of the simulations at various redshifts, with the lower (orange) and upper (red) redshifts bracketing the observational ranges.  For reference, the dotted cyan curve in each panel represents the simulated $z=0.1$ GSMF.  For $\log_{10}[M_*/{\rm M}_\odot] \ga 10.5$ the predictions are in reasonably good agreement with the observed evolution, with a deficit of the most massive galaxies at the highest redshifts.  At lower stellar masses (i.e., near the resolution limit), the simulations overpredict the abundance, particularly at $1 < z < 2$.  
}
\end{figure*}

Below $z\approx2$ (lookback time of $\approx10$ Gyr) the observational estimates of $\rho_*$ are expected to be robust to completeness issues.  Over this range of redshifts, the \calsim~simulations reproduce the observed total stellar mass density evolution well when we adopt the 30 kpc aperture (i.e., appropriate for comparisons to the observations).  We note that approximatetly 30\% of total stellar mass density within the 30 kpc aperture is contributed by low-mass galaxies with $\log_{10}[M_*/{\rm M}_\odot] \le 10.5$ locally, increasing up to two-thirds of the total by $z\sim2$.  A comparison with the total stellar mass density without imposing an aperture shows that at late times there is a significant contribution from stellar mass distributed over large spatial scales (e.g., in the ICL), which is qualitatively consistent with that found in other recent simulations (e.g., \citealt{Puchwein2013,Furlong2015}).

At higher redshifts ($z\ga1$), the predicted trend (solid red curve) lies slightly above the observations.  However, we note that observational surveys can probe only relatively massive galaxies at high redshift, and one should account for this selection effect when comparing the simulations and observations.  We can see that with a simple accounting of the stellar mass limits of \citet{Muzzin2013} (dot-dashed curve; i.e., we integrate the simulations down to the galaxy stellar mass completeness limits of \citealt{Muzzin2013}) that most of the small discrepancy at high-z is indeed probably due to observational completeness issues.

We now turn to the evolution of the GSMF.  Note that the cosmic stellar mass density at a given redshift, $\rho_*(z)$, is derived by simply integrating over the GSMF at that redshift.  It is therefore interesting to see if the simulations, which reproduce the integrated stellar mass density reasonably well, also reproduce the detailed distribution of galaxy masses as a function of redshift.

In Fig.~\ref{fig:GSMF_evol} we compare the predicted GSMF with observations over a range of redshifts.  The black data points represent the UltraVISTA/zCOSMOS measurements of \citet{Muzzin2013}.  Note that observations measure the GSMF in redshift intervals (e.g., $0.2 < z < 0.5$) whereas the simulations sample the GSMF at discrete redshifts (in snapshots).  We therefore plot the predicted GSMF at two different redshifts that bracket the observational ranges (orange and red curves represent the lower and upper redshifts, respectively).  For reference, the dotted cyan curve in each panel represents the simulation GSMF at $z=0.1$.

For stellar masses of $\log_{10}[M_*/{\rm M}_\odot] \ga 10.5$ the predictions are in reasonably good agreement with the observed evolution.  There is an indication of a deficit of massive galaxies at the highest redshifts.  (At fixed abundance, this implies the most-massive simulated galaxies at high redshift are up to 0.2 dex less massive than observed.).  At $\log_{10}[M_*/{\rm M}_\odot] \la 10.5$, the simulations strongly overpredict the observed abundance at $z\ga1$.

Examining the various panels (compare the offset of the solid red and orange curves with respect to the dotted cyan curve), one can see that the abundance of the simulated low-mass galaxies has not changed significantly since $z\approx2$.  A likely explanation for this behaviour is that these systems suffer from inefficient feedback which is plausibly due to poor sampling/mass resolution in the simulations.  That is, before feedback can have a significant impact on its surroundings there must be sufficient sources of feedback present.  The first generation of star formation in the simulations therefore has no chance of being regulated by feedback and therefore if the mass resolution is too low this will result in overcooling near the resolution limit (see \citealt{Schaye2015} for further discussion).

The magnitude of the offsets between the simulations and observations at low and high masses is, however, relatively modest and the level of agreement over this mass range is as good as that reported for other recent (generally much higher resolution) simulation studies such as EAGLE (see \citealt{Furlong2015}) and Illustris (see \citealt{Genel2014}).

\begin{figure}
\includegraphics[width=0.995\columnwidth]{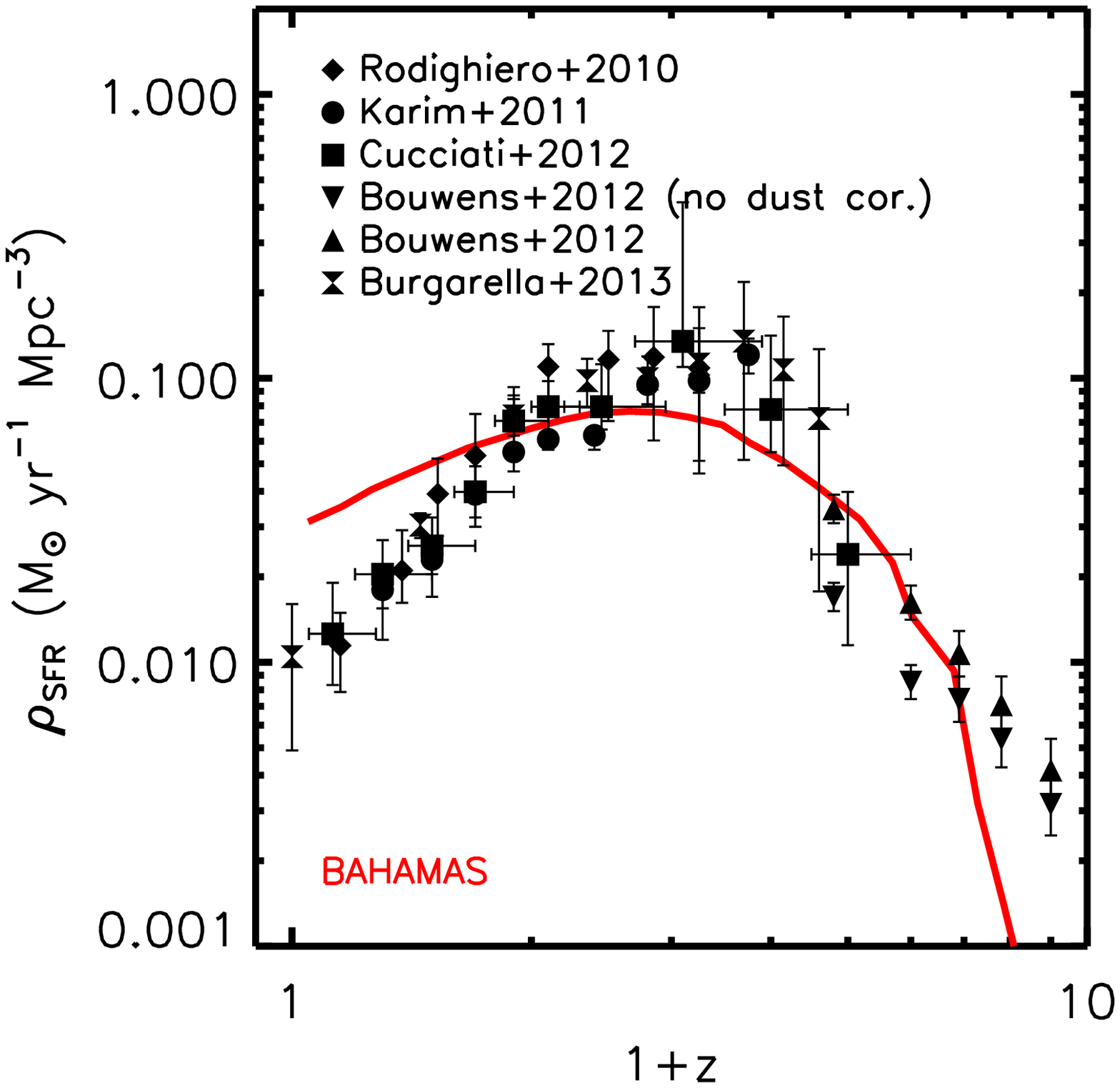}
\caption{\label{fig:rhosfr_evol}
Evolution of the cosmic star formation rate density (SFRD).  The solid red curve represents the predictions of the simulations.  For comparison, we show recent measurements from \citet{Rodighiero2010} (IR), \citet{Karim2011} (radio), \citet{Cucciati2012} (UV), \citet{Bouwens2012} (UV), and \citet{Burgarella2013} (IR+UV).  The simulations qualitatively reproduce the observed cosmic SFRD trend, but  they underpredict the peak at $z\approx2$ somewhat and overpredict the SFRD at late times.  Evidently these differences are not sufficiently large to have had significant effects on the predicted stellar mass evolution (see Figs.~\ref{fig:rhostar_evol} and \ref{fig:GSMF_evol}).
}
\end{figure}

\subsection{Evolution of star formation rates}

We now turn to the evolution of star formation rates.  In Fig.~\ref{fig:rhosfr_evol} we show the evolution of the cosmic star formation rate density (SFRD), defined as the total star formation rate (i.e., summed over all star forming gas particles in the simulation) per unit comoving volume.  We show the results for the fiducial 30 kpc aperture (solid red curve), but we note that changing the aperture has essentially no effect on the result since all of the star formation is located near the centers of dark matter (sub)haloes.  For comparison, we show a range of recent measurements which use different SFR tracers, including \citet{Rodighiero2010} (IR), \citet{Karim2011} (radio), \citet{Cucciati2012} (UV), \citet{Bouwens2012} (UV), and \citet{Burgarella2013} (IR+UV).  We have adjusted the measured SFRs to correspond to a Chabrier IMF.  Generally speaking there is good consistency between the different studies in spite of the fact that they use different tracers, although it should be noted that the employed scalings between luminosity and SFR have all been calibrated on essentially the same local galaxies \citep{Kennicutt1998}.

On a qualitative level, the simulations show a similar trend to the observations, with rates increasing strongly between $z\approx9$ and $z\approx3$, effectively plateauing between $z\approx3$ and $z\approx1$, and then declining towards the present day.  In detail, however, the simulations underpredict the peak of the SFRD somewhat and significantly overpredict the SFRD at late times.  Interestingly, there are no large offsets with respect to the observed evolution of the stellar mass density (see Fig.~\ref{fig:rhostar_evol}).  This may be because the SFRs are just generally lower at low redshifts and the issue only arises fairly late ($z\la0.5$), so that the increase in stellar mass over that already formed prior to $z\approx0.5$ is relatively small.  While rectifying this problem would be desirable, it is not essential for our purposes.  Our aim is to calibrate the feedback so that the simulated haloes have approximately the correct stellar and hot gas mass fractions in order to ensure that the effects of feedback on the underlying total matter distribution have been (approximately) correctly captured.

\begin{figure*}
\includegraphics[width=1.99\columnwidth]{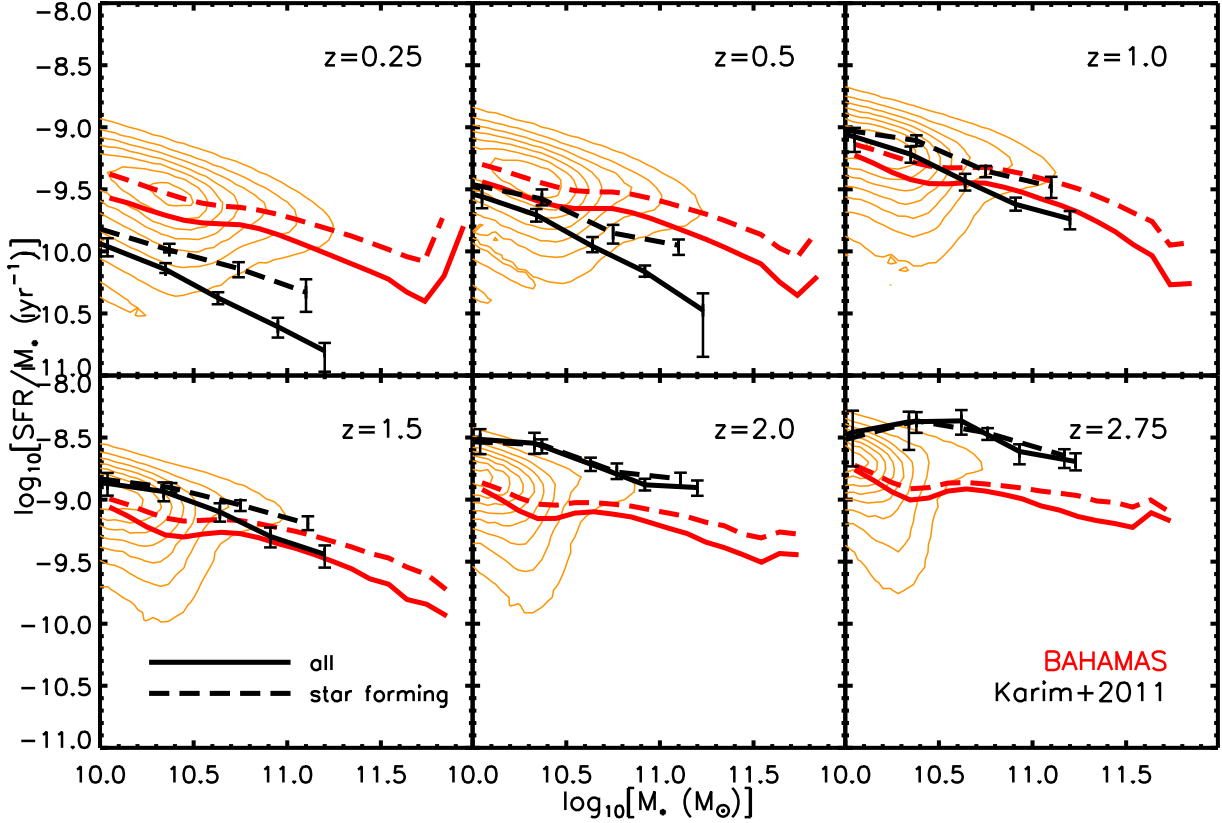}
\caption{\label{fig:sSFR_evol}
Evolution of the sSFR-$M_*$ relation.  The thick black curves correspond to the mean relations of \citet{Karim2011} derived from stacking of radio data (solid is for all galaxies and dashed corresponds to star forming galaxies).  The thick red curves represent the predicted mean relations for all (solid) and star forming (dashed) galaxies, while the thin orange contours delineate the (log of the) number density of galaxies in bins of (sSFR,$M_*$).  While generally reproducing the observed mild slope between the sSFR and $M_*$ (as well as the offset between the relations for the star forming and total galaxy populations), the simulations underpredict the rate of evolution of the amplitude of the star forming main sequence, consistent with the cosmic SFRD comparison in Fig.~\ref{fig:rhosfr_evol}.
}
\end{figure*}

Finally, we turn to the evolution of the distribution of SFRs.  Specifically, in Fig.~\ref{fig:sSFR_evol} we compare the observed and predicted mean sSFR in bins of stellar mass with the radio stacking results of \citet{Karim2011}.  \citet{Karim2011} derived two estimates of the mean sSFR in bins of stellar mass, one corresponding to the total (mass-selected) sample (black solid curves) and the other corresponding to just the star forming population (black dashed curves).  We compute the corresponding curves for the simulations (thick red curves), using a threshold of $10^{-11}$ yr$^{-1}$ in sSFR to separate between star forming and not.  In addition to computing the mean sSFR-$M_*$ relation for the simulations, we also show the distribution as a set of orange contours, which trace the log of the number density of simulated galaxies in (logarithmic) bins of sSFR and $M_*$.

The simulations successfully reproduce the observed mild trend (slope) between the sSFR and $M_*$ (which does not evolve significantly with redshift), as well as the magnitude of the offset between the relations of the star forming and total populations.  However, it is evident that they underpredict the rate of evolution of the amplitude of the relation for the star forming main sequence compared to what is measured observationally.  This is fully consistent with (and in fact drives) the differences between the predicted and observed cosmic SFRDs in Fig.~\ref{fig:rhosfr_evol}.  

\begin{figure*}
\includegraphics[width=0.995\columnwidth]{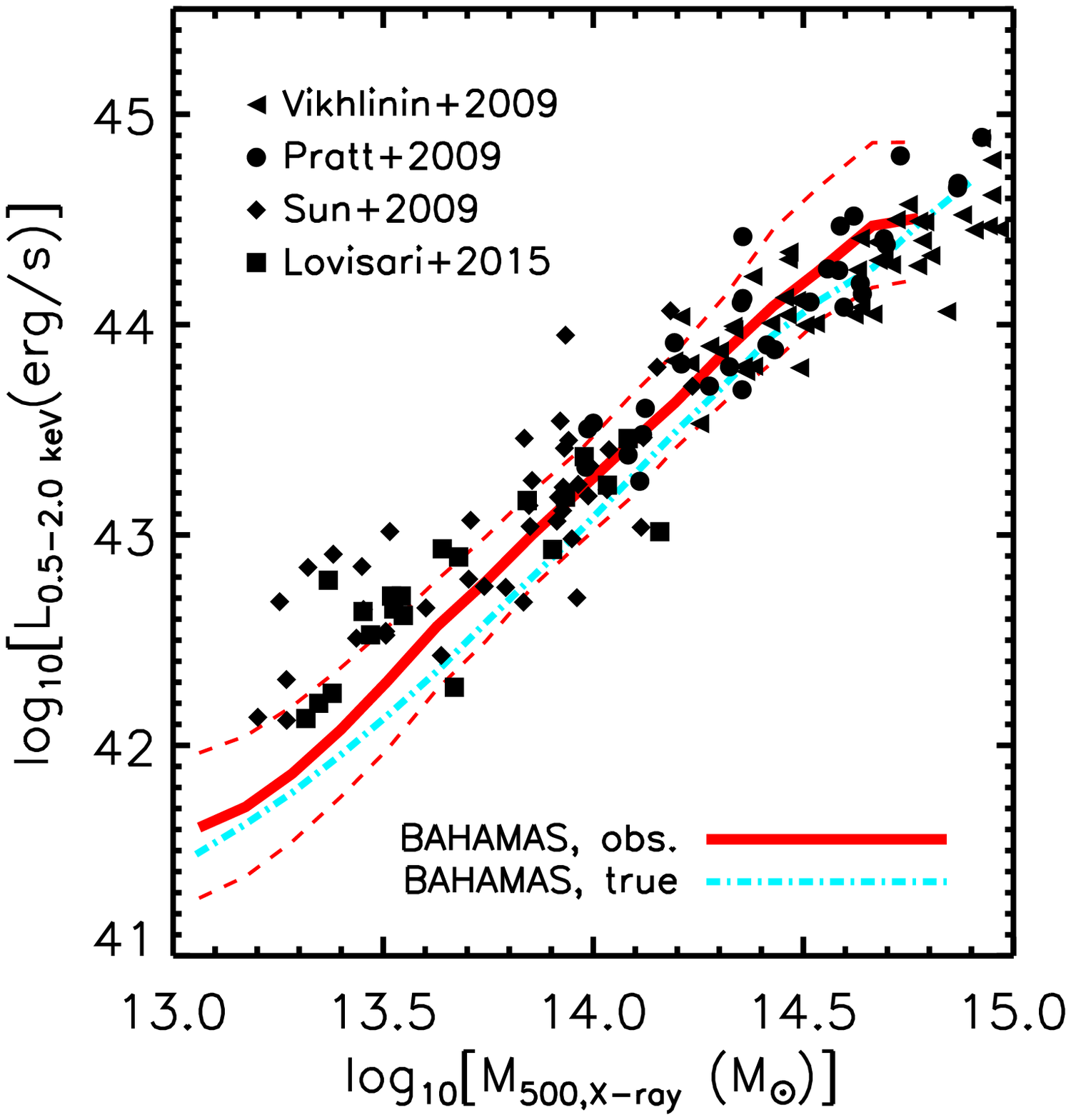}
\includegraphics[width=0.995\columnwidth]{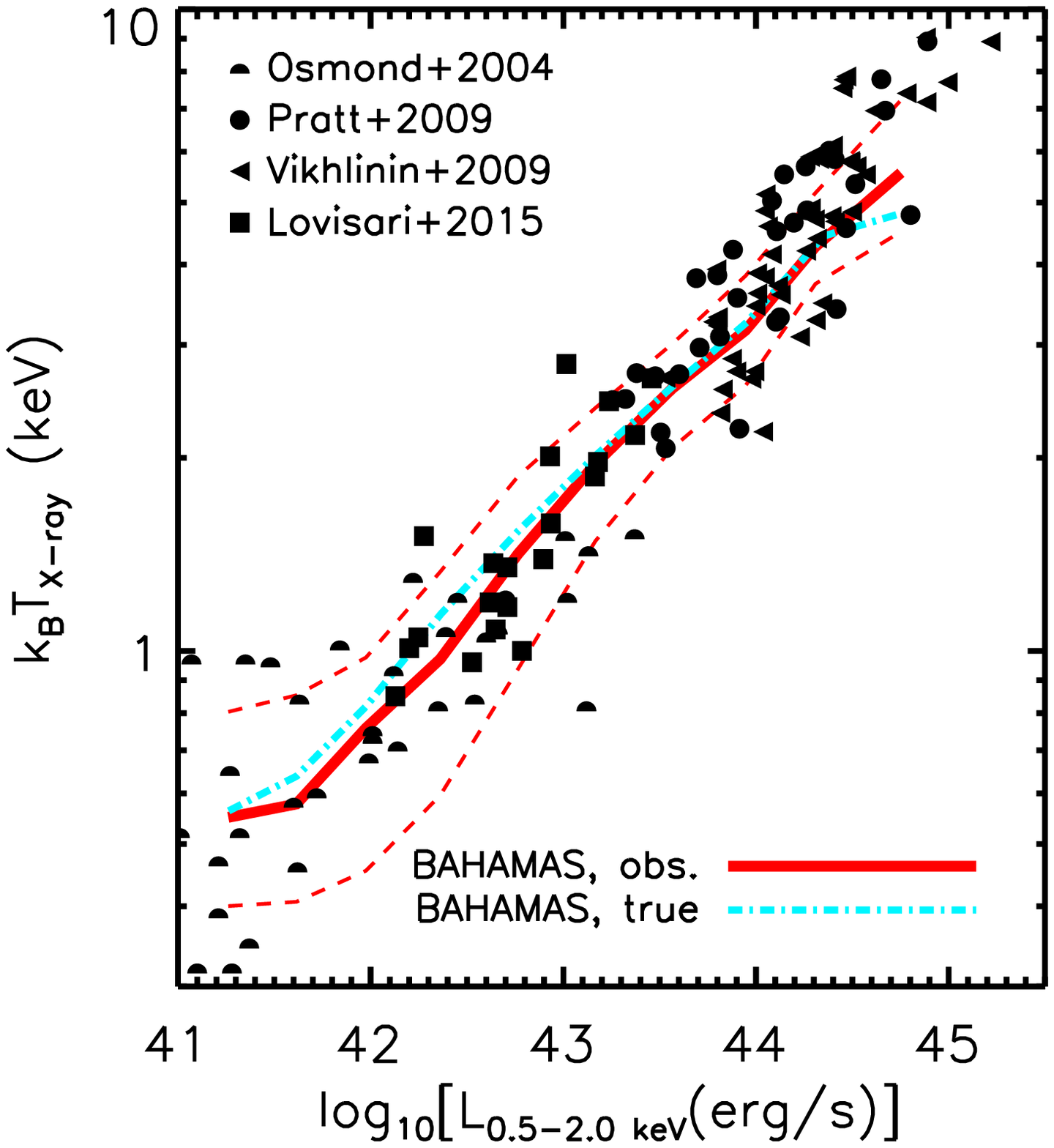}
\caption{\label{fig:xray_scalings}
The X-ray luminosity$-$halo mass and X-ray luminosity$-$temperature relations of groups and clusters, compared to local X-ray samples.  The red curves (solid represents the median and dashed enclose 68\% of the population) represent the relations derived from a synthetic X-ray analysis of a mass-limited sample (all haloes with $M_{500,{\rm true}} > 10^{13} \ {\rm M}_\odot$).  The cyan curve represents the true relation (i.e., not processed through synthetic X-ray observations).  The two observed scaling relations (both their median and scatter) are recovered very well.
}
\end{figure*}

A simple explanation for why the simulations overpredict the star formation rates at late times, is that the calibration to the $z=0$ GSMF forces them to compensate for the lower than observed star formation rates near the peak of the cosmic SFRD at $z\approx2$.  If the star formation rates at higher redshifts are lower than observed (due, e.g., to an incomplete/inaccurate feedback model and/or relatively poor resolution), then there must be more late time star formation to end up with the correct distribution of stellar masses today.  However, this discussion leaves aside the apparent relatively good agreement with the observed evolution of the stellar mass density in Fig.~\ref{fig:rhostar_evol}.  Clearly, further exploration of these issues is warranted and we leave this for future work, noting that the level of overall agreement between the predicted and observed evolution of stellar masses and star formation rates is comparable with other recent simulation campaigns done in much smaller volumes but with significantly higher numerical resolution.

\section{The hot gas$-$halo connection}

In this section we explore the hot gas properties of massive dark matter haloes, making comparisons to recent X-ray and thermal Sunyaev-Zel'dovich (SZ) effect measurements.    Although we have made comparisons involving all of the observables explored in L14 using the new calibrated \calsim~model, we present only a subset of them here (arguably the most important tests).  We note, however, that the new calibrated model performs at least as well as the successful `AGN 8.0' model of L14 for all of the other observables explored in that study but with the important advantage of also reproducing key properties of the galaxy population, as demonstrated in Sections 3 and 4.  A complementary test of the model, to close the loop, is to compare the predicted and observed relations between galaxies and their hot gas haloes, which we will present in Section 6.

\subsection{Synthetic X-ray observations}

We use our synthetic observation pipeline (described in detail in L14) to post-process the simulations to make like-with-like comparisons to X-ray observations.  We provide a brief description of the pipeline here and refer the reader to L14 for a more detailed description.

For each gas particle within a group/cluster we compute a 0.5--10.0 keV band X-ray spectrum using the Astrophysical Plasma Emission Code \citep[APEC;][]{Smith2001} with updated atomic data and calculations from the AtomDB v2.0.2 \citep{Foster2012}. The spectrum of each gas particle is computed using the particle's density, temperature, and full abundance information.  Note that we exclude cold gas below $10^5$ K which contributes negligibly to the total X-ray emission.  We also exclude any (hot or cold) gas which is bound to satellites, as observers also typically excise substructures from X-ray data.  Note that the smallest subhaloes that can be resolved in the simulations have total masses $\sim 10^{11}~\textrm{M}_\odot$.

We measure gas density, temperature, and metallicity profiles for each simulated system in an observationally-motivated way, by fitting single-temperature APEC models with a metallicity that is a fixed fraction of Solar to spatially-resolved X-ray spectra in radial bins.  The radial bins are spaced logarithmically and we use between 10-20 bins within $r_{500}$, similar to what is possible for relatively deep {\it Chandra} observations of nearby systems.  To more closely mimic the actual data quality and analysis, the cluster and model spectra are multiplied by the effective area energy curve of {\it Chandra}, subjected to Galactic absorption due to HI with a typical column density of $2\times10^{20}$ cm$^2$, and re-binned to an energy resolution of 150 eV. The single-temperature model spectra are fitted to the cluster spectra using the \textsc{mpfit} least-squares package in \textsc{idl} \citep{Markwardt2009}. 

In addition to deriving profiles, we also derive global system X-ray temperatures and metallicities by following the above procedure but using only a single radial bin: either [0--1]$r_{500}$ (`uncorrected') or [0.15--1]$r_{500}$ (`cooling flow-corrected').  System X-ray luminosities within $r_{500}$ are computed in the soft $0.5-2.0$ keV band by summing the luminosities of the individual particles within that radius.

When making comparisons to X-ray-derived mass measurements, we employ a hydrostatic mass analysis of the simulated systems using the measured gas density and temperature profiles inferred from the synthetic X-ray analysis described above. Specifically, we fit the density and temperature profiles using the functional forms proposed by \citet{Vikhlinin2006} and assume hydrostatic equilibrium to derive the hydrostatic mass profile. We will use the subscript `X-ray' to denote quantities inferred from synthetic observations under the assumption of hydrostatic equilibrium.  Consistent with the findings of previous studies (e.g., \citealt{Rasia2006,Nagai2007,Battaglia2013}; L14; \citealt{Biffi2016,Henson2016}), we measure a median hydrostatic X-ray to true mass ratio of $0.84$ within $r_{500}$ for all systems with a true mass exceeding $10^{13} \ {\rm M}_\odot$.  Note, however, that the intrinsic scatter about this ratio is significant, with a standard deviation of $\approx40\%$.  Generally one can therefore not simply adopt a single value for the bias, as this will neglect the scattering between mass bins.

\subsection{X-ray scaling relations}

In Fig.~\ref{fig:xray_scalings} we compare the predicted X-ray luminosity$-$halo mass (left panel) and X-ray luminosity$-$temperature (right panel) relations with that of local X-ray-selected groups and clusters \citep{Osmond2004,Vikhlinin2009,Pratt2009,Lovisari2015}.  X-ray luminosities are computed in the 0.5-2.0 keV band.  Note that even though we compute the X-ray quantities in an observational manner, we do not {\it select}\footnote{Occasionally observational X-ray studies focus on systems with a ``relaxed'' X-ray morphology; i.e., systems that appear to be more or less circularly symmetric.  We have elected not to select a relaxed subset of simulated clusters for comparison to the observations for the following reasons: i) there is no unique and well-defined observational definition of what it means to be relaxed (e.g., how close to symmetric must the X-ray morphology be?); ii) in any case, observational studies do not just select based on relaxation state but also on other important criteria (e.g., surface brightness); iii) the relation between observational diagnostics and simulation diagnostics of relaxation is murky; and iv) we have found that when adopting a simple ``simulator's'' relaxation diagnostic (specifically, the kinetic-to-thermal energy ratio of the ICM) that there were only very minor differences in the resulting scaling relations and profiles when selecting the relaxed subsample compared to selecting all systems.  For these reasons, we have not focused on a relaxed subsample for comparison to the observations.} the simulated clusters in the same way as the observed systems, which may be particularly relevant for group samples \citep{Lovisari2015}, where generally only the X-ray-brightest systems will have estimates of mass and temperature available.  We plot the results for all simulated systems with a hydrostatic X-ray mass $M_{\rm 500,X-ray} > 10^{13} \ {\rm M}_\odot$, of which there are 51,964 systems distributed over the four independent cosmological volumes.

The \calsim~simulation reproduces the two observed X-ray scaling relations well over approximately 3 orders of magnitude in X-ray luminosity (or two in halo mass and 1.5 in temperature), although there is an indication of a mild over-prediction of the X-ray luminosities at the very highest masses and temperatures (see also \citealt{Barnes2016}).  The instrinsic scatter of the relations are also reasonably well recovered.
 
\begin{figure}
\includegraphics[width=0.995\columnwidth]{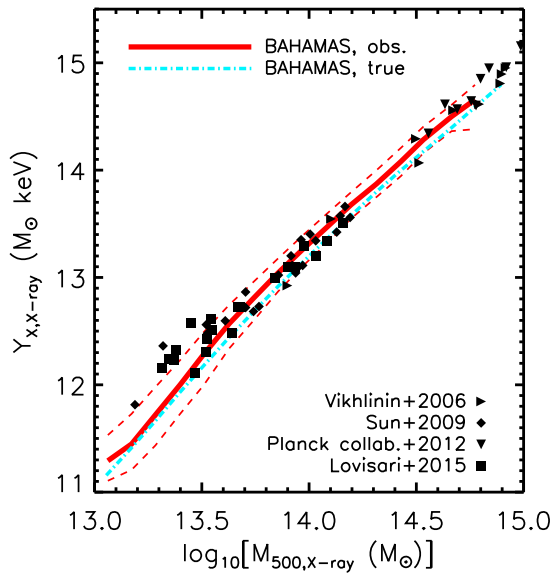}
\caption{\label{fig:xray_scalings2}
The $Y_X$$-$halo mass relations of groups and clusters, compared to local X-ray samples ($Y_X$ is defined as the product of the gas mass and core-excised temperature of the ICM).  The red curves (solid represents the median and dashed enclose 68\% of the population) represent the relations derived from a synthetic X-ray analysis of a mass-limited sample (all haloes with $M_{500,{\rm true}} > 10^{13} \ {\rm M}_\odot$).  The cyan curve represents the true relation (i.e., not processed through synthetic X-ray observations).  The observed scaling relation (including median and scatter) is recovered well.
}
\end{figure}

In Fig.~\ref{fig:xray_scalings2} we compare the predicted $Y_X$$-$halo mass relation with that of local X-ray-selected groups and clusters \citep{Vikhlinin2006,Sun2009,Pratt2009,Planck2012,Lovisari2015}.  Note that $Y_X$ is defined as the product of the (hot) gas mass within $r_{500}$ and the (core-excised) temperature measured within $[0.15-1.0]r_{500}$ and is often adopted as a total mass proxy due to its low intrinsic scatter with halo mass \citep{Kravtsov2006}.  

The simulations reproduced the observed $Y_X$$-$halo mass relation over approximately two orders of magnitude in halo mass (3 in $Y_X$).  There is perhaps an indication of a slight underestimate of the $Y_X$ for the lowest-mass groups compared to \citet{Lovisari2015} but, as already noted, we have not selected the simulated groups/clusters in an observational way.  

Power law fits (including errors and intrinsic scatter) to the above X-ray scaling relations, as well as to other combinations of these variables, can be found in \citet{Barnes2016}, who do a combined analysis of \calsim~and the \mac~suite of zoomed high-mass cluster simulations (which uses the \calsim~calibrated feedback model and adopts the same cosmology).

Evidently, calibrating the feedback to reproduce the observed gas mass fraction (see Fig.~\ref{fig:calibrated}) is all that is required for the model to reproduce these and other related scaling relations simultaneously (the simulations reproduce the various combinations of hydrostatic mass, temperature, gas mass, X-ray luminosity and $Y_X$).  This is non-trivial, particularly in the case of X-ray luminosity, since it is primarily set by the density of gas in the very central regions, whereas most of the gas mass within $r_{500}$ is at much larger radii.  

Using the OWLS AGN model \citet{McCarthy2011}, presented a simple picture for why there is such a close physical connection between the small-scale (e.g., X-ray luminosity) and large-scale (e.g., total gas mass) properties of the ICM.  Specifically, they showed that the vast majority of the gas expulsion done by AGN feedback occurred at high redshift in the progenitors of groups and clusters, during the peak of cosmic black hole accretion/growth.  This `quasar mode' feedback efficiently ejects the lowest-entropy (highest-density) gas from the progenitors, which otherwise would have significantly cooled and formed stars and/or ended up in the highest-density regions of the ICM (convective stability demands this).  By contrast, the hot gas that ends up forming the ICM in groups and clusters at the present time is that which was {\it not} significantly affected by this mode of feedback.  Its density distribution (and therefore X-ray luminosity) is set primarily by the entropy acquired via gravitational shock heating during accretion, with late time `radio mode' AGN feedback effectively preserving this configuration.  Thus, it is the effectiveness of the quasar mode feedback which dictates precisely how much of the low-entropy gas ends up in the central ICM today.  By calibrating the feedback to reproduce the overall gas fractions, we are effectively calibrating the amount of low-entropy gas that gets removed from the system and this is likely why the central regions are also faithfully reproduced by the simulations (we show radial distributions below).

\subsection{SZ scaling relations}

\begin{figure}
\includegraphics[width=0.995\columnwidth]{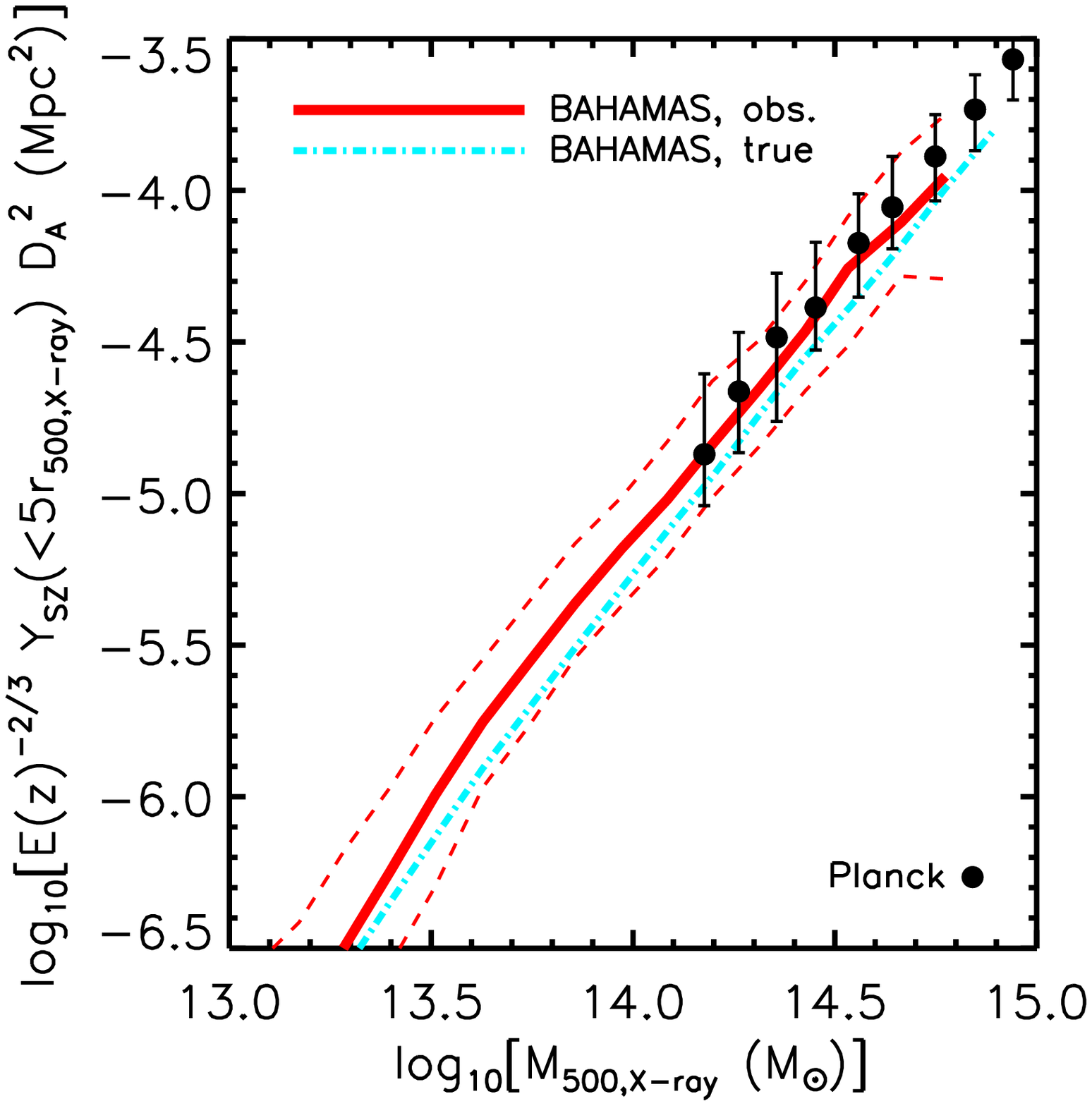}
\caption{\label{fig:tsz_scaling}
The predicted local SZ flux$-$halo mass relation compared with that derived for local clusters ($z<0.25$) from the \planck~second catalog of SZ sources (the PSZ2 Union catalog, \citealt{Planck2015b}).  The red curves (solid represents the median and the dashed enclose 68\% of the population) represents the relation derived from the simulation  of a mass-limited sample (all haloes with $M_{500,{\rm true}} > 10^{13} \ {\rm M}_\odot$).  In accordance with the observational analysis, we use (synthetic) X-ray observations to derive the halo mass and the aperture within which the SZ flux is calculated.  The dot-dashed cyan curve represents the true SZ flux$-$halo mass relation from the simulation, using the true $5 r_{500}$ aperture to derive the SZ flux.  The observed relation is reproduced well by the simulations when they are analysed in a like-with-like fashion to the observational data.
}
\end{figure}

We now move on to a comparison of the hot gas properties at larger scales, specifically with SZ effect observations from \planck.  In Fig.~\ref{fig:tsz_scaling} we compare the predicted integrated SZ flux ($Y_{\rm SZ}$)$-$halo mass relation with that derived from the most recent version of the \planck~second catalog of SZ sources\footnote{http://pla.esac.esa.int/pla/} (Union catalog v2.08; \citealt{Planck2015b}).  

As a reminder, the integrated SZ flux within a 3-D radius $R$ is defined as:
\begin{equation}
Y_{\rm SZ}(<R) \ D_{A}(z)^{2} =\frac{\sigma_T}{m_ec^2}\int_0^R P_e(r)dV
\end{equation}
\noindent where $D_A$ is the angular diameter distance of the cluster, $\sigma_T$ is the Thomson cross-section, $c$ the speed of light, $m_e$ the electron rest-mass and $P_e=n_ek_BT_e$ is the electron pressure with $k_B$ being the Boltzmann constant.  Thus, the integrated Compton y parameter (SZ `flux') is directly proportional to the total thermal energy of the hot gas.

From the \planck~catalog we select local clusters with $z < 0.25$, which are not heavily IR contaminated by cold gas clumps in the Galaxy (IR FLAG$ = 0$), that have a neural network quality flag of Q NEURAL$ > 0.4$ (the recommended quality threshold), and that have a $M_{500}$ estimate.  These cuts reduce the original number of 1653 clusters down to a sample of 616 clusters.  The catalog provides estimates of the integrated SZ flux within $5 r_{500}$ [$Y_{\rm SZ}(<5r_{500})$ in arcmin$^2$].  Note that the mass estimate, and the corresponding adopted aperture $5 r_{500}$ within which the SZ flux is measured, are derived by adopting the X-ray $Y_X$$-$$M_{500}$ scaling relation of \citet{Arnaud2010}.  We scale the observed SZ fluxes by the square of the angular diameter distance of each cluster to remove the explicit redshift dependence of the SZ `flux'.  Furthermore, we apply a self-similar scaling of $E(z)^{-2/3}$ [where $E(z) \equiv H(z)/H_0 = \sqrt{\Omega_m(1+z)^3+\Omega_\Lambda}$~] to account for the variation in the mean density of clusters as a function of redshift, which is just due to the evolution of the background critical density.  Note, however that, since we have selected only local clusters, scaling the SZ fluxes by $D_A^2 E(z)^{-2/3}$ has only a small effect on the resulting $Y_{\rm SZ}(<5r_{500})$$-$$M_{500}$ relation, apart from the overall amplitude shift.  The filled black circles with error bars in Fig.~\ref{fig:tsz_scaling} correspond to the median and 1-sigma intrisinc scatter of the \planck~clusters.

\begin{figure*}
\includegraphics[width=0.995\columnwidth]{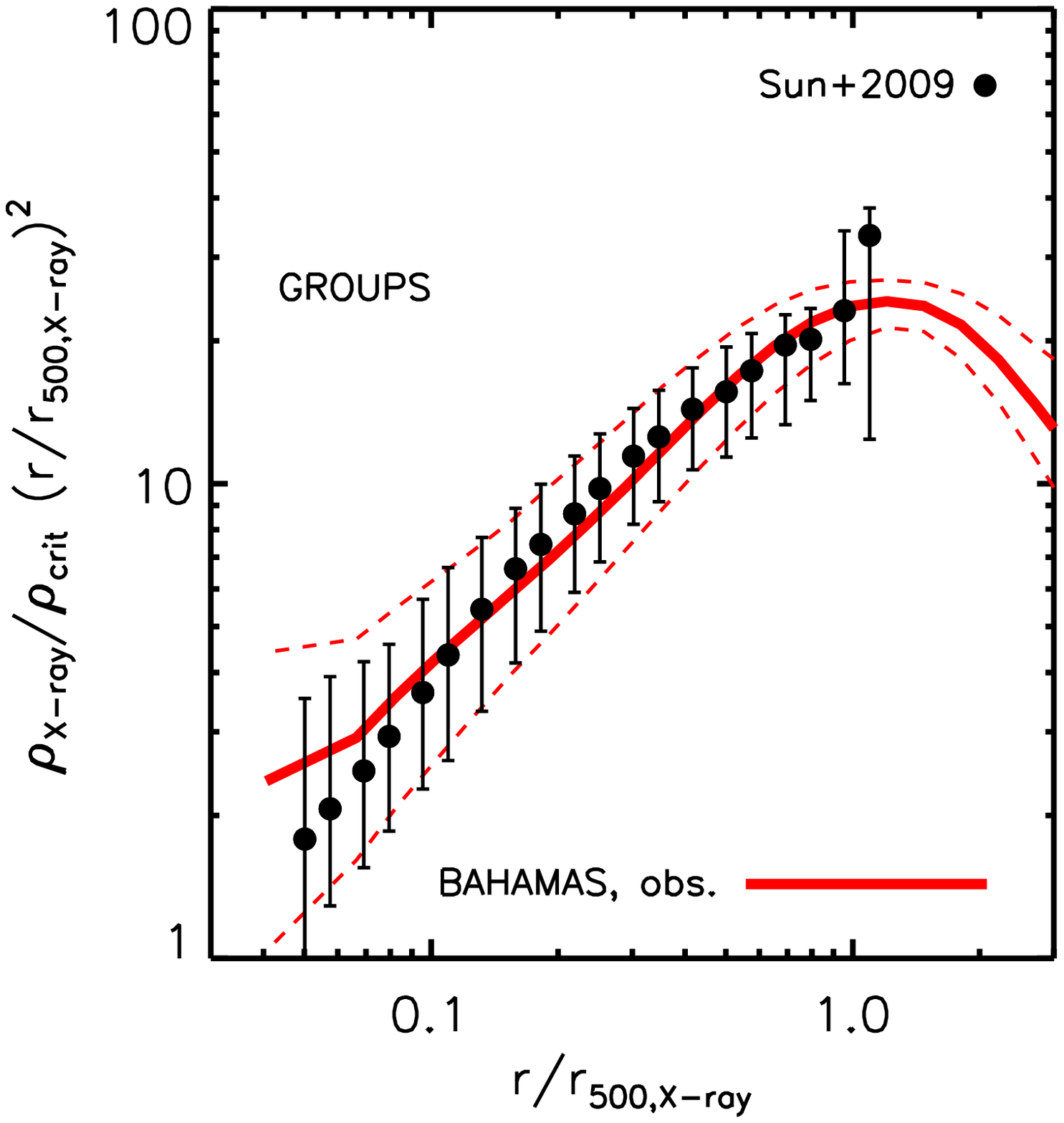}
\includegraphics[width=0.995\columnwidth]{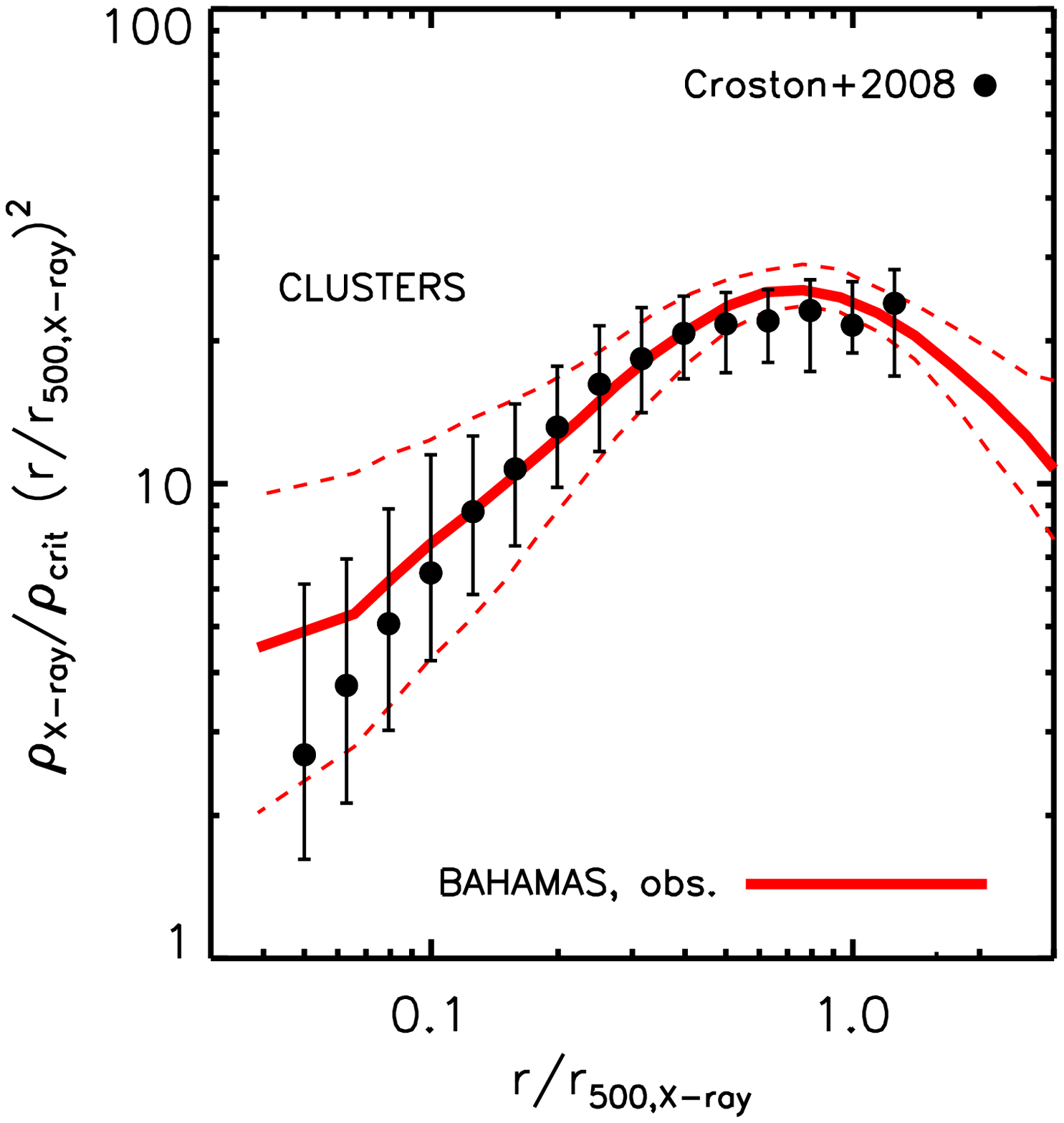}
\caption{\label{fig:rho_profs}
Hot gas density profiles of galaxy groups (left) and clusters (right) compared with local X-ray samples.  The red curves (solid represents the median and the dashed enclose 68\% of the population) represent the predicted gas density profiles for a sample of systems which have the same median halo mass as the observational samples (\citealt{Sun2009} in the case of groups and \citealt{Croston2008} for clusters).  The filled black circles with error bars represent the median and 1-sigma intrinsic scatter for the observational samples.  The observed profiles, including the intrinsic scatter, are reproduced remarkably well.
}
\end{figure*}

\begin{figure*}
\includegraphics[width=0.995\columnwidth]{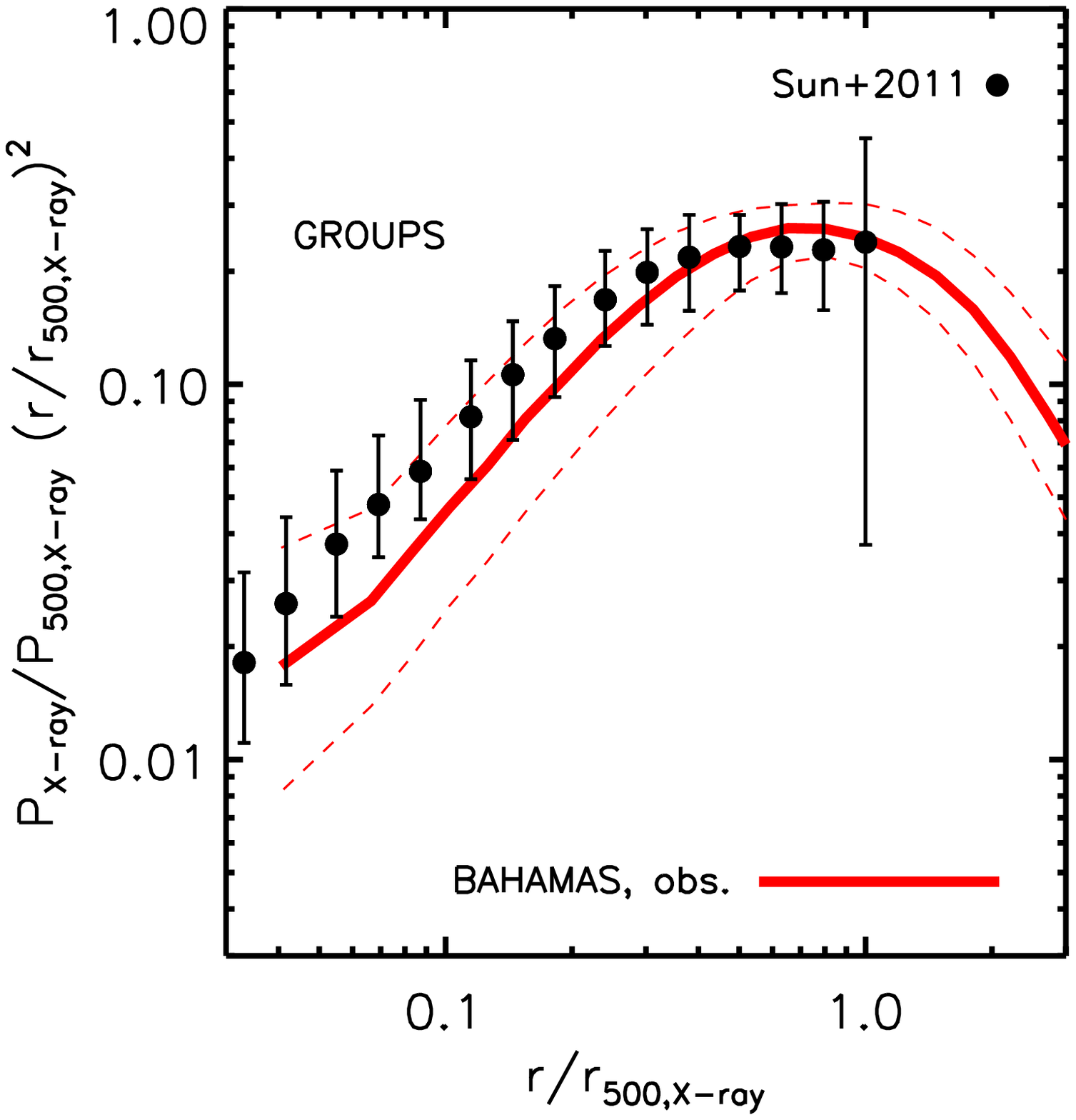}
\includegraphics[width=0.995\columnwidth]{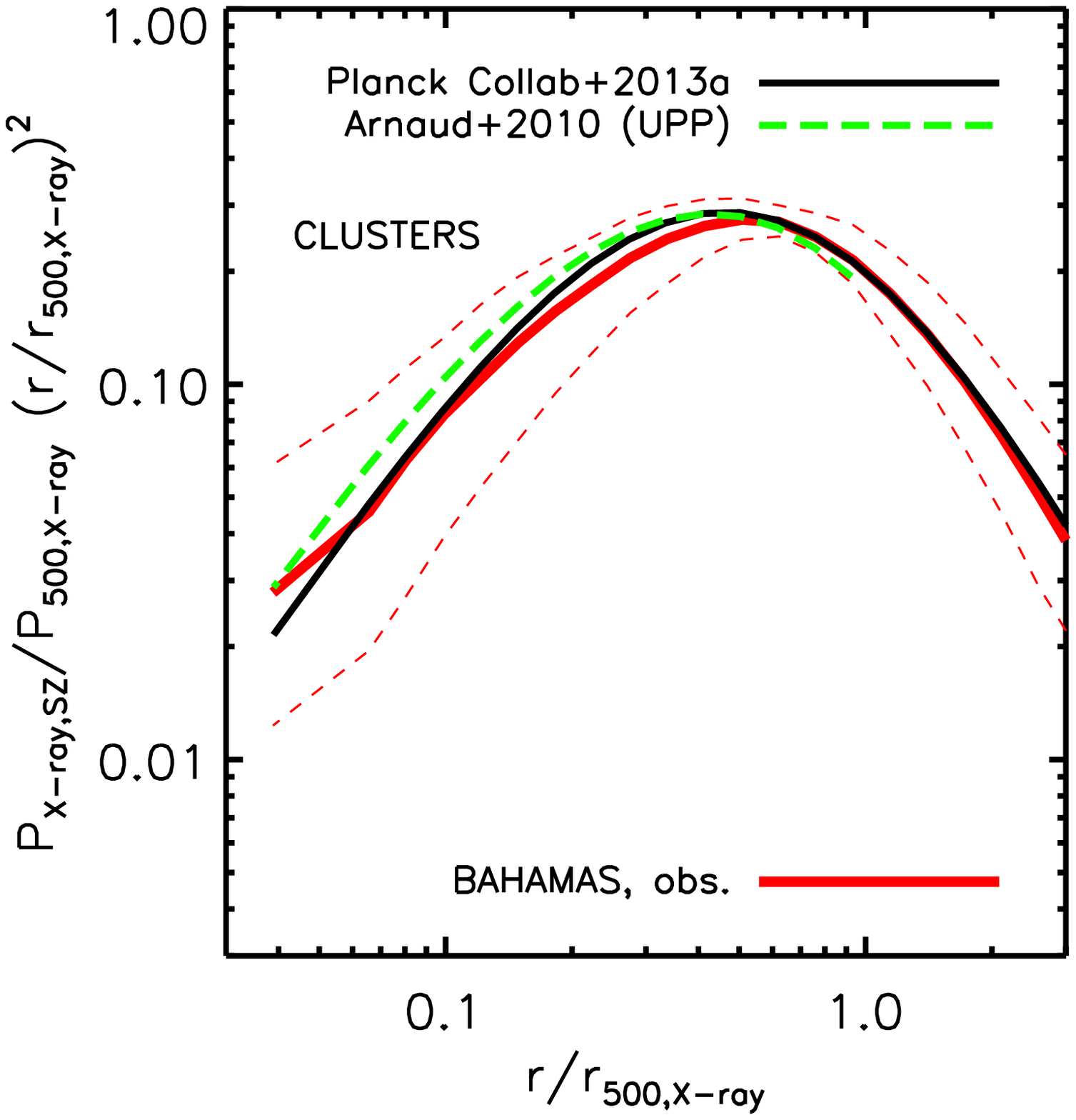}
\caption{\label{fig:pres_profs}
Hot gas pressure profiles of galaxy groups (left) and clusters (right) compared with local samples.  The red curves (solid=median and dashed enclose 68\% of the population) represent the predicted gas pressure profiles for a sample of systems which have the same median halo mass as the observational samples (\citealt{Sun2011} in the case of groups and \citealt{Arnaud2010} for clusters).  In the left panel the filled black circles with error bars represent the median and 1-sigma intrinsic scatter for the group sample of Sun et al.  In the right panel the dashed green curve represents the `universal pressure profile' of \citet{Arnaud2010}, derived from a local X-ray sample, while the solid black curve represents the best-fit to combined X-ray+SZ stacked observations of local clusters \citep{Planck2013a}.  Note that in the latter case the inclusion of SZ data allows one to measure the pressure profiles out to much larger radii than is possible with typical X-ray observations.  The simulations slightly underpredict the gas pressure in the central regions of groups, but otherwise they reproduce the pressure distribution of the hot gas quite well.}
\end{figure*}

To make a fair comparison to the observations, we use our synthetic X-ray pipeline to estimate a hydrostatic mass $M_{500}$ and use the corresponding value of $5 r_{500}$ as the aperture within which we compute the integrated SZ flux.  Note that it is traditionally more common for observational SZ studies to quote values of $Y_{\rm SZ}$ within $r_{500}$ rather than within the actual measurement aperture (which is $5 r_{500}$ in the case of \planck), mainly for historical reasons (e.g., comparison to X-ray properties).  However, \citet{LeBrun2015} have shown that the conversion between the measured flux and that within $r_{500}$ can be sensitive to the assumed radial pressure distribution.  For the case of low-resolution \planck~measurements in particular, the potential bias introduced in the conversion can be severe for groups and low-mass clusters.  We therefore avoid making comparisons to derived fluxes within $r_{500}$, which is generally not resolved by \planck, and instead compare fluxes within the actual apertures.

The \planck~sample is primarily composed of massive systems, with $M_{500}$ greater than a few times $10^{14} \ {\rm M}_\odot$.  Over this range, the predicted median SZ flux$-$$M_{500}$ relation from the simulation agrees well with that derived from the simulations, and the intrinsic scatters about the observed and predicted relations are comparable.

\subsection{Hot gas profiles}

We have shown that the \calsim~simulations reproduce the integrated ICM properties of local clusters.  What about the radial distribution of the hot gas?  In Fig.~\ref{fig:rho_profs} we compare the predicted and observed hot gas density profiles of groups (left panel) and clusters (right panel).  We use our synthetic X-ray pipeline to derive spatially-resolved gas density profiles and hydrostatic modelling to measure $r_{500}$ and $M_{500}$ for the simulated systems.  Note that since the gas content is a relatively strong function of halo mass for both the real and simulated systems (see Fig.~\ref{fig:calibrated}) it is important to compare objects of the same mass.  We therefore impose minimum and maximum halo mass cuts so that the median hydrostatic mass $M_{500}$ matches that of the observed samples we are comparing to.  Specifically, for comparison to the \citet{Croston2008} REXCESS cluster sample {we select all simulated clusters with a hydrostatic X-ray mass of $M_{\rm 500,X-ray} > 2\times10^{14} \ {\rm M}_\odot$ (of which there are 166 from the four independent volumes), yielding a median X-ray mass of $\approx2.6\times10^{14}~{\rm M}_\odot$.  For comparison to the \citet{Sun2009} group sample we select all simulated clusters with a hydrostatic X-ray mass of $5.25\times10^{13} < M_{\rm 500,X-ray}/{\rm M}_\odot < 2\times10^{14}$ (of which there are 526 from the four independent cosmological volumes), yielding a median mass of $\approx7.9\times10^{13} \ {\rm M}_\odot$.  Note that to reduce the dynamic range on the y-axis, we scale the gas density profiles by $r^2$.

The simulations reproduce the observed gas density profiles (median and intrinsic scatter) for both the group and cluster samples remarkably well over the full range of observed radii.  While reasonable agreement should be expected at large radii, given that this is where most of the gas mass is located and that the feedback has been calibrated to reproduce the gas fractions within $r_{500}$, the agreement down to small radii (including the system-to-system scatter) was certainly not guaranteed.  

We can also compare to the observed (electron) pressure distribution of the hot gas, the volume integral of which gives the integrated SZ flux.  In Fig.~\ref{fig:pres_profs} we compare the predicted and observed hot gas pressure profiles of groups (left panel) and clusters (right panel).  As in the comparison to the gas density profiles, we measure the pressure in the simulations in an observational way (i.e., by deriving the electron density and temperature through synthetic spatially-resolved X-ray spectroscopy) and select a subset of systems that have the same median mass as the observational samples (\citealt{Sun2011} in the case of groups and \citealt{Arnaud2010} and \citealt{Planck2013a} for clusters).  Note that the \citet{Planck2013a} result is based on a combined SZ+X-ray stacking analysis of nearby systems which are reasonably well resolved by \planck.  We scale the pressures by $r^2$ to reduce the dynamic range on the y-axis.

The agreement in the cluster regime is very good over the full range of radii (which extends well beyond $r_{500}$ for the observations thanks to the SZ stacking).  A similar level of agreement is also seen for the group comparison at radii beyond $\approx0.3 r_{500}$.  Inside $\approx0.3 r_{300}$ the simulations slightly underpredict the measurements of \citet{Sun2011}, as also found previously by \citet{McCarthy2014}.  And yet there is excellent agreement with the gas density profiles of \citet{Sun2009} (which is based on the same group sample and data).  The density and pressure are not physically independent from each other; hydrostatic equilibrium relates the two via the total mass density distribution.  Thus, matching one thermodynamic variable but not the other implies that either the total mass distributions for the simulated and observed groups differ, or else that the level of non-thermal pressure support in the centers of the simulated groups exceeds that of the groups in the observational sample (so that the simulated clusters maintain a somewhat lower central thermal temperature).  However, it should be borne in mind that the level of deviation we are talking about, in terms of the central pressure distribution of groups, is relatively minor (less than $50\%$) and that there is still significant overlap in the simulated and observed populations (i.e., the intrinsic scatters overlap each other).

Recently, \citet{Barnes2016} have compared the combined \calsim+\mac~suite with the observed radial profiles (pressure, density, etc.) of SPT-selected massive clusters at $z\sim1$, finding excellent agreement with the observed evolution.

Note that in the above analysis, we have focused on the gas density and pressure profiles, rather than on the temperature and entropy profiles, which are also commonly presented in observational studies.  Through the ideal gas law and the adiabatic equation of state, however, only two of the four thermodynamic variables are independent.  We have focused on the gas density because of its link to the redistribution of mass in haloes (which has implications for large-scale structure cosmology) and on the pressure, since we have independent constraints on this quantity from Sunyaev-Zel'dovich effect observations.  One can infer from Figs.~\ref{fig:rho_profs} and \ref{fig:pres_profs}, however, that the simulations reproduce the observed temperature and entropy profiles very well, except in the very inner regions of groups, where the simulated temperatures and entropies have a median value that lies slightly below what is observed (i.e., consistent with the pressure).

\section{The galaxy $-$- hot gas connection}

In Sections 3 and 5, we have shown that the \calsim~simulations reproduce key observed relations between stellar properties and total halo mass and between hot gas properties and total halo mass, respectively, of local systems.  On this basis, one might conclude that the simulations should therefore also match relations between stellar and hot gas properties.  However, this is not guaranteed for a number of reasons.  Current X-ray and SZ studies of individual systems are generally confined to relatively massive groups and clusters ($M_{500} \ga 5\times10^{13} \ {\rm M}_\odot$); we do not yet know whether the simulations faithfully reproduce the hot gas properties of more typical lower-mass systems.  Capturing these systems correctly is important since they are relevant for cosmological studies of large-scale structure, such as cosmic shear and the SZ power spectrum, where haloes of masses $\sim10^{13} \ {\rm M}_\odot$ contribute significantly to the observed signal (e.g., \citealt{Battaglia2012,McCarthy2014,Battaglia2015,Hojjati2015}).  Furthermore, in the case of groups, only the X-ray-brightest systems can be studied in any detail on a per system basis, which could significantly bias our view of the hot gas component of these systems if there is significant intrinsic scatter in the hot gas properties of massive dark matter haloes.  In addition, while we have produced synthetic X-ray observations and derived hydrostatic masses for consistent comparison with X-ray and SZ observations, there could be an inconsistency in this comparison if the level of hydrostatic bias present in the simulations differs significantly from that of real systems.

To overcome these issues we would ideally like to make comparisons between the predictions of the simulations and observations of the relations between the hot gas and stellar properties of {\it representative} populations over a wide range of (true) halo masses.  However, as already noted, hot gas studies of individual systems are limited to relatively massive haloes.  Thus, to proceed further down the mass function stacking/binning is required to boost the SZ/X-ray signal-to-noise ratio.  The advent of surveys that cover a large fraction of the sky (and thus provide hundreds of thousands of potential stacking targets), such as \planck~(SZ), {\it ROSAT} (X-ray), and SDSS (optical) now make this possible.  

\begin{figure*}
\includegraphics[width=0.995\columnwidth]{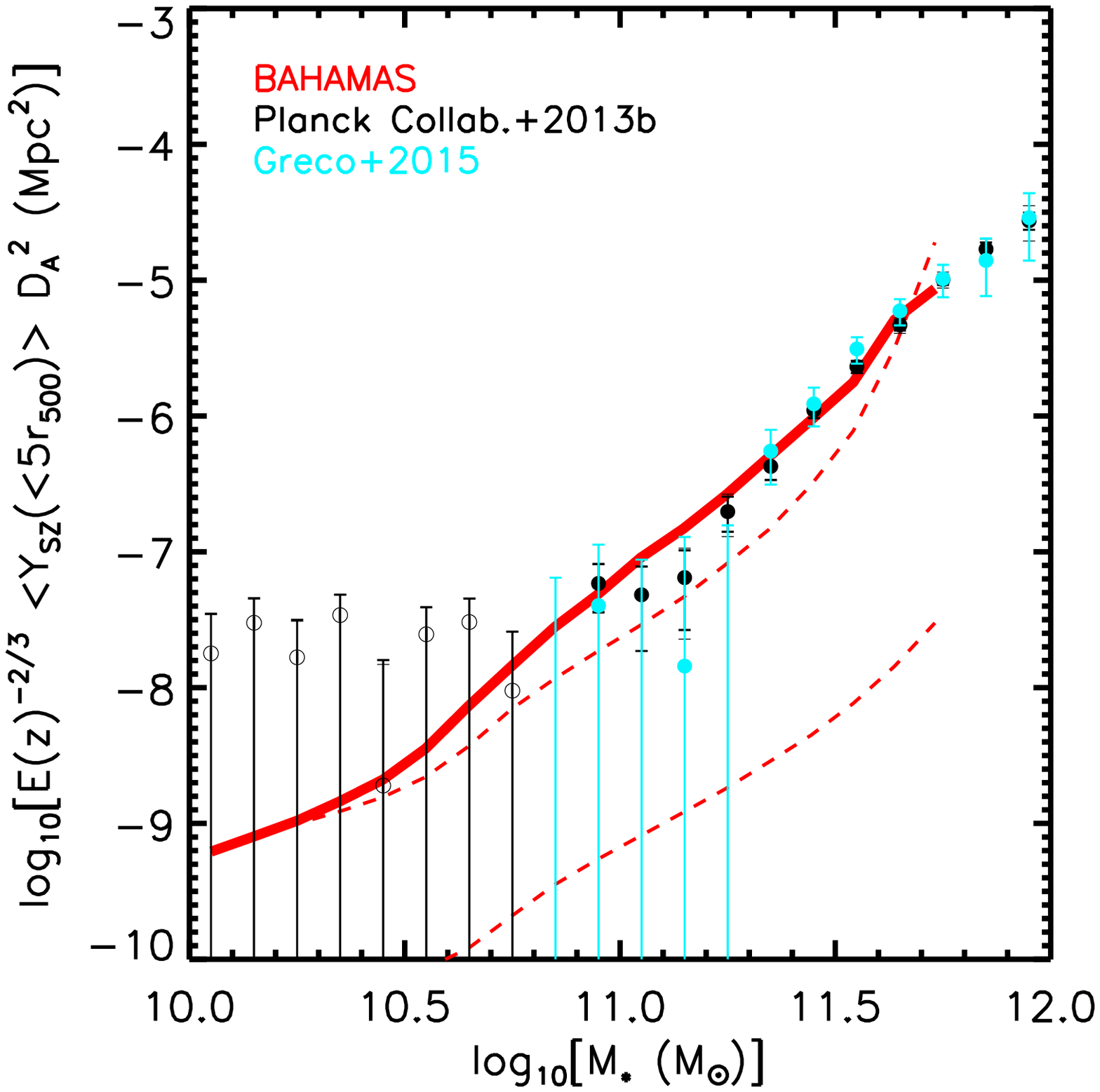}
\includegraphics[width=0.995\columnwidth]{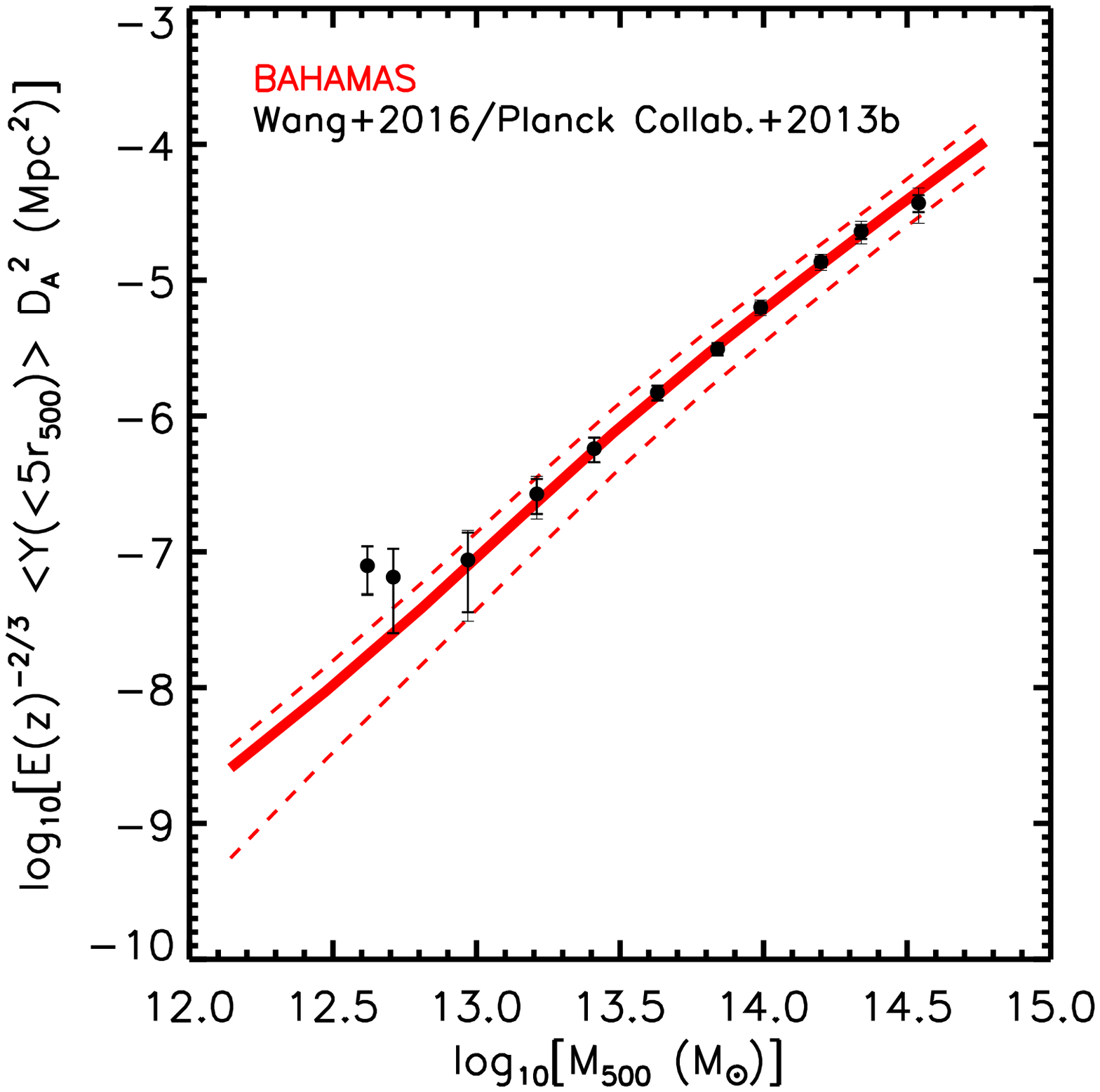}
\caption{\label{fig:ysz_stacked}
The stacked SZ flux$-$stellar mass and SZ flux$-$halo mass relations, compared with the results of stacking SDSS locally brightest galaxies (LBGs) from \citet{Planck2013b}.  The solid red curves give the mean SZ flux in bins of stellar mass (left panel) and halo mass (right panel), while the dashed red curves enclose the central 68\% of the population.  True halo masses are used for the simulated relation (i.e., not processed through synthetic X-ray observations), as the halo masses for the observed relation (in the right panel) have been determined from stacked weak lensing analyses \citep{Wang2016} which are assumed here to be unbiased.  Both trends are recovered remarkably well.  Note that the mean SZ flux$-$stellar mass relation has a much higher amplitude than the median relation, due to the steeper than linear relation of SZ flux with halo mass and the scatter in the stellar mass$-$halo mass relation (i.e., high-mass haloes in a given stellar mass bin dominate the recovered mean SZ flux).}
\end{figure*}

Two recent studies that have exploited these surveys for this purpose are \citet{Planck2013b} and \citet{Anderson2015} (see also \citealt{Greco2015}).  \citet{Planck2013b} defined a sample of SDSS `locally brighest galaxies' (LBGs) - galaxies which are brighter than some apparent magnitude limit ($r<17.7$) and are intrinsically brighter than all other galaxies within a projected 1 Mpc aperture and within 1000 km/s in redshift space.  With the aid of a semi-analytic galaxy formation model \citep{Guo2011,Guo2013} they demonstrated that these selection criteria are quite good at minimizing the contamination due to satellite (sub)haloes.  These authors then stacked the \planck~SZ signal in bins of stellar mass, robustly detecting the hot gas down to a stellar mass of $\sim10^{11} \ {\rm M}_\odot$ (but see \citealt{Greco2015} who argue that dust emission may contaminate several of the lowest mass bins), allowing them to measure the integrated $Y_{\rm SZ}$$-$stellar mass relation above this mass.  Using the same sample of LBGs, \citet{Anderson2015} stacked {\it ROSAT} All-Sky Survey data in bins of stellar mass and obtained a clear detection of the hot gas down to a similar limiting stellar mass.  They measured the stacked X-ray luminosity$-$stellar mass relation.  Using the Guo et al.\ model, the two studies were then able to determine the relations between the SZ flux and X-ray luminosity and halo mass, finding that simple power laws describe the relations remarkably well, with the SZ$-$halo mass relation having close to a self-similar scaling\footnote{As already noted, the integrated SZ flux, although quoted within an aperture of $r_{500}$, is really measured within a much larger aperture of $5 r_{500}$ due to the limited spatial resolution of \planck.  This strongly reduces the sensitivity of the SZ signal to non-gravitational processes that occur within dark matter haloes (e.g., AGN feedback), yielding a close to self-similar scaling.  See \citet{LeBrun2015} for further discussion.} while the X-ray luminosity$-$halo mass relation is significantly steeper than self-similar.

Taken together, these results imply that the gas must be more spatially extended/puffed up in groups relative to clusters, which is consistent with the findings of previous studies of X-ray bright systems (e.g., \citealt{Sun2009}) and the predictions of simulations with efficient AGN feedback (e.g., \citealt{McCarthy2010,Battaglia2010,Planelles2014}; L14).  Recently, \citet{Wang2016} performed a stacked weak lensing analysis of the LBG sample, allowing for a direct (i.e., nearly model independent) measurement of the relations between the SZ flux, X-ray luminosity, and total halo mass (assuming the weak lensing masses to be unbiased).  Here we compare the predictions of the \calsim~simulations with the results of these studies. 

\begin{figure*}
\includegraphics[width=0.995\columnwidth]{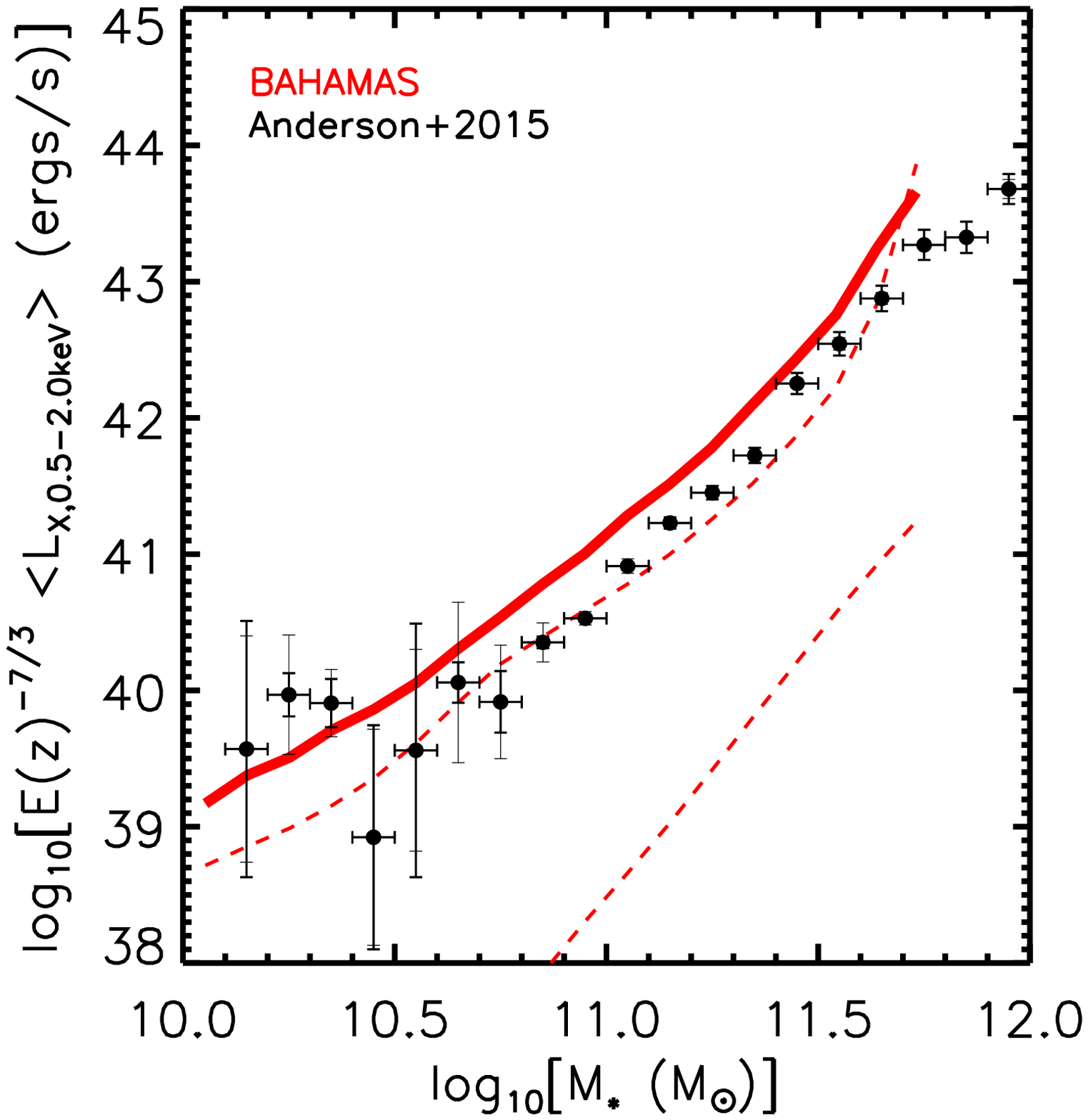}
\includegraphics[width=0.995\columnwidth]{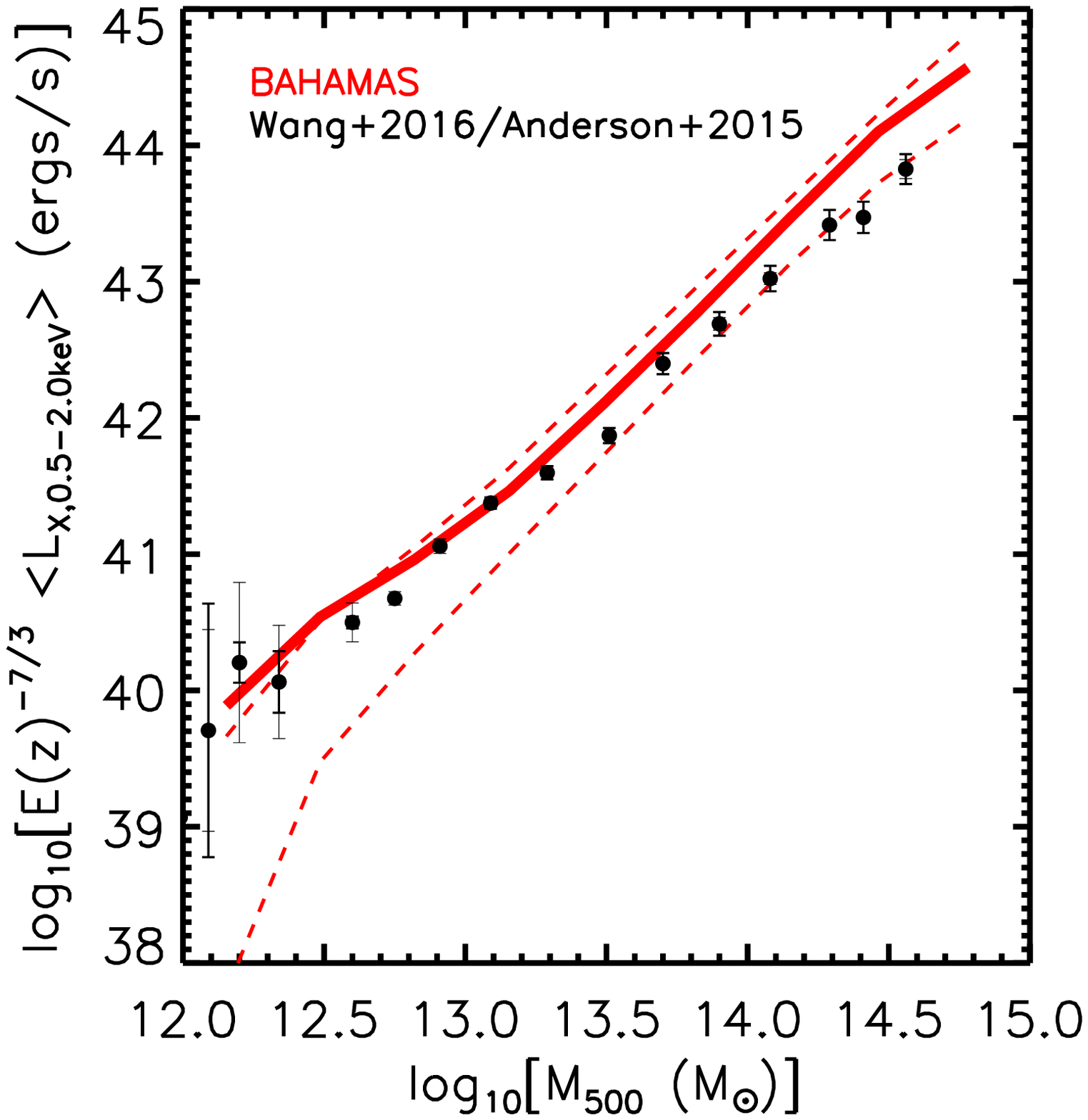}
\caption{\label{fig:lx_stacked}
The stacked $L_X$$-$$M_*$ and $L_X$$-$$M_{500}$ relations, compared with the {\it ROSAT} stacking results of SDSS LBGs by \citet{Anderson2015}. The solid red curves give the mean total soft X-ray luminosity in bins of stellar mass (left panel) and halo mass (right panel), while the dashed red curves enclose the central 68\% of the population.  True halo masses are used for the simulated relation (i.e., not processed through synthetic X-ray observations), as the halo masses for the observed relation (in right panel) have been determined from stacked weak lensing analyses \citep{Wang2016} which are assumed here to be unbiased.  Both simulated relations are shifted slightly in amplitude with respect to the observed relations, perhaps indicating an inconsistency in the relation derived from X-ray and optically-selected samples (note that there is no such amplitude offset in Fig.~\ref{fig:xray_scalings}).
}
\end{figure*}

In the left panel of Fig.~\ref{fig:ysz_stacked} we compare the predicted mean SZ flux$-$stellar mass relation, i.e., $\mean{Y(<5r500)}(M_*)$, with that measured by \citet{Planck2013b}.  Note that we have converted the SZ fluxes reported in \citet{Planck2013b} from $Y(<r_{500})$ back into the actual measured flux $Y(<5r_{500})$ by multiplying by a constant factor of $1.796$ (this corresponds to the ratio of $Y(<5r_{500})/Y(<r_{500})$ assuming the universal pressure profile of \citealt{Arnaud2010}, derived by \citealt{Planck2013b}) and, in doing so, effectively removed the dependence of the measured flux on the spatial template (the universal pressure profile) in their matched filter.  For the simulations we directly measure the integrated flux within $5 r_{500}$.  The agreement between the predicted and observed relations is good.  This is remarkable considering how important the role of intrinsic scatter is: the dashed red curves enclose 68\% of the population and the median SZ flux$-$stellar mass relation has a much lower amplitude than the mean relation (which is what is recovered by stacking analyses).  The origin of the large difference between the predicted median and mean relations is the large range of halo masses that corresponds to any given stellar mass bin, coupled with the fact that the SZ flux scales steeply with halo mass as $M_{500}^{5/3}$.  The net result is that the high (halo) mass tail in any stellar mass bin has a disproportionately large effect on the stacked relation.  The fact that the predicted and observed stacked relations agree as well as they do is therefore another indication that the scatter in the stellar mass$-$halo mass relation is realistic.

In the right panel of Fig.~\ref{fig:ysz_stacked} we compare to the SZ flux$-$halo mass relation of \citet{Planck2013b}.  We have boosted the amplitude of the observed relation by a constant factor 1.35 to take into account the difference between the mean effective halo mass estimated using the Guo et al.~model (derived in \citealt{Planck2013b}) and that measured empirically for the same LBG sample by \citet{Wang2016} via stacked weak lensing analyses.  The agreement in slope and amplitude is remarkably good.

Note that in Fig.~\ref{fig:ysz_stacked} we are using true estimates of $M_{500}$ and $r_{500}$, whereas in Fig.~\ref{fig:tsz_scaling} we used hydrostatic masses and their corresponding apertures for a consistent comparison with individual \planck~clusters where the masses were estimated using an X-ray hydrostatic mass scaling relation.  The fact that the predicted relations agree with observed relations in amplitude in the two comparisons may therefore suggest that the level of hydrostatic bias in the simulations (which has a median value of nearly 20\% within $r_{500}$ - see Section 5.1; see also \citealt{Henson2016}) is also realistic.  This statement assumes that the stacked weak lensing measurements are effectively unbiased.

In the left panel of Fig.~\ref{fig:lx_stacked} we compare the predicted mean X-ray luminosity$-$stellar mass relation, i.e., $\mean{L_X(<r_{500})}(M_*)$, with that measured by \citet{Anderson2015}.  Specifically, we compare to their measured (rest-frame) 0.5-2.0 keV `total' luminosities (Table 3 of that study).  For the simulations, we compute the luminosities within the same band, summing the luminosities of all of the individual particles within $r_{500}$.  (The choice of aperture has a very small effect on the result, since the X-ray emission is dominated by the central regions.)  The predicted mean relation has a very similar slope to the observed relation, but is offset slightly in amplitude by $\approx0.3$ dex.  Given the sensitivity of the predicted mean relation to the scatter in the stellar mass$-$halo mass relation and the sensitivity of the $L_X$$-$halo mass relation to subgrid physics (see, e.g., L14, \citealt{LeBrun2015}), this is still an impressive level of agreement.  

\begin{figure*}
\includegraphics[width=0.995\columnwidth]{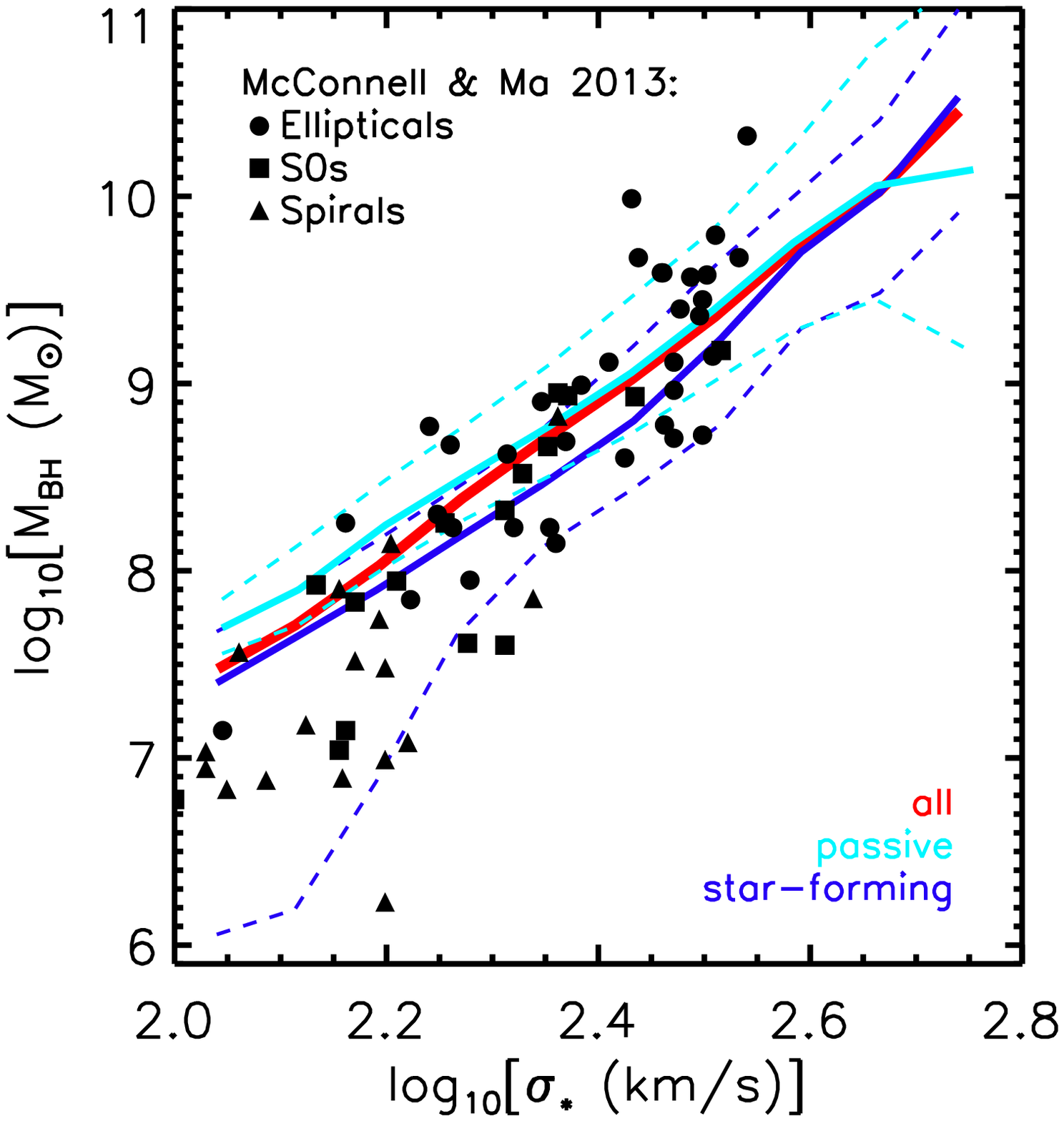}
\includegraphics[width=0.995\columnwidth]{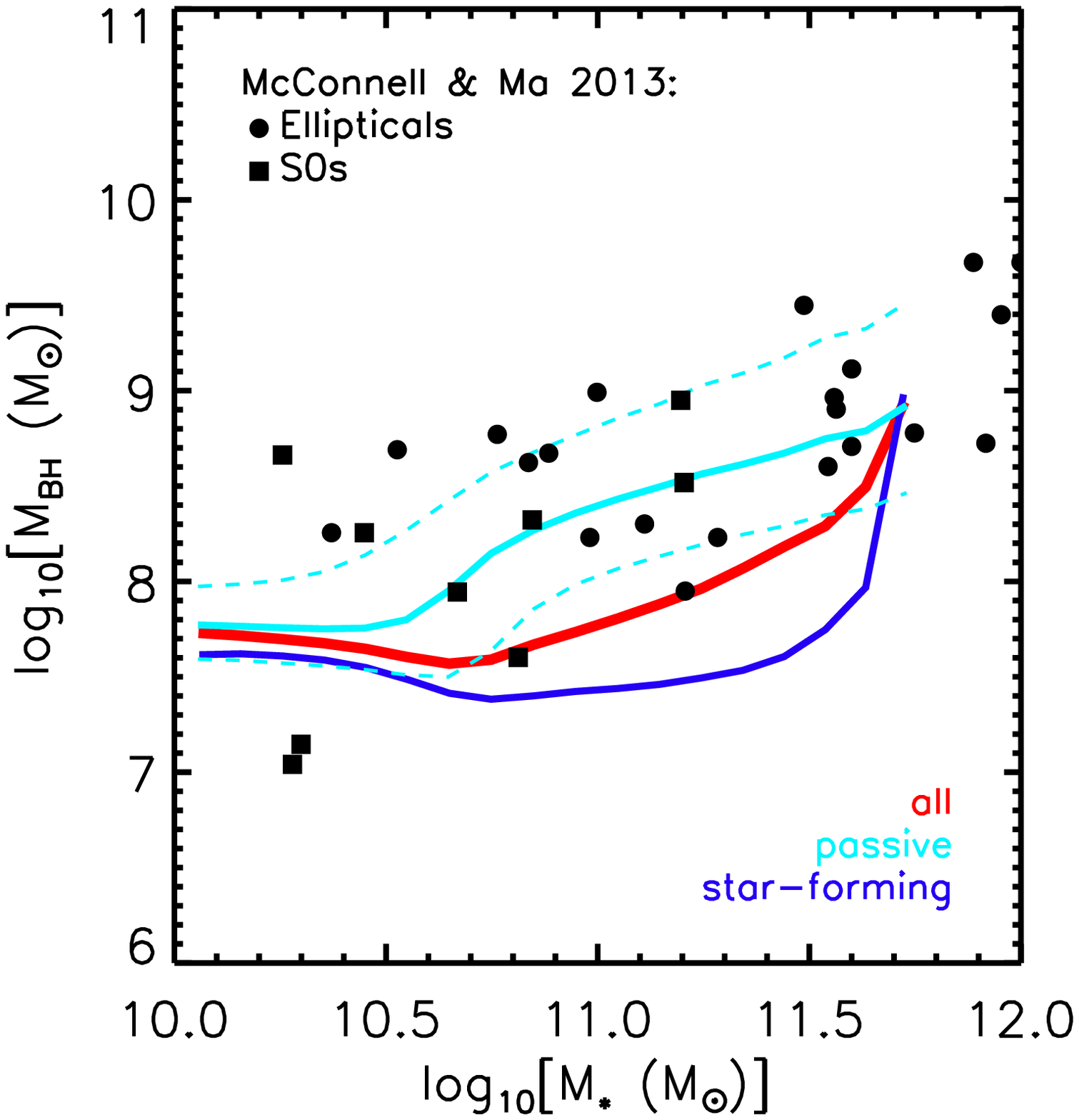}
\caption{\label{fig:bh_dyn}
Relations between central black hole mass and stellar velocity dispersion (left panel) and stellar mass (right panel).  The red curves show the trend for the full simulated population (includes highly star forming galaxies), while the cyan and blue curves show the relations for passive and star-forming galaxies separately.  Overall, the predicted relations agree broadly with the observed relations when an appropriate selection is applied to the simulated galaxies.   
}
\end{figure*}

We might, however, have expected even better agreement given the excellent consistency of the SZ relations in Figs.~\ref{fig:tsz_scaling} and \ref{fig:ysz_stacked} (which suggest that the stellar mass$-$halo mass relation and the level of hydrostatic bias in the simulations are realistic) and the agreement with the $L_X$$-$$M_{500,{\rm X-ray}}$ relation of individual X-ray-selected systems (see the left panel of Fig.~\ref{fig:xray_scalings}).  Examining the right panel of Fig.~\ref{fig:lx_stacked}, which compares the predicted and observed mean X-ray luminosity$-$halo mass relations (where we have boosted the amplitude of the Anderson et al.~ result, by 40\%, in accordance with \citealt{Wang2016}), shows a similar offset in amplitude to the X-ray luminosity$-$stellar mass relation.  This then suggests that there is an inconsistency in the measurements of the X-ray luminosity$-$halo mass relations of individual, X-ray-selected systems and that inferred from the optically-selected stacking analysis of \citet{Anderson2015}, which cannot easily be remedied by appealing to differences in the halo mass definition or estimation (i.e., the differences appear due to the X-ray luminosity estimation).  Detailed intercomparisons of the observational samples would clearly be beneficial in the future.

\section{Properties of black holes and quasars}

We finish the comparison to observational data by exploring here how well the simulations reproduce the observed local scalings between black hole mass and galaxy properties, as well as the observed evolution of the quasar luminosity function.  The latter is a particularly interesting test, since we have previously shown \citep{McCarthy2011} that high-$z$ quasars do the lion's share of the work in setting the present-day properties of the hot gas.

In Fig.~\ref{fig:bh_dyn} we examine the relations between black hole mass and galaxy velocity dispersion (left panel) and stellar mass (right panel) and compare to observational data compiled by \citet{McConnell2013}.  For the simulations, we analyse central galaxies/subhaloes only.  The velocity dispersion is computed as the RMS of the 1D peculiar velocity of star particles within a 30 kpc aperture (we average the three independent 1D velocity dispersions), while the stellar mass is that within a 3-D aperture of 30 kpc.  Note that for the observed systems in the right panel what is measured is not strictly the total stellar mass, but the dynamical bulge mass.  However, since this comparison is limited to observed early-type galaxies, the stellar mass in the bulge component should be dominant.  Furthermore, dark matter is not expected to contribute significantly to the dynamical mass at such small radii (e.g., within the bulge half-light radius).

\begin{figure*}
\includegraphics[width=1.99\columnwidth]{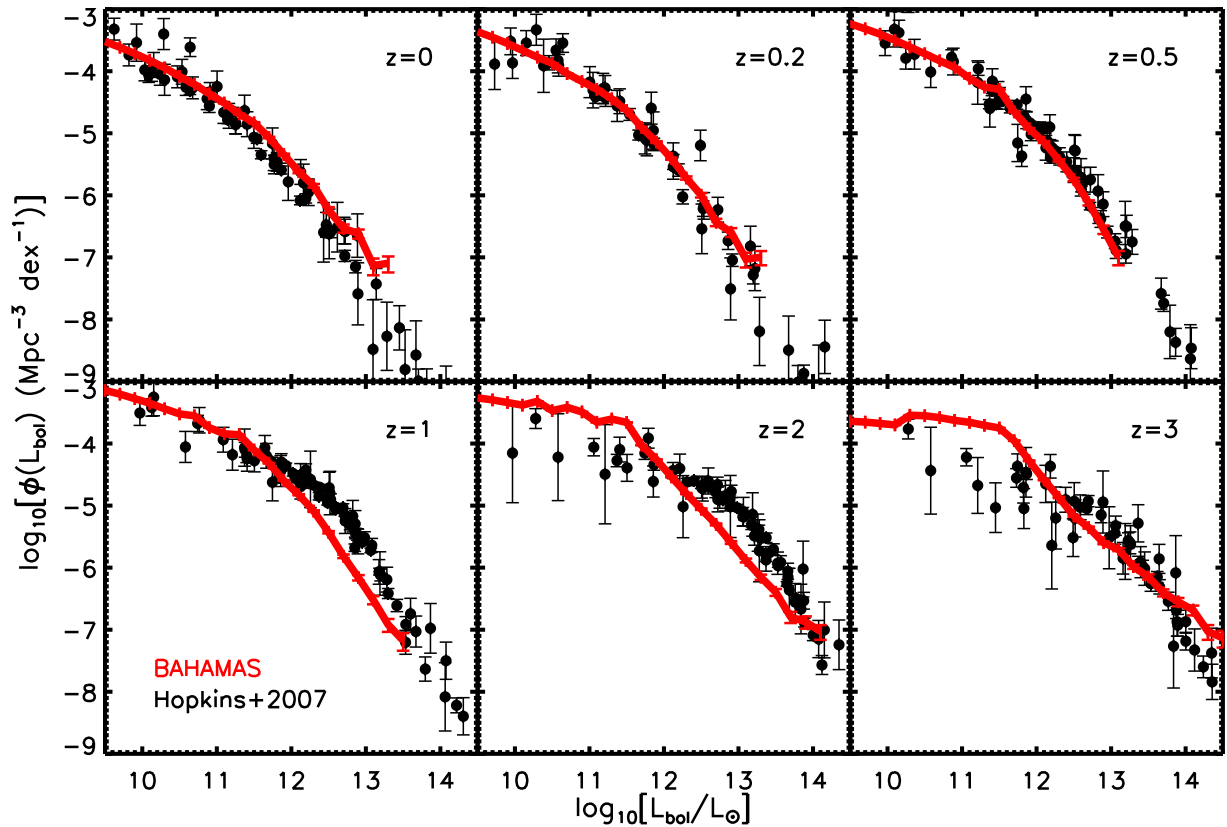}
\caption{\label{fig:qlf}
Evolution of the bolometric quasar luminosity function, compared with the analysis of \citet{Hopkins2007}.  There is reasonable qualitative agreement.  In detail, the simulations slightly overpredict the abundance of bright quasars at low-$z$ and underpredict the knee at $z\sim1-2$, similar to the cosmic SFRD history (see Fig.~\ref{fig:rhosfr_evol}).
}
\end{figure*}

In the left panel of Fig.~\ref{fig:bh_dyn} we see that the overall amplitude of the predicted $M_{\rm BH}$$-$$\sigma_*$ is in reasonable agreement with the observational data.  This is consistent with the previous findings of \citet{Booth2009} and L14.  \citet{Booth2009,Booth2010} showed that the amplitude of the BH scaling relations scales with the (inverse of the) feedback efficiency $\epsilon_f$ and found that a value of $\epsilon_f=0.15$ yields a good match to the observed black hole masses, which we verify here.  The agreement with the data is therefore not a success of the model, but a result of calibration.  

Examining the trend more closely (solid red curve), however, suggests that there may be a slight difference in slope, particularly at low $\sigma_*$.  It has been noted in previous observational studies that the black hole mass is not solely a function of the velocity dispersion, but also depends on the morphology of the galaxy (e.g., \citealt{Graham2001,Graham2007}).  Indeed, the spiral galaxies (triangles) in \citet{McConnell2013} appear to follow a relation that has a lower amplitude compared to that which the ellipticals follow.  S0s lie somewhere in between.  Due to the relatively low resolution of the \calsim~simulations, we cannot reliably split the simulated galaxies into discs or ellipticals.  However, we can split them by star-forming / passive based on their present (specific) star formation rates, which may be a reasonable proxy for morphology (see Fig.~\ref{fig:fstar_centrals} and associated text).  When we do so, we do see a minor bifurcation of the median $M_{\rm BH}$$-$$\sigma_*$ relations (solid blue and cyan curves), in the same sense as the observations.  Interestingly, the {\it scatter} towards lower BH masses is significantly larger for the star-forming population and encompasses most of the observed systems.

In the right panel of Fig.~\ref{fig:bh_dyn} we see that the choice of galaxy type (passive or star-forming) has an even larger effect on the predicted $M_{\rm BH}$$-$$M_*$ relation.  In particular, BHs are on average more massive in passive galaxies compared to star-forming galaxies of the same stellar mass\footnote{Note that this result mirrors the trends seen in Fig.~\ref{fig:fstar_centrals} (right panel), where it was demonstrated that the mean halo masses of passive galaxies are higher than those of star-forming galaxies of the same stellar mass.  The similarity of these trends indicates that the BH mass is more tightly correlated with the total halo mass than with the stellar mass in the simulations, as was explicitly shown previously by \citet{Booth2010}.}.  Only the early-type galaxies in \citet{McConnell2013} have mass estimates.  When we select passive simulated galaxies for comparison, we find the predicted relation to be in reasonable agreement with the observed relation.

Finally, in Fig.~\ref{fig:qlf} we compare the predicted evolution of the quasar luminosity function to that derived from a large suite of multiwavelength observations compiled and modelled by \citet{Hopkins2007}.  For the simulations we compute the bolometric luminosity of each BH particle as $L_{\rm bol} = \epsilon_r (1-\epsilon_f) \dot{M}_{\rm BH} c^2$, where $\dot{M}_{\rm BH}$ is the instantaneous accretion rate and the $(1-\epsilon_f)$ factor takes into account that a fraction of the radiated rest mass energy goes into doing feedback.  Note that we select all BH particles and hence do not impose any constraints on Eddington ratio.

\citet{Hopkins2007} compiled a large suite of observational determinations of the quasar luminosity function at many different wavelengths.  Using spectral modelling they combined the results into a consistent determination of the bolometric luminosity function, that accounts for the effects of absorption, over a wide range in redshift.  Although \citet{Hopkins2007} also give luminosity functions in different bands (X-ray, IR, etc.), the simulations can only reliably predict the bolometric luminosities, since we do not model the detailed accretion disc physics nor the interaction of the emitted radiation with local gas.

Overall, the agreement between the predicted and observed luminosities functions is quite good.  In particular, for $z < 1$ the simulations reproduce the data very well.  When considering the simplicity of the accretion model in the simulations (as well as the simplified treaments of feedback physics), the level of agreement seems all the more impressive.  However, in detail the predicted luminosities appear too low near the knee of the luminosity function at $z\approx1-2$.  This may signal that the gas fractions of the galaxies hosting these quasars are somewhat lower than in reality.  Alternatively, the bolometric corrections applied to the observational data may be overestimated.  Indeed, in a future study (Koulouridis et al., in prep) we will show that adopting more recent bolometric correction determinations results in a much improved match to the observed evolution of the quasar luminosity function.

\section{Summary and Discussion}

In this study we have presented a new set of large-volume cosmological smoothed particle hydrodynamics simulations called \calsim~(for BAryons and HAloes of MAssive Systems).  The simulations presented here use $2\times1024^3$ particles in a 400 Mpc/$h$ on a side, and assume either a \wmap~9-year or a \planck~2013 cosmology.  

These simulations are a direct descendant of the OverWhelmingly Large Simulations (OWLS) and cosmo-OWLS projects \citep{Schaye2010,LeBrun2014,McCarthy2014}.  Where the \calsim~project differs from these studies is in its explicit attempt to calibrate the feedback parameters to match some key observables (as also employed in the recent EAGLE simulation program), whereas the OWLS and cosmo-OWLS simply explored the effects of varying the parameters.  We have used the knowledge gleaned from the experimentation in the OWLS and cosmo-OWLS projects to derive a simple feedback model calibrated to reproduce the present-day baryon content of massive systems, specifically the galaxy stellar mass function (over the range $\log_{10}[M_*/{\rm M}_\odot] = 10.0-12.0$) and the hot gas mass fraction$-$halo mass relation of galaxy groups and clusters (over the range $\log_{10}[M_{500}/{\rm M}_\odot] = 13.0-15.0$; see Fig.~\ref{fig:calibrated}).  We note that the black hole feedback efficiency was also calibrated (by \citealt{Booth2009}) to reproduce the amplitude of the black hole mass$-$stellar mass relation, by adjusting the feedback efficiency, $\epsilon_f$.  (The feedback efficiency is unimportant for anything other than the black hole masses though.)

We have focused on the baryon content because our (eventual) aim is to use the simulations to aid the cosmological interpretation of large-scale structure tests, such as cosmic shear, cluster counts, Sunyaev-Zel'dovich map statistics, etc., which probe into the non-linear regime and may be sensitive to the back reaction of baryons on the dark matter.  We thus want to ensure that the degree of back reaction, as well as the relations between observables and total matter, are properly captured by the simulations.  

We point out that the stellar and AGN feedback models were not calibrated simultaneously.  The stellar feedback wind velocity was adjusted to reproduce the observed abundance of the lowest-mass galaxies we examine (Fig.~\ref{fig:gsmf_vary_vwind}), while the mass of gas heated by AGN was adjusted to better reproduce the knee of the galaxy stellar mass function (Fig.~\ref{fig:gsmf_nheat}).  The AGN heating temperature was separately calibrated to reproduce the amplitude of the hot gas mass fraction$-$halo mass relation of local groups and clusters (Fig.~\ref{fig:fgas_vary_theat}).  It is interesting that it is possible to construct models that reproduce the observed galaxy stellar mass function but that fail to reproduce the observed hot gas properties of massive galaxies/clusters and vice-versa (i.e. the stellar and hot gas properties are `decoupled').

It should be noted that we did not examine {\it any} other observables during the calibration process.  Furthermore, we highlight the simplicity of our final calibrated model: the parameters governing the efficiencies of the AGN and stellar feedback are constants.  We speculate that such a simple model is possible here because of the relatively low resolution we are working at (a consequence of the large volumes we are simulating) and that we are not attempting to match galaxies with $M_* < 10^{10} \ {\rm M}_\odot$.  We do not claim to have found a {\it unique} solution even at this resolution, but we have identified a simple model which satisfies our requirements (the baryon content of massive systems) and whose realism can be tested against independent observations.

After calibrating the model, we compared the predictions of the model to a wide range of observational constraints, both locally and at higher redshifts.  From these comparisons we deduce the following:

\begin{itemize}
\item{The simulations reproduce the observed stellar mass fractions of central galaxies, including the dependence on galaxy type (see Fig.~\ref{fig:fstar_centrals}). They also reproduce the amplitude of the relation between the integrated stellar mass fraction (i.e., including satellites and intracluster light) and halo mass for local groups and clusters (Fig.~\ref{fig:fstar_integrated}).  The shape of the dependence of the integrated stellar mass fraction on halo mass, however, differs significantly between different observational studies, so the level of agreement is less clear here (the simulations fall in the middle of the observed trends in terms of slope).}
\item{The Halo Occupation Distribution-inferred fractional contribution of centrals and satellites to the integrated stellar mass fractions as a function of halo mass \citep{Zu2015} is reproduced remarkably well (Fig.~\ref{fig:fstar_sat_cen}), as is the observed spatial distribution of stellar mass (in satellites) in massive clusters (Fig.~\ref{fig:rho_star_prof}).}
\item{The observed dynamics of satellite galaxies as a function of halo mass (from maxBCG) is also recovered, with strong evidence for a negative velocity bias of the satellite galaxies with respect to the underlying dark matter distribution (Fig.~\ref{fig:sigma_mass})}.
\item{The simulations reproduce the observed local stellar mass autocorrelation function from SDSS well on large scales ($>$ 1 Mpc), while on small scales the level of agreement depends on the choice of aperture (Fig.~\ref{fig:stellar_acf}).  To our knowledge this is the first time cosmological hydrodynamical simulations have been shown to reproduce the observed clustering of stellar mass.}
\item{We have compared the simulations to the observed evolution of the cosmic stellar mass density (Fig.~\ref{fig:rhostar_evol}), as well as the observed evolution of the galaxy stellar mass function (Fig.~\ref{fig:GSMF_evol}).  The integrated stellar mass density is recovered reasonably well.  For $z \ga 0.5$ the simulations overpredict the abundance of galaxies with stellar masses $\sim10^{10}  \ {\rm M}_\odot$ (about the resolution limit) and they slightly underpredict the abundance of the most massive galaxies at the highest redshifts ($z\approx3$).}
\item{The simulations qualitatively reproduce the observed evolution of the star formation rates (Figs.~\ref{fig:rhosfr_evol} and \ref{fig:sSFR_evol}), but they somewhat underpredict the peak at $z\approx2-4$ and significantly overpredict the star formation rates of galaxies at $z \la 0.5$.}
\item{We have compared the simulations to the observed X-ray and SZ scalings of local groups and clusters (in a like-with-like fashion using virtual X-ray observations), including the X-ray luminosity$-$ and $Y_X$$-$halo mass scalings (Fig.~\ref{fig:xray_scalings}) and the integrated SZ flux$-$halo mass relation (Fig.~\ref{fig:tsz_scaling}).  The simulations reproduce these relations well.  They also reproduce the observed radial density and pressure profiles for groups and clusters (Figs.~\ref{fig:rho_profs} and \ref{fig:pres_profs}), apart from slightly underestimating (by $\la0.15$ dex) the pressure/temperature in the inner regions of groups. }
\item{The simulations match the observed stacked mean SZ flux$-$stellar mass and SZ flux$-$halo mass relations for optically-selected SDSS `locally brightest galaxies' (LBGs) of \citet{Planck2013b} and \citet{Wang2016} very well (Fig.~\ref{fig:ysz_stacked}).  The agreement with the former suggests that the scatter in the stellar mass$-$halo mass relation (0.24 dex median) is realistic, as the mean SZ-stellar mass relation is strongly affected by the tail of the stellar mass$-$halo mass relation.  Furthermore, the consistency of the SZ flux$-$halo mass estimates using X-ray hydrostatic masses (Fig.~\ref{fig:tsz_scaling}) and stacked weak lensing (Fig.~\ref{fig:ysz_stacked}) suggests that the level of `hydrostatic bias' in the simulations (median of $\approx20\%$) is also realistic.}
\item{The simulations predict stacked mean X-ray luminosity$-$stellar mass and $-$halo mass relations that are very similar to those recently measured by \citet{Anderson2015} and \citet{Wang2016}, with a nearly identical slope but a slight amplitude offset of $\approx0.3$ dex (Fig.~\ref{fig:lx_stacked}).  No such offset is seen in the comparison to individual X-ray systems however (see Fig.~\ref{fig:xray_scalings}), suggesting that there is some difference in the observed X-ray luminosities.}
\item{The observed local relations between black hole mass and velocity dispersion and stellar mass are reasonably well recovered (Fig.~\ref{fig:bh_dyn}).}
\item{Lastly, the observed evolution of the bolometric quasar luminosity function is reproduced for $z \la 1$.  The simulations underpredict the knee of the luminosity function at $z\approx1-2$.}
\end{itemize}

To our knowledge, \calsim~represents the first set of cosmological hydrodynamical simulations to simultaneously reproduce the observed hot gas and stellar properties of massive systems with such precision and for such a wide range of observables.  The level of agreement is even more remarkable given the simplicity of the model and the fact that we did not calibrate on (or even examine) anything other than the local galaxy stellar mass function and the X-ray-based gas mass fractions of local groups and clusters.  Nevertheless, there is still significant room for improvement, particularly at higher redshifts (e.g., evolution of star formation rates and quasar luminosities).  

With a realistic model for the dominant baryonic components of massive haloes in a self-consistent cosmological context in hand, we are now in a position to make strong predictions for a variety of large-scale structure measurements, including the SZ power spectrum, cluster number counts, cosmic shear, CMB lensing, galaxy-galaxy lensing, redshift-space distortions, etc. and to examine in detail the recently reported claims of tensions between the cosmological constraints from these measurements with the \planck~constraints from the analysis of the primary CMB.  The latter will be the subject of our next study.

Finally, we note that \calsim~has recently been complemented by a suite of zoom simulations of very massive clusters (called \mac), which was run using the same feedback model calibrated here and run at the same numerical resolution (see \citealt{Barnes2016,Henson2016}).  The combination of \calsim~and \mac~allows one to probe a very large dynamic range in halo mass, which is not accessible to either suite individually.

\section*{Acknowledgements}

The authors would like to thank Johnny Greco, Cheng Li, Ming Sun, and Ying Zu for providing their observational data in electronic format.  IGM thanks Simon White for helpful discussions.  IGM is supported by a STFC Advanced Fellowship.  JS acknowledges support from ERC grant 278594 - GasAroundGalaxies and from the Netherlands Organisation for Scientific Research (NWO) through VICI grant 639.043.409.  SB was supported by NASA through Einstein Postdoctoral Fellowship Award Number PF5-160133.  AMCLB acknowledges support from the French Agence Nationale de la Recherche under grant ANR-11-BS56-015.

This work used the DiRAC Data Centric system at Durham University, operated by the Institute for Computational Cosmology on behalf of the STFC DiRAC HPC Facility (www.dirac.ac.uk). This equipment was funded by BIS National E-infrastructure capital grant ST/K00042X/1, STFC capital grants ST/H008519/1 and ST/K00087X/1, STFC DiRAC Operations grant ST/K003267/1 and Durham University. DiRAC is part of the National E-Infrastructure.

\appendix
\section{Calibrating the AGN feedback model}

\begin{figure*}
\includegraphics[width=0.98\columnwidth]{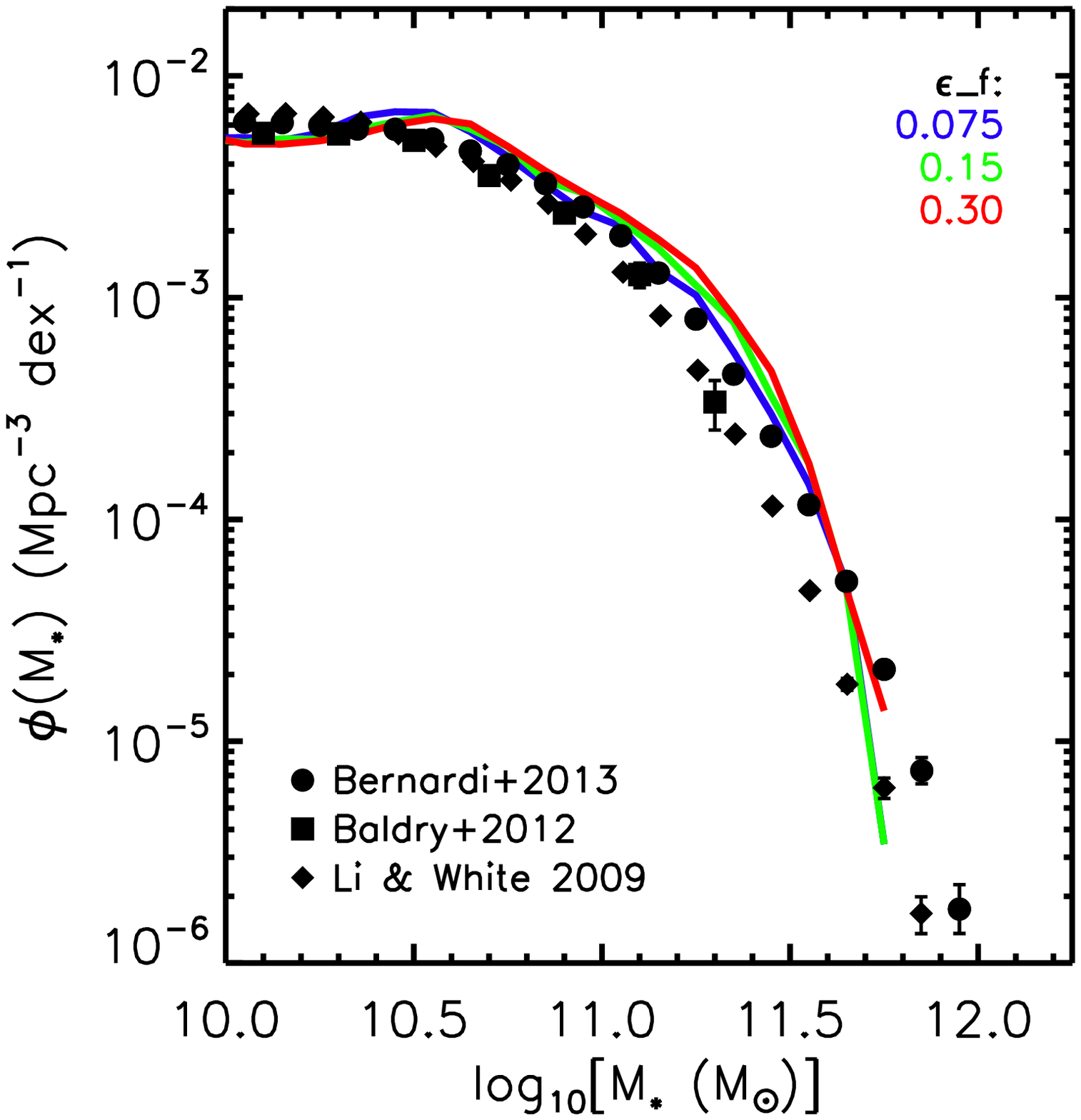}
\includegraphics[width=0.98\columnwidth]{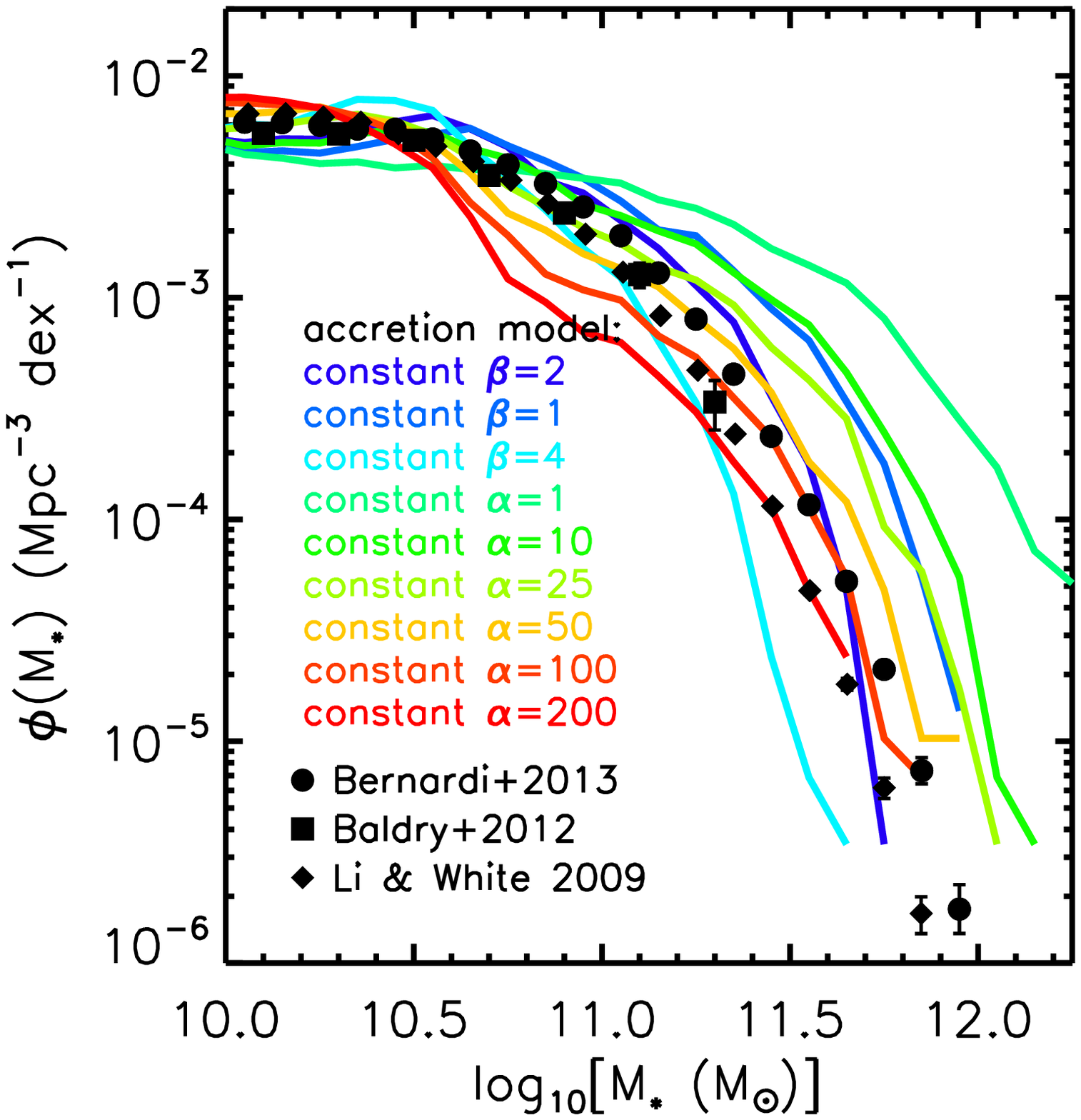}
\includegraphics[width=0.98\columnwidth]{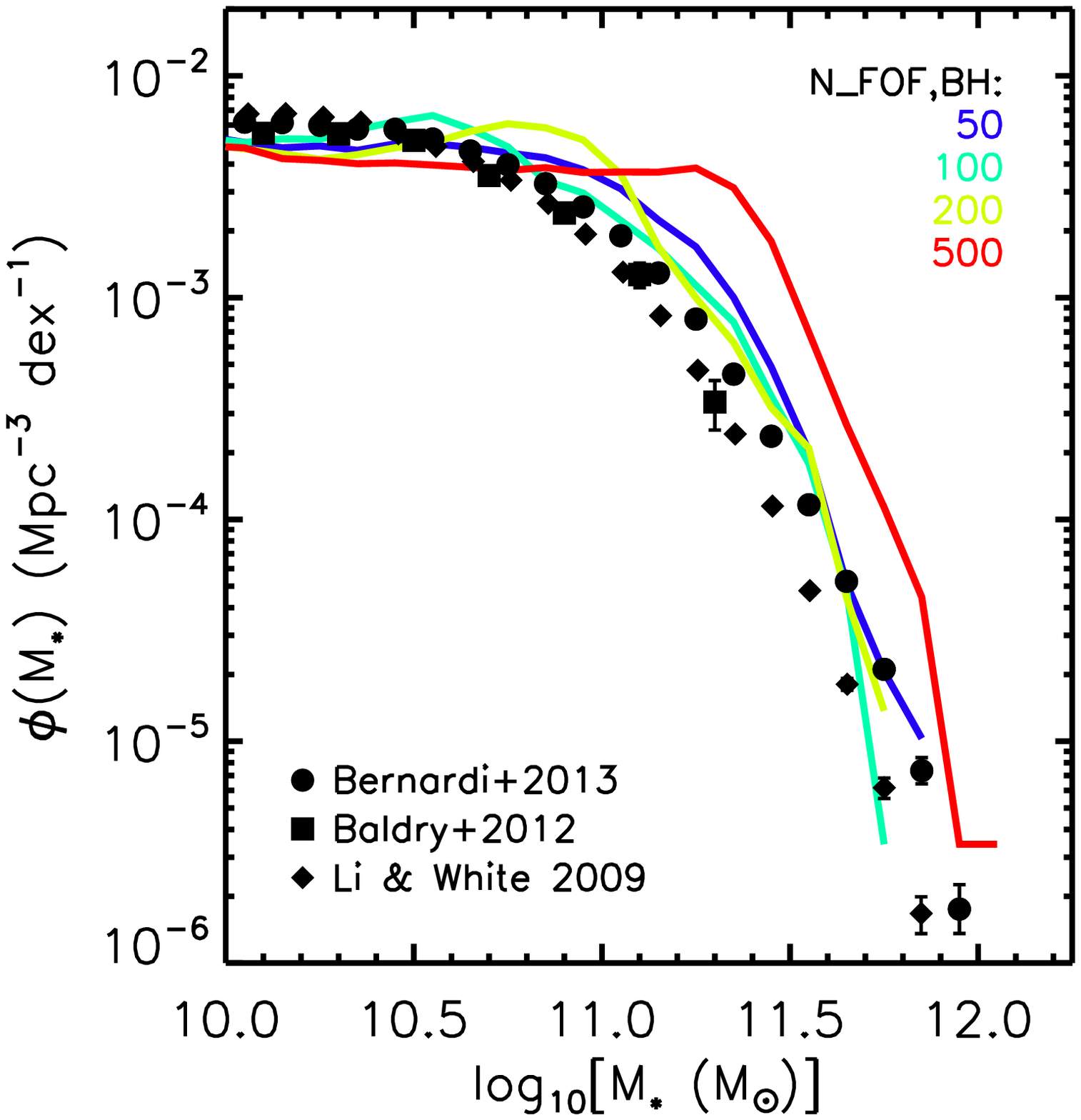}
\includegraphics[width=0.98\columnwidth]{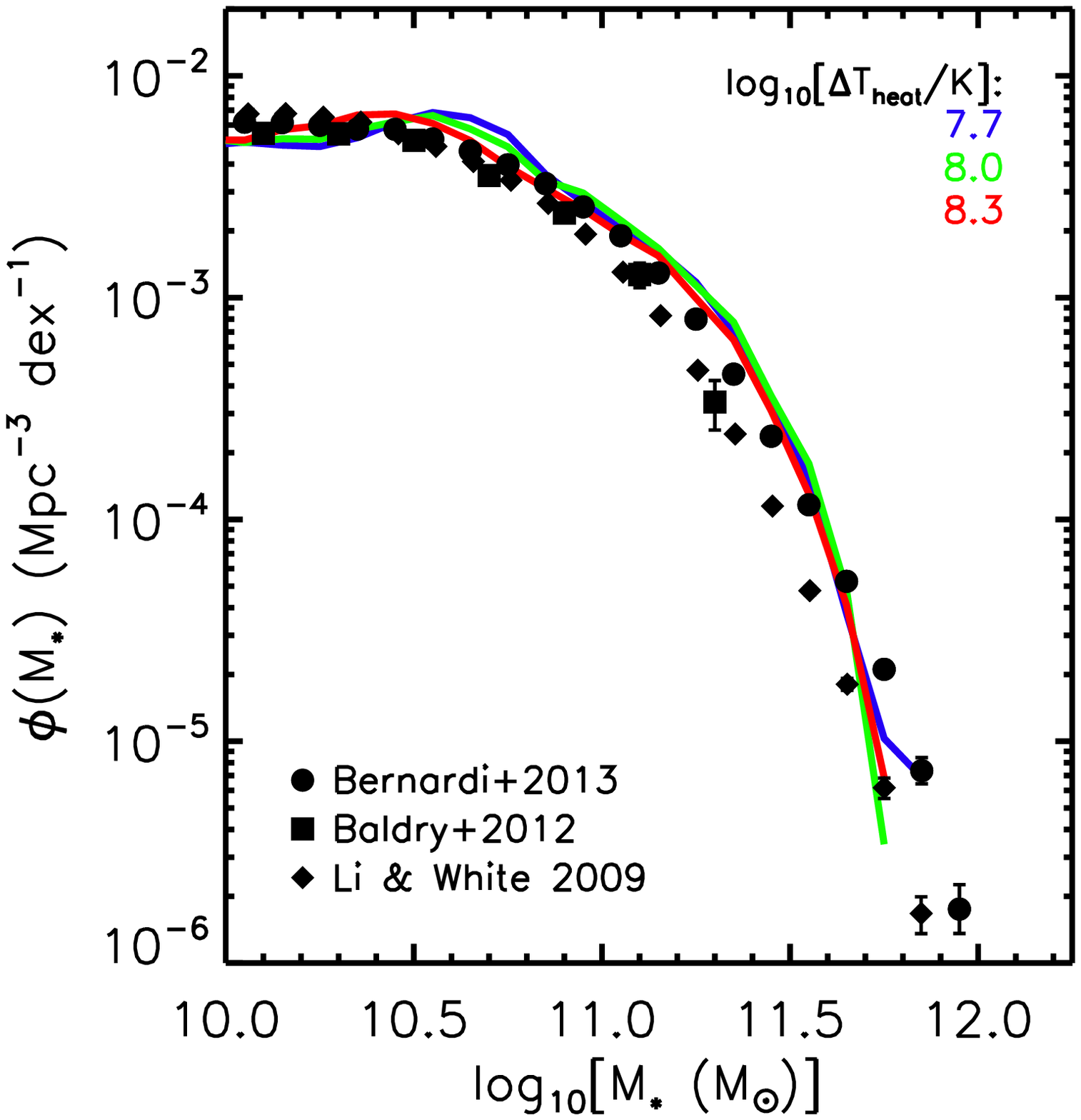}
\caption{\label{fig:gsmf_vary_bh}
The effect of varying the AGN feedback efficiency $\epsilon_f$ (top left panel), BH accretion model (top right panel), the minimum halo mass for BH seed injection (bottom left panel), and the AGN heating temperature (bottom right panel) on the local GSMF.  The GSMF is insensitive to the choice of $\epsilon_f$ and heating temperature, as found previously (e.g., Booth \& Schaye 2009; Le Brun et al.\ 2014; Schaye et al.\ 2015).  Varying the minimum halo mass for BH seed injection affects the knee of the GSMF.  The relatively low resolution of these simulations prevents us from injecting the BHs at lower masses, while going to higher masses worsens the agreement with the observations.  The GSMF does depend on the choice of accretion model, with constant $\beta$ models performing better in terms of the shape of the GSMF.  For convenience/simplicity we adopt the Booth \& Schaye (2009) fiducial model (constant $\beta=2$).
}
\end{figure*}

In Fig.~\ref{fig:gsmf_vary_bh} we explore the dependencies of the predicted galaxy stellar mass function on the various parameters of the subgrid AGN feedback model of \citet{Booth2009}.  In particular, we explore varying the feedback efficiency ($\epsilon_f$), the accretion model `boost' factor (and its possible dependence on density), the minimum halo mass for BH seed injection, and the AGN heating temperature (note that the effects of varying the mass of heated gas is explored in Fig.~\ref{fig:gsmf_nheat} in the main text).

The predictions are generally insensitive to the feedback efficiency and the AGN heating temperature, but {\it are} sensitive to the minimum halo mass into which BH seeds are injected as well as to the value of the accretion `boost' factor and its assumed dependence on the local gas density.  The relatively low resolution of these simulations prevents us from injecting the BHs at lower masses, while going to higher masses worsens the agreement with the observations.  In terms of the accretion boost factor, models with a dependence on the local gas density (i.e., constant $\beta$ models) tend to perform better than constant $\alpha$ models in terms of the shape of the GSMF.  We could therefore adjust $\beta$ to help better reproduce the GSMF.  However, we have instead elected to vary the mass of heated gas and leave $\beta=2$ as in \citet{Booth2009}.

\begin{figure}
\includegraphics[width=0.98\columnwidth]{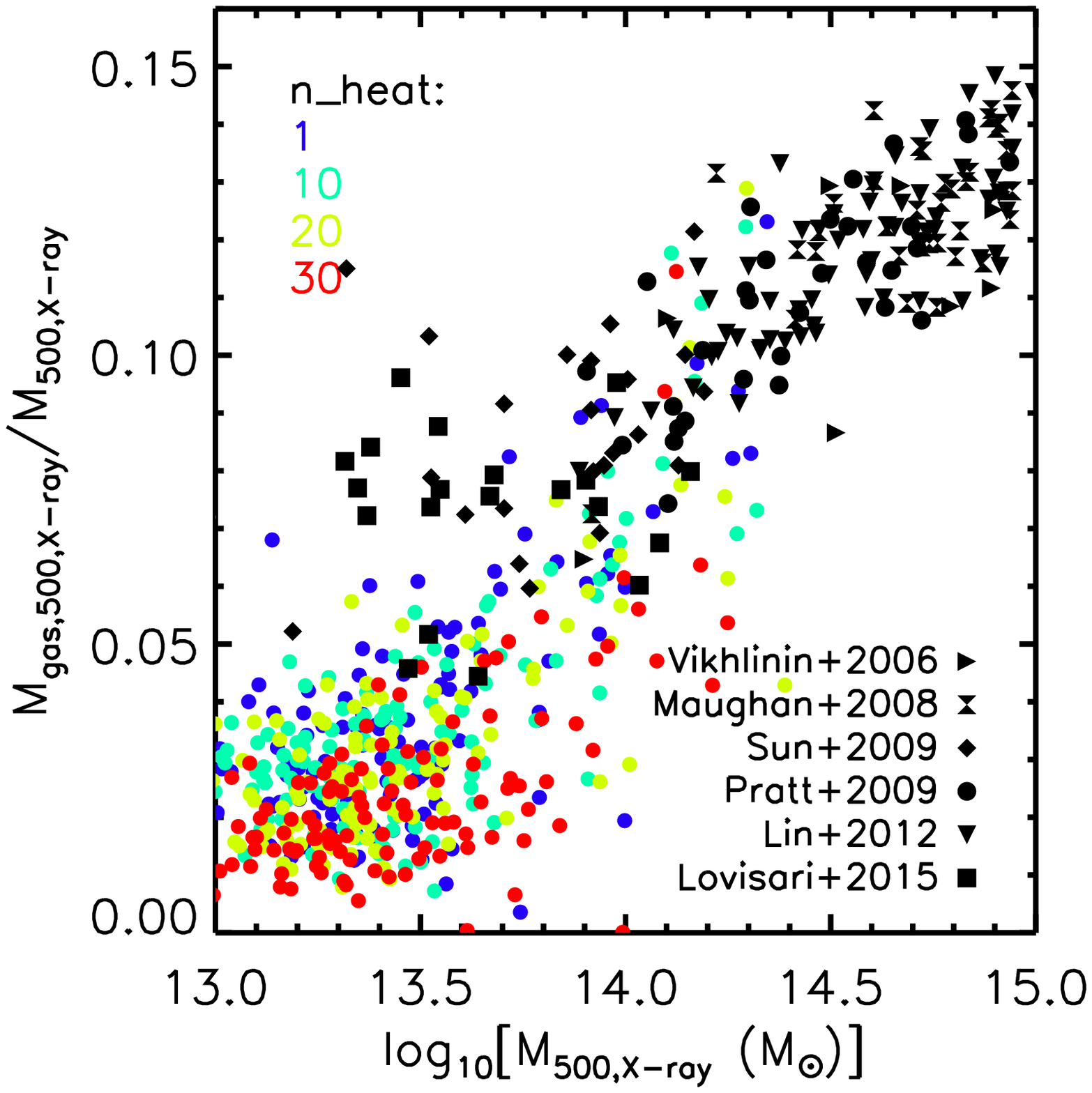}
\caption{\label{fig:fgas_vary_nheat}
The effect of varying the mass of gas heated (characterised by $n_{\rm heat}$, the number of gas particles heated per feedback event) by AGN feedback on the local gas mass fraction$-$halo mass relation.  Here we adopt a wind velocity of 300 km/s for feedback from star formation and an AGN heating temperature of $\Delta T_{\rm heat}=10^8$ K.  Varying the mass of heated gas has a modest effect on the gas fractions of low-mass systems, while leaving the gas fractions of high-mass systems relatively unaffected.  Note that even changing the heated gas mass by a factor of 30 has a smaller effect on the gas fractions than varying the heating temperature by only 0.2 dex (see Fig.~\ref{fig:fgas_vary_theat}).
}
\end{figure}

In Fig.~\ref{fig:fgas_vary_nheat} we show the effect of varying the mass of heated gas on the gas mass fractions of groups and clusters.  At high masses, where the gas fractions are rising steadily towards the universal mean, the effect of varying the mass of heated gas is mininal.  This is because, to zeroth order, whether gas remains bound to the system is set by the ratio of $\Delta T_{\rm heat}/T_{\rm vir}$.  If the heating temperature is relatively low, gas will not be ejected from the system no matter how much is heated.  On the other hand, if the heating temperature is sufficiently high to result in significant expulsion (as for the gas in systems with $M_{500}\la10^{14} \ {\rm M}_\odot$ here), then the choice of heated gas mass does have a slight effect on the gas fractions.  But note that the sensitivity to the choice of the mass of heated gas is much lower than to the choice of the heating temperature: a change of a factor 30 in the heated gas mass affects the gas fractions at about the same level (or slightly less than) as a change of only 0.2 dex in the heating temperature.  We have therefore elected to fix the heated mass of gas using the knee of the GSMF and to calibrate the gas fractions using the heating temperature.

\section{Importance of aperture stellar masses}

\begin{figure}
\includegraphics[width=0.98\columnwidth]{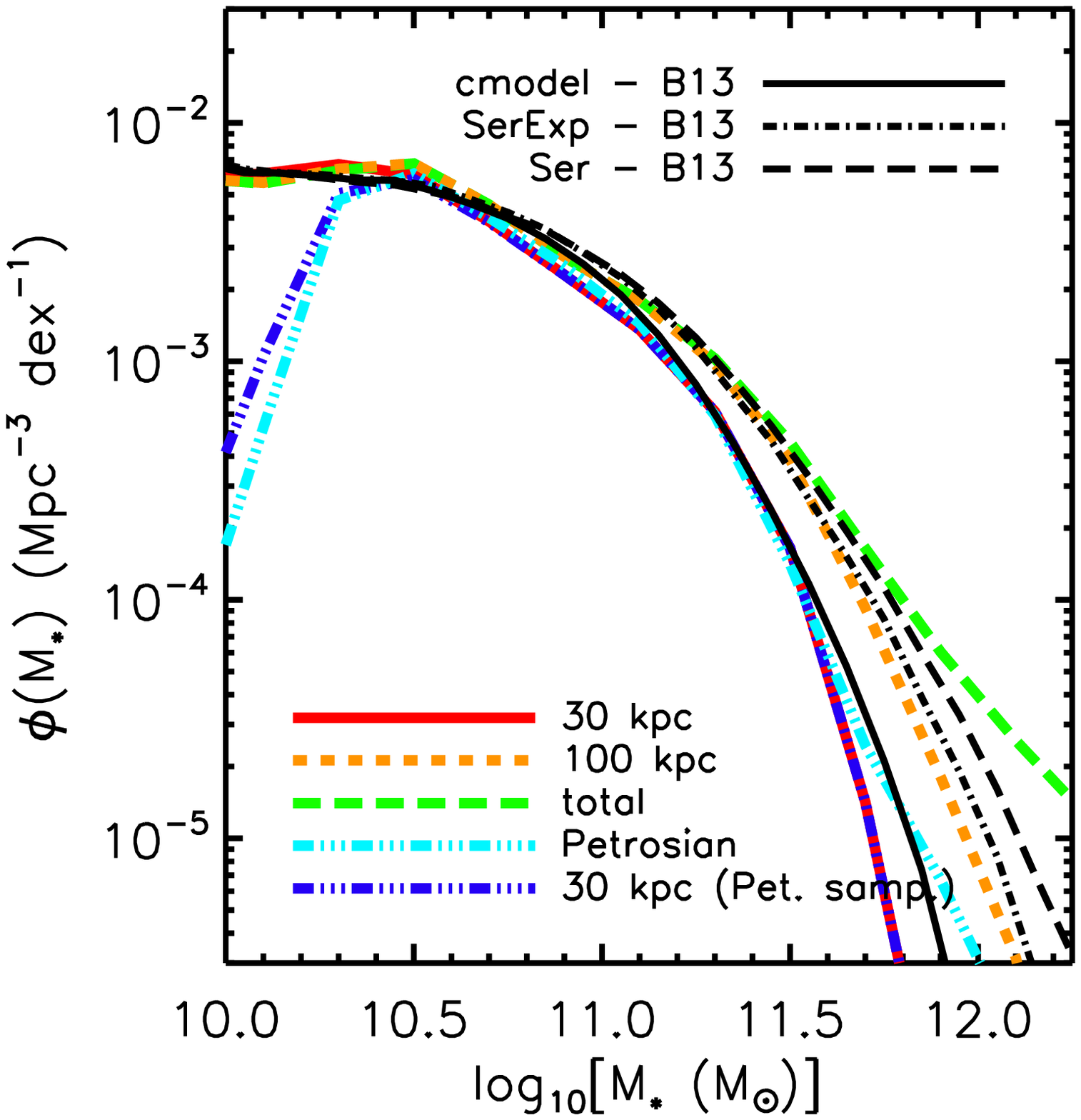}
\caption{\label{fig:gsmf_ap}
Dependence of the predicted present-day galaxy stellar mass function on the adopted aperture.  Shown are the results for 3-D apertures of 30 and 100 physical kpc (solid red and short-dashed orange curves), as well as the total bound stellar mass (long-dashed green curve).  The dot-dashed cyan curve shows the resulting GSMF when an observationally-motivated 2-D Petrosian analysis is applied to the simulated galaxies.  The dot-dashed blue curve shows the GSMF when we use the same galaxy selection criteria as in the Petrosian analysis but instead use a 3-D 30 kpc aperture (see text for details).   Adopting a 3-D 30 kpc aperture results in a GSMF that is quite similar to that derived by employing a Petrosian analysis, in agreement with the findings of \citet{Schaye2015}.  For comparison, we show the SDSS galaxy stellar mass functions measured by \citet{Bernardi2013} assuming different parametric forms for the light profile, which has a large effect at the massive end due to the presence of an extended component (i.e., intracluster light).  The `cmodel' case corresponds to typical SDSS `pipeline' results.  
}
\end{figure}

In Fig.~\ref{fig:gsmf_ap} we show the dependence of the predicted galaxy stellar mass function on the adopted aperture.  Shown are the results for apertures of 30 and 100 physical kpc (solid red and short-dashed orange curves), as well as the total bound stellar mass (long-dashed green curve).  The choice of aperture becomes significant for $\log_{10}{M_*/{\rm M}_\odot} \ga 10.7$, as also found by \citet{Schaye2015}.  

Observationally, one does not typically measure the luminosity/stellar mass within a fixed 3-D radius but instead either defines the luminosity/mass within some adapative 2-D radius chosen to enclose a certain fraction of the total light (as in the Petrosian system) or, alternatively, a parametric model is fitted to the light profile and integrated out to some sufficiently large radius to get a converged total luminosity/mass.  In the case of SDSS-based studies, both Petrosian and so-called `cmodel' luminosities (the cmodel corresponds to the best-fit linear combination of an exponential + de Vaucouleurs model to the light profiles) are widely used.  \citet{Bernardi2013} have shown that these two methods give very similar results.  \citet{Schaye2015} have shown that a 3-D 30 kpc aperture yields stellar mass estimates for the simulated galaxies that are very similar to those derived by applying a Petrosian analysis to 2-D images of the simulated galaxies.  Here we repeat this test for the \calsim~simulations.

The dot-dashed cyan curve in Fig.~\ref{fig:gsmf_ap} represents the results of applying a Petrosian analysis of the simulated galaxies.  Specifically, we compute stellar surface mass density profiles for each simulated galaxy and define the Petrosian radius as that which the local surface mass density is 0.2 times the mean surface mass density within that radius.  The associated Petrosian stellar mass is derived by summing the mass within twice the Petrosian radius.  The correspondence between the Petrosian-based GSMF and that derived using a 30 kpc aperture is quite good, in agreement with the results of \citet{Schaye2015}.  There is a small departure in the results of the two analyses at the largest masses, where adopting the Petrosian estimate results in a somewhat higher stellar mass at fixed abundance (or higher abundance at fixed stellar mass).  At low stellar masses of $\log_{10}{M_*/{\rm M}_\odot} \la 10.4$ there is a strong decline but this is artifical; we have found that we cannot reliably estimate the Petrosian radius for systems with less than $\sim50$ stellar particles, so we just exclude these systems from our Petrosian analysis.  The dot-dashed blue curve shows the GSMF when we apply the same selection to our 30 kpc analysis, which shows good agreement with the Petrosian-based results.

It should be noted, however, that \citet{Bernardi2013} (see also \citealt{Kravtsov2014}) have shown that both Petrosian and cmodel luminosities typically underestimate the total light of the most massive galaxies, which are preferentially found at the centers of galaxy groups and clusters.  Such systems often have an important extended component (commonly referred to as `intracluster light') that is not captured well by the cmodel or Petrosian estimates.  \citet{Bernardi2013} have explored the effect of adopting alternative parameterisations of the stellar light profile on the total stellar mass estimation.  The most flexible/accurate parametric model they consider is a combined Sersic+exponential model (SerExp).  When adopting this parameterisation they indeed find that the stellar masses are boosted, by up to 0.2 dex (compare the solid and dashed black curves in Fig.~\ref{fig:gsmf_ap}).   We find a similar boost for the simulated galaxies when we apply a larger physical aperture of 100 kpc, which is comparable to the half-light radii of the most massive systems when the ICL component is included in the fit (e.g., \citealt{Stott2011}).

\section{Resolution study}

Here we examine the numerical convergence of the baryon content of haloes (see Fig.~\ref{fig:res}).  For this test, we compare three simulations run in 100 Mpc/$h$ boxes.  One is run at the same resolution as the production runs presented in the main paper (L100N256, equivalent to L400N1024).  The other two are run with 8 times better mass resolution (L100N512, which we refer to as `high res.').   For one of the high-res. runs we have left the subgrid feedback parameters unchanged with respect to the fiducial resolution model, representing a `strong' convergence test in the terminology of \cite{Schaye2015}.  In the other high-res. run we have adjusted the feedback parameters (discussed below) to re-establish a virtually identical fit to the galaxy stellar mass function as obtained in the fiducial resolution model, representing a `weak' convergence test.  

In the strong convergence test, when we increase the mass resolution by a factor of 8, we also increase the number of particles heated by AGN by this same factor, therefore preserving the {\it mass} of gas that is heated.  Furthermore, black holes are seeded in haloes with a minimum number of FOF particles equal to 800 instead of 100, preserving the minimum halo mass where BHs are injected.  We also increase the number of SPH smoothing neighbours by a factor of 8 (to 384), so that gas properties are determined by smoothing over a similar mass/volume as in the fiducial resolution case.  (This also results in the metal enrichment being distributed over a similar mass/volume in the fiducial and high-res. simulations, as the enrichment is distributed over the SPH smoothing kernel.)  The AGN heating temperature and feedback efficiency are the same in both the fiducial and high-res. simulations, as are the stellar feedback wind velocity and mass loading.  

Note that it is perhaps more traditional to fix the number of resolution elements (particles) in the kernel rather than masses/volumes in convergence studies.  This is reasonable in the context of dark matter only simulations, where (scale-free) gravity is the only relevant force.  However, in hydrodynamical simulations many critical scales exist, the most important of which are the scales associated with subgrid feedback.  If the subgrid physics is tied directly to the resolution of the simulation (e.g., feedback heats a fixed number of resolution elements), then by changing the resolution of the simulation we are changing these scales.  If one changes these important scales, then one should not necessarily expect to achieve good convergence, as the energy is being injected in a different way.  This motivates us to fix the physical feedback scales when changing the resolution of our simulations.  Note that it has previously been explicitly demonstrated that fixing the physical scales (mass/volume) associated with AGN feedback results in much better strong convergence than that obtained by fixing the number of resolution elements subjected to AGN feedback \citep{Bourne2015}, which is what one expects based on the above discussion.

A comparison of the solid and dashed curves in Fig.~\ref{fig:res} suggests that the strong convergence is relatively good in the simulations.  This is particularly the case at high masses ($\log_{10}[M_{200}/{\rm M}_\odot] > 13.0$), where both the stellar and baryon fractions do not change by more than $\approx 10\%$ when changing the resolution.  However, at lower masses the convergence is less good, in that by increasing the resolution the stellar masses of the lowest-mass ($\log_{10}[M_{200}/{\rm M}_\odot] \la 12.0$) haloes increases by up to 40\%.  This is associated with a significant decrease in the gas fractions at slightly higher halo masses ($\log_{10}[M_{200}/{\rm M}_\odot] \approx 12.0-12.5$).  The increase in the stellar mass fractions of the lowest-mass haloes is not unexpected, since one requires several generations of star particles to be formed before stellar feedback becomes effective (e.g., \citealt{McCarthy2012,Schaye2015}).

One can increase the efficiency of the feedback to restore the agreement with the observed stellar mass fractions at the lowest masses (dot-dashed red curve).  To achieve this, we have simply increased the stellar feedback mass-loading (from 2 to 6) and lowered the minimum halo mass for BH seeding (to a factor of 8 below the fiducial resolution run).  Restoring this agreement to the stellar mass fractions also yields gas fractions for the high-mass haloes ($\log{[M_{200}/{\rm M}_\odot]} > 13.0$) in excellent agreement with the fiducial resolution.  However, the predicted gas fractions of lower-mass haloes ($\log{[M_{200}/{\rm M}_\odot]} \approx 12.0-12.5$) are still lower than that for the fiducial resolution run.  We speculate that this is because the increased feedback has resulted in more gas being ejected from these haloes, but that the increased feedback is insufficient to induce extra ejection from more massive group/cluster haloes.

We conclude that while the strong convergence is already quite good for massive haloes, this is not the case for lower-mass haloes.  While we are interested mainly in high-mass haloes and large-scale structure, it is also important to get the stellar mass fractions of lower-mass systems correct, as galaxies are often used as tracers of large-scale structure.  We have shown that a simple re-adjustment of the feedback parameters can re-establish the agreement with the stellar mass fractions of low-mass haloes and that this actually improves the convergence of the baryon fractions of massive haloes, so that it is virtually identical to those for our observationally-calibrated fiducial resolution run.

\begin{figure}
\includegraphics[width=0.98\columnwidth]{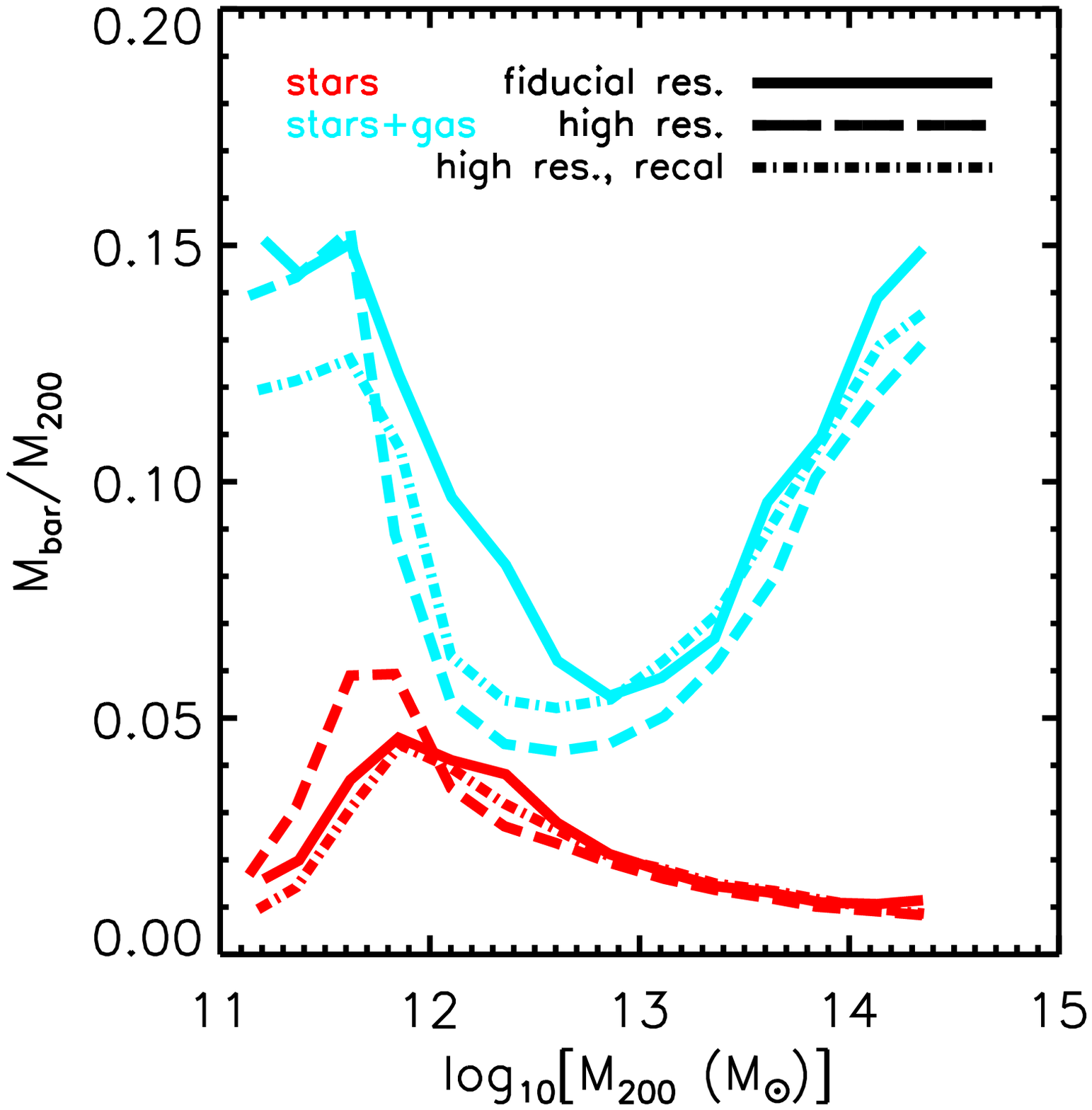}
\caption{\label{fig:res}
The total stellar (red curves) and baryon (cyan curves) mass fractions of central haloes vs. total halo mass as a function of resolution.  The solid curves correspond to the fiducial resolution used in the production runs presented in the main study.  The dashed curves correspond to a run with a factor of 8 better mass resolution while leaving the feedback parameters unchanged (see text).  This represents a strong convergence test.  The dot-dashed curves correspond to a high-resolution simulation where the stellar and AGN feedback have been recalibrated to reproduce the observed galaxy stellar mass function similarly well as in the fiducial resolution run.  
}
\end{figure}

\label{lastpage}

\end{document}